%% file: article.tex
\documentclass[12pt]{spieman}  % 12pt font required by SPIE;
\usepackage{amsmath,amsfonts,amssymb}
\usepackage{graphicx}
\usepackage{setspace}
\usepackage{tocloft}
\usepackage{framed,multirow}

\usepackage{amssymb}
\usepackage{latexsym}

\usepackage{url}
\usepackage{xcolor}
\usepackage{hyperref}

\usepackage{float}
\usepackage{amsmath,amssymb,amsfonts}
\usepackage{graphicx}
\usepackage{textcomp}
\usepackage{subcaption}
\usepackage{tikz}
\usepackage{lscape}
\usepackage{cite}
\usepackage{tabu} 
\usepackage{booktabs}% http://ctan.org/pkg/booktabs
\usepackage{dsfont}
\usepackage{algorithm}

\usepackage{algpseudocode}
\usetikzlibrary{tikzmark,calc,fit}

\title{Demystifying the Effect of Receptive Field Size in U-Net Models for Medical Image Segmentation}

\author[a]{Vincent Loos}
\author[b]{Rohit Pardasani}
\author[a,c,*]{Navchetan Awasthi}
\affil[a]{Faculty of Science, Mathematics and Computer Science,
Informatics Institute, University of Amsterdam, Amsterdam 1090 GH, The Netherlands}
\affil[b]{General Electric Healthcare, Bengaluru, Karnataka 560066, India}
\affil[c]{Department of Biomedical Engineering and Physics, Amsterdam UMC, Amsterdam 1081 HV, The Netherlands}

\cftpagenumbersoff{figure}
\cftpagenumbersoff{table} 
\begin{document} 
\maketitle

\begin{abstract}\\
{ \bf Purpose}: Medical image segmentation is a critical task in healthcare applications, and U-Nets have demonstrated promising results in this domain. This work delves into the understudied aspect of receptive field (RF) size and its impact on the U-Net and Attention U-Net architectures used for medical imaging segmentation. \\
{ \bf Approach}: This work explores several critical elements including the relationship between RF size, characteristics of the region of interest, and model performance, as well as the balance between RF size and computational costs for U-Net and Attention U-Net methods for different datasets. This work also proposes a mathematical notation for representing the theoretical receptive field (TRF) of a given layer in a network and proposes two new metrics namely - effective receptive field (ERF) rate and the Object rate to quantify the fraction
of significantly contributing pixels within the ERF against the TRF area and assessing the
relative size of the segmentation object compared to the TRF size respectively. \\
{ \bf Results}: The results demonstrate that there exists an optimal TRF size that successfully strikes a balance between capturing a wider global context and maintaining computational efficiency, thereby optimizing model performance. Interestingly, a distinct correlation is observed between the data complexity and the required TRF size; segmentation based solely on contrast achieved peak performance even with smaller TRF sizes, whereas more complex segmentation tasks necessitated larger TRFs. Attention U-Net models consistently outperformed their U-Net counterparts, highlighting the value of attention mechanisms regardless of TRF size. \\
{ \bf Conclusions}: These novel insights present an invaluable resource for developing more efficient U-Net-based architectures for medical imaging and pave the way for future exploration of other segmentation architectures. A tool is also developed that calculates the TRF for a U-Net (and Attention U-Net) model, and also
suggest an appropriate TRF size for a given model and dataset.
\end{abstract}

% Include a list of up to six keywords after the abstract

% Include email contact information for corresponding author
{\noindent \footnotesize\textbf{*}Navchetan Awasthi,  \linkable{n.awasthi@uva.nl} }

\begin{spacing}{2}   % use double spacing for rest of manuscript
\section{Introduction}

% start of new introduction
Medical imaging, a cornerstone of modern healthcare, provides non-invasive means for diagnosing and monitoring a wide range of diseases. However, the interpretation of medical images often requires expert knowledge and can be time-consuming, leading to a growing interest in automated analysis methods \cite{litjens2017}.

\begin{figure*} [ht!]
    \centering
    \includegraphics[height=10 cm]{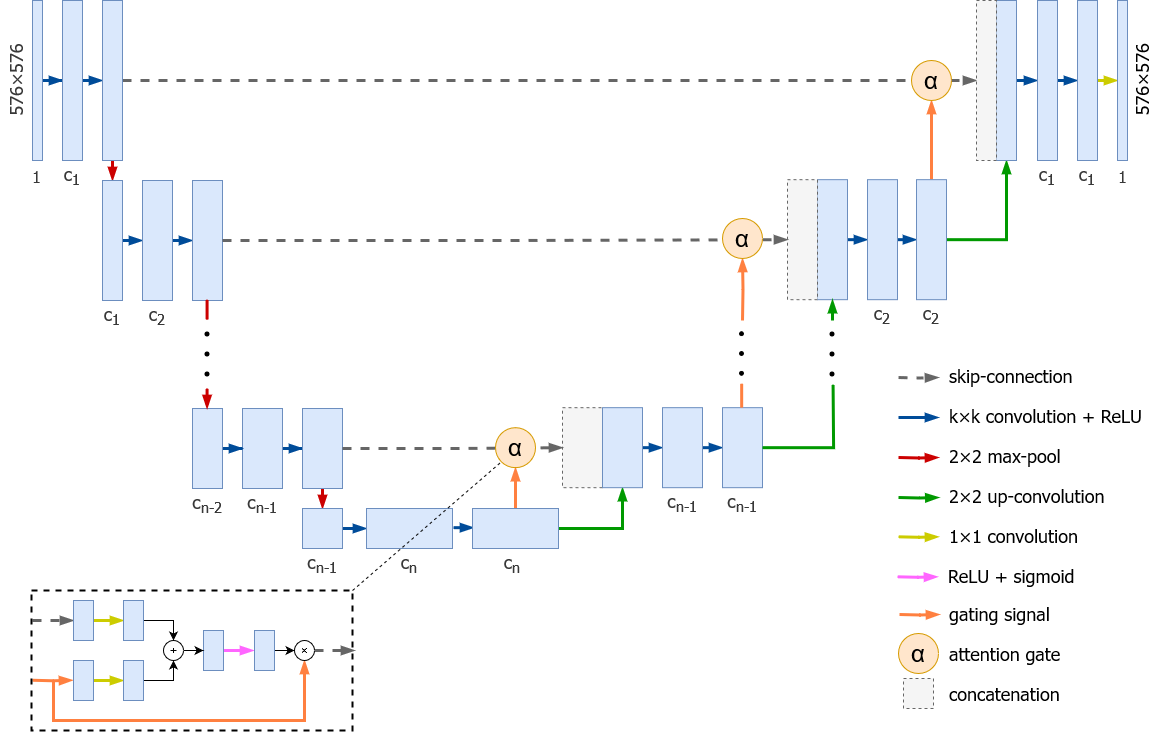}
    \caption{Variable attention U-Net in which the depth ($n$), kernel size of the convolution layers ($k$), and number of channels ($c$) can be tuned to alter the size of the TRF. It can be converted to a regular U-Net by simply removing the attention gates and gating signals.}
    \label{var-unet}
\end{figure*} 

Semantic segmentation, a key task in computer vision, plays a crucial role in this context. It involves the categorization of pixels in an image into predefined classes, enabling the delineation of anatomical structures and pathological regions in medical images \cite{hesamian2019deep}. The U-Net architecture, a convolutional neural network (CNN) designed specifically for biomedical image segmentation, has emerged as a popular choice for semantic segmentation tasks in medical imaging \cite{ronneberger2015unet}. As illustrated in Figure \ref{var-unet}, it employs an encoder-decoder structure. The encoder progressively reduces the spatial dimensionality while increasing the feature representation, capturing the global context of the image. The decoder, on the other hand, gradually recovers the spatial information, enabling precise localization \cite{williams2023unified}. The U-Net is renowned for its accuracy in semantic segmentation tasks \cite{ronneberger2015unet}. An extended version, Attention U-Net, integrates an attention mechanism to enhance feature capturing to improve overall performance \cite{oktay2018attentionunet}.

\begin{figure}[h]
    \centering
    \includegraphics[width=0.6\linewidth]{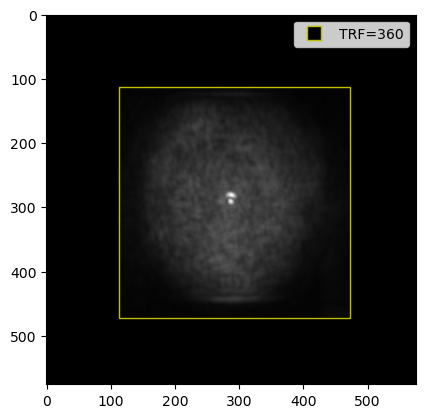}
    \caption{Example of Theoretical Receptive Field (TRF) and Effective Receptive Field (ERF) in an image. The yellow square denotes the TRF, the maximum input area influencing the output pixel located at the centre of the square. The gray pixels, representing the ERF, show the actual input area affecting a neuron's activation, with intensity indicating the impact level.}
    \label{fig:trf-erf-example}
\end{figure}

Within these networks, the concept of the receptive field (RF) is crucial. It refers to the region in the input space that affects a feature in a CNN \cite{luo2017understanding}. There are two kinds of receptive fields: the theoretical receptive field (TRF) and the effective receptive field (ERF). The TRF is defined as the maximum region of the input image that influences a specific pixel of the output, considering only the receptive field from the preceding layers that are relevant to the current layer \cite{araujo2019computing}. This is in contrast to the ERF, which is the actual region of the input image that contributes to the activation of a particular neuron in the network, taking into account the impact of operations such as pooling \cite{luo2017understanding}. An example of the TRF and ERF is illustrated in Figure \ref{fig:trf-erf-example}.

% Despite the importance of the RF size, its impact on the performance of U-Net-based architectures, particularly in the context of medical imaging, has not been thoroughly investigated.

% This paper aims to fill this gap by conducting an in-depth study of the influence of RF size on the performance of U-Net-based architectures in semantic segmentation for ultrasound imaging tasks. We propose a series of U-Net and Attention U-Net architectures with varying TRF sizes and evaluate their performance on several benchmark ultrasound segmentation datasets. Our ultimate goal is to provide valuable insights that can guide the design of U-Net-based architectures for ultrasound image segmentation, with the size of the receptive field as a critical parameter.
Previous studies have started investigating the role of RF size on U-Net performance for image segmentation tasks, but not all aspects have been explored. In one such study, \cite{behboodi2020} focused on ultrasound image segmentation, demonstrating that the RF size has a more critical role than the network's depth or the number of parameters. They suggested that a computationally efficient shallow network could replace a deep one without performance loss by manipulating the RF size. However, their study was limited to the U-Net architecture and a single dataset, comparing only a deep and a shallow network, leaving room for a more exhaustive investigation.

In another study, \cite{sytwu2022} delved into the influence of RF size and network complexity on a CNN's performance for transmission electron microscopy (TEM) image analysis. They found that the  RF size's influence varied with TEM image resolution and contrast characteristics. For low-resolution TEM images, where contrast is crucial, RF size had a minimal influence. But for high-resolution TEM images, where identification is less dependent on contrast changes, the RF size was vital, especially for low contrast images. However, they only considered TEM images and regular U-Nets, leaving the applicability of these findings to other medical imaging tasks unexplored.

Our study builds upon these insights by examining the influence of RF size on U-Net and Attention U-Net architectures across multiple medical image segmentation datasets with certain characteristics in the region of interest (RoI). We compare ten different U-Net architectures shown in Table-\ref{tbl:configs} with varied RF sizes and equal total parameters, thus isolating RF size's impact on performance while taking into consideration specific dataset characteristics. Moreover, we repeat the experiments on eight different synthetic datasets and six real-world medical datasets. We also extend our investigation to the Attention U-Net architecture, thereby expanding the study beyond regular U-Net architecture. Our aim is to offer critical insights for U-Net-based architectures' design, considering RF size as a key parameter.

Specifically, this paper makes the following contributions to the field of medical image segmentation with U-Nets:
\begin{enumerate}
    \item We provide a comprehensive analysis of the role of the RF size in the performance of U-Net and Attention U-Net architectures, demonstrating its significance in capturing the necessary context for accurate segmentation.
    \item We propose a mathematical notation to represent the TRF at a given layer within a network, utilizing a 4-dimensional tensor and provide the mathematical framework for calculating TRF for the different deep learning layers.
    \item We propose two new metrics called the \textit{ERF rate} and the \textit{Object rate} to quantify the fraction of significantly contributing pixels within the ERF against the TRF area and assessing the relative size of the segmentation object compared to the TRF size respectively.
    \item We explore the trade-off between RF size and computational cost for a variety of medical imaging datasets and synthetic datasets.
    \item We compare the performance of U-Net and Attention U-Net architectures for the same RF size, highlighting the effectiveness of the attention mechanism in improving the model's overall performance.
    \item We present a nuanced analysis of the performance trends across datasets with different characteristics in the RoI, particularly its size and contrast to the surrounding area.
    \item We provide a tool that calculates the TRF for a U-Net (and Attention U-Net) model, and also suggest an appropriate TRF size for a given model and dataset.
\end{enumerate}
\section{Methodology}
This study explores the role of the receptive field in the performance of U-Net and Attention U-Net models in semantic segmentation tasks. Through a series of experiments with varying TRF sizes, we evaluated these models on diverse range of datasets. In this section, we further provide comprehensive descriptions of the model architectures, as well as the TRF and ERF computation.

\subsection{U-Net Design and TRF Tuning}
The configuration of the hyper-parameters of a U-Net model significantly impacts the size of its TRF. As established by \cite{yu2016multiscale}, the TRF size is primarily determined by the number of pooling layers and the convolutional kernel sizes. To elaborate, Figure \ref{var-unet} illustrates a variable attention U-Net diagram, where the TRF size can be adjusted in two different ways. Firstly, when the vertical depth ($d$) of the network is increased, one encoder and one decoder block is added before and after the bottleneck respectively. This increases the number of pooling layers and therefore increases the TRF size. Changing the network depth on its own does not result in a significant impact on the model's performance \cite{behboodi2020}. Secondly, the TRF size can be varied by changing the kernel sizes of the convolutional layers within the network. The mathematical details of the effects of various layers on the TRF size are provided in Section \ref{sec:compute_trf}.

It should be noted that adjusting these hyper-parameters also impacts the total number of parameters in the model. To ensure a fair comparison between the performance of various configurations, the total number of parameters must remain approximately equal. According to \cite{SARIGUL2019279}, this can be achieved by modifying the number of output channels in each convolution layer within the network blocks. Table \ref{tbl:configs} provides an overview of all configurations utilized in this study. It is important to mention that the parameter count is based on the standard U-Net architecture. The Attention U-Net introduces additional parameters due to the inclusion of an attention block at each layer. However, this increase varies approximately on the order of 100,000, which is relatively insignificant and can be considered negligible in this context.

% The U-Net model's hyper-parameters configuration, especially the number of pooling layers and convolutional kernel sizes, significantly influence its TRF size \cite{yu2016multiscale}. The TRF can be adjusted by increasing network depth, i.e. adding encoder and decoder blocks, or modifying convolutional kernel sizes \cite{behboodi2020}. While adjusting these parameters also impacts the model's total parameter count, equal comparisons can be maintained by modifying output channel numbers in each convolution layer \cite{SARIGUL2019279}. The Attention U-Net introduces additional parameters of around the order of 100,000, but the increase is negligible in this context. Table \ref{tbl:configs} provides an overview of all configurations utilized in this study.

\begin{table}[h]
\centering
\caption{All U-Net configurations with different TRF sizes. The TRF size is influenced by the convolutional kernel size ($k$) and the vertical depth of the network ($d$).\label{tbl:configs}}
\resizebox{0.48\textwidth}{!}{
\begin{tabular}{@{}cccc@{}}
\toprule
\multicolumn{1}{l}{\textbf{TRF size}} & \multicolumn{1}{l}{\textbf{$k,d$}} & \multicolumn{1}{l}{\textbf{Out channels per layer}} & \multicolumn{1}{l}{\textbf{\# Parameters}} \\ \midrule
54 & 3, 2 & [230, 456, 765, 1245] & 31,013,720 \\
100 & 3, 3 & [145, 256, 512, 1024] & 31,012,268 \\
146 & 3, 4 & [133, 244, 355, 791] & 31,032,960 \\
204 & 4, 3 & [64, 128, 256, 512, 1024] & 31,042,369 \\
230 & 3, 6 & [63, 170, 256, 512] & 31,031,345 \\
298 & 4, 4 & [25, 44, 110, 451, 756] & 31,043,816 \\
360 & 3, 8 & [47, 83, 180, 360] & 31,062,482 \\
412 & 5, 3 & [63, 64, 115, 255, 512, 1024] & 31,043,945 \\
486 & 4, 6 & [28, 58, 146, 270, 510] & 31,027,119 \\
570 & 4, 7 & [24, 55, 101, 223, 481] & 31,041,124 \\ \bottomrule
\end{tabular}
}
\end{table}

\subsection{Computing the TRF}
\label{sec:compute_trf}
Formally, the TRF refers to the maximum region of the input image $\mathbf{X}\in[0,1]^{h\times w}$ that potentially influences a specific pixel in the output layer. To represent the TRF at layer $d$ in a U-Net architecture of depth $D$, we introduce a 4-dimensional tensor $\mathbf{T}^{(d)}\in\mathds{R}^{h\times w\times 2\times 2}$. Here, the first two dimensions correspond to the $y$ and $x$ axes of the input image, respectively, while the third and fourth dimensions represent the top-left (\textit{t}-\textit{l}) and bottom-right (\textit{b}-\textit{r}) coordinates of the TRF at layer $d$. For a given pixel located at position $(i,j)$ in the output layer $D$, the TRF can be expressed as a $2\times 2$ matrix in which the first row corresponds to the top-left corner of the TRF, and the second row corresponds to the bottom-right corner of the TRF:

\begin{equation}\label{eq:trf_out}
    \mathbf{T}_{i,j}^{(D)} = \begin{bmatrix}t_{i,j}^{(D)} & l_{i,j}^{(D)} \\ b_{i,j}^{(D)} & r_{i,j}^{(D)}    \end{bmatrix}
\end{equation}
Empirically, it has been observed that all pixels have an equal TRF size in the output layer, except those located around the border because of the padded zeroes. Based on this observation, a single (maximum) TRF value can be assigned to the entire U-Net. In the remainder of this paper, we define the TRF size of a U-Net as the size of the TRF of the center pixel $(u,v)=(h/2,w/2)$ in the output layer:

\begin{equation}\label{eq:trf_size}
    \text{TRF} = \sqrt{\left(\mathbf{T}_{u,v,1,0}^{(D)}-\mathbf{T}_{u,v,0,0}^{(D)}\right)\cdot\left(\mathbf{T}_{u,v,1,1}^{(D)}-\mathbf{T}_{u,v,0,1}^{(D)}\right)}
\end{equation}

% \begin{figure} [h]
%     \centering
%     \includegraphics[width=0.6\linewidth]{images/trf_size_pixels.png}
%     \caption{TRF size of each pixel in the output image for the default U-Net configuration.}
%     \label{fig:trf-size-pixels}
% \end{figure} 
To compute the values of the TRF matrix in Equation \ref{eq:trf_out}, we traverse the network from the first to the final layer, tracking the TRF of each pixel at every layer based on the previous layer's pixels until reaching the output layer \cite{araujo2019computing}. Therefore, the TRF of the pixel at position $(i,j)$ in a layer at depth $d$ can be expressed as:

\begin{equation}\label{eq:trf_layer}
    \mathbf{T}_{i,j}^{(d)} =\begin{bmatrix}
        t_{i,j}^{(d)} & l_{i,j}^{(d)} \\ b_{i,j}^{(d)} & r_{i,j}^{(d)}
    \end{bmatrix}
\end{equation}

In the input layer $0$, the TRF of each pixel corresponds to the pixel itself:
\begin{equation}
    \mathbf{T}_{i,j}^{(0)} =\begin{bmatrix}
        i & j \\ i & j
    \end{bmatrix}
\end{equation}
The computation of the TRF in subsequent layers depends on the U-Net's configuration. Here are all possible layers that a U-Net may include.

\subsubsection{Convolution}
% A 2D convolution layer in the U-Net architecture applies a filter, also known as a kernel, to a 2D input image. The filter is convolved with the entire image in a sliding window fashion, computing the dot product between the filter and a local region of the input image at each position \cite{convolutions}.

% The kernel size ($k$) determines the size of the filter used for convolution. A larger kernel size allows the convolutional layer to capture more complex features and spatial structures, while a smaller kernel size allows the layer to capture more fine-grained details. The stride ($s$) determines the number of pixels by which the kernel is shifted at each step of the convolution operation. Padding ($p$) is a technique used to ensure that the output feature map has the same spatial dimensions as the input image. In this study, the padding was set to \textit{same}, which means that the input image is padded with zeros around the edges so that the output feature map has the same height and width as the input image. This technique is critical because it enables the U-Net to perform convolution operations on the border pixels of the input image, which are critical for capturing features that span across the image's edges.
In a 2D convolution layer, a filter or kernel is applied to a 2D image, performing a dot product at each position \cite{convolutions}. The kernel size ($k$) impacts the detail level captured, while the stride ($s$) affects the kernel shift amount. Padding ($p$), set to `same' in this study, ensures the output feature map matches the input image dimensions, permitting edge-based convolution operations.

If the padding is set to \textit{same}, the number of arrays that must be padded on every side of the $\mathbf{T}^{(d-1)}$ tensor to simulate a convolution while maintaining the previous layer's dimensions can be calculated. For a $h\times w$ layer, the padding values along the $y$ and $x$ axes are computed as follows:
\begin{align}
    p_y &= \left\lfloor\frac{(h - 1) \cdot s + k - h}{2}\right\rfloor\\
    p_x &= \left\lfloor\frac{(w - 1) \cdot s + k - w}{2}\right\rfloor
\end{align}

Therefore, along the first and second axes of the 4-dimensional tensor $\mathbf{T}^{(d-1)}$, the tensor is padded with $p_y$ and $p_x$, 2-dimensional tensors that contain the same values as the edges along the first and second axes of $\mathbf{T}^{(d-1)}$. Let $\mathbf{P}^{(d-1)}$ denote this padded tensor. For each position $(i,j)$, the top-left and bottom-right pixels from the previous layer's TRF can be fetched from $\mathbf{P}^{(d-1)}$ at the indices $(i\cdot s,j\cdot s)$ and $(i\cdot s+k-1,j\cdot s+k-1)$, respectively. Thus, the TRF at position $(i,j)$ for a convolutional layer at depth $d$ can be denoted as:
\begin{equation}
    \mathbf{T}_{i,j}^{(d)}=
    \begin{bmatrix}
        \mathbf{P}_{i\cdot s,j\cdot s,0,0}^{(d-1)} &
        \mathbf{P}_{i\cdot s,j\cdot s,0,1}^{(d-1)} \\\\
        \mathbf{P}_{i\cdot s+k-1,j\cdot s+k-1,1,0}^{(d-1)} &
        \mathbf{P}_{i\cdot s+k-1,j\cdot s+k-1,1,1}^{(d-1)}
    \end{bmatrix}
\end{equation}

\subsubsection{Max pooling}
% 2D max pooling is a technique for feature map reduction. It involves applying a rectangular kernel to the feature map, selecting the maximum value within each region, and producing a smaller feature map \cite{convolutions}. The size of the rectangular kernel ($k$) specifies the size of the window that slides over the input feature map during the pooling operation. The stride ($s$) determines the amount of movement of the kernel window between two consecutive pooling operations. Typically, the stride is set to the kernel size ($s=k$) as done in this study. This design choice has the advantage of ensuring that the kernel is positioned right next to the previous operation's position, thereby simplifying the computation of the TRF.
2D max pooling is a feature map reduction method where a rectangular kernel selects maximum values within regions, creating a smaller feature map \cite{convolutions}. The kernel size ($k$) defines the sliding window size over the input, and the stride ($s$) -- in our study equal to $k$ in order to simplify the computation -- controls the window's movement.

For a given position $(i,j)$, the topmost and leftmost pixels from the previous layer's TRF can be accessed from the $\mathbf{T}^{(d-1)}$ tensor at the index of $(i\cdot k,j\cdot k)$, while the bottom-most and rightmost pixels can be accessed at the index of $(i\cdot k+k-1,j\cdot k+k-1)$. As such, the TRF at position $(i,j)$ for a max pooling layer at depth $d$ can be expressed as follows:
\begin{equation}
    \mathbf{T}_{i,j}^{(d)}=
    \begin{bmatrix}
        \mathbf{T}_{i\cdot k,j\cdot k,0,0}^{(d-1)} &
        \mathbf{T}_{i\cdot k,j\cdot k,0,1}^{(d-1)} \\\\
        \mathbf{T}_{i\cdot k+k-1,j\cdot k+k-1,1,0}^{(d-1)} &
        \mathbf{T}_{i\cdot k+k-1,j\cdot k+k-1,1,1}^{(d-1)}   
    \end{bmatrix}
\end{equation}

\subsubsection{Upsampling}
Upsampling is a technique used to increase the spatial resolution of feature maps. In particular, it is implemented through transposed convolution or deconvolution, which is the reverse operation of convolution. During the transposed convolution operation, a kernel of size $k$ is applied to the input feature map to generate an output feature map with a higher spatial resolution. The stride $s$ determines the amount of shift in the output feature map for each kernel application \cite{convolutions}. When the stride is set to $k$, the size of the output feature map is equal to the size of the input feature map multiplied by the stride.

However, when the stride $s$ is different from the kernel size $k$, there may be overlaps in the values of the output feature map. Therefore, an iterative method is required to identify the corners of the TRF for each pixel in the output feature map. Specifically, Algorithm \ref{alg:upsample-trf} is applied to each pixel $(i,j)$ in the input map, computing the range in which the pixel is copied to the output feature map by multiplying the top and left indices with the stride and the bottom and right indices with the stride and then adding the kernel size. The algorithm then iterates over the pixels $(m,n)$ in the output feature map within this range. If there is no overlap, the indices from the previous layer at $(i,j)$ are simply copied. Otherwise, for the top and left of the TRF, the algorithm takes the minimum of the current index and a potentially overlapping index, while for the bottom and right TRF, it takes the maximum.
\begin{algorithm}[h]
\caption{TRF at layer $d$ and pixel $(i,j)$ after upsampling}\label{alg:upsample-trf}
\begin{algorithmic}
\For{$m \gets i \cdot s$ to $i \cdot s + k$}
\For{$n \gets j \cdot s$ to $j \cdot s + k$}\vspace{1mm}
\If{$\mathbf{T}_{m,n}^{(d)}$ \textbf{is} None} \Comment{If no overlap (yet)}\vspace{1mm}
\State $\mathbf{T}_{m,n}^{(d)} \gets \mathbf{T}_{i,j}^{(d-1)}$ \Comment{Copy from previous layer}
\State \textbf{continue} \Comment{Go to next pixel}
\EndIf\vspace{1mm}
\\\Comment{If there is overlap:}
\If{$\mathbf{T}_{i,j,0,0}^{(d-1)} \leq \mathbf{T}_{m,n,0,0}^{(d)}$} \Comment{Get smallest value}\vspace{1mm}
\State $ \mathbf{T}_{m,n,0,0}^{(d)} \gets \mathbf{T}_{i,j,0,0}^{(d-1)}$ \Comment{Update top}
\EndIf\vspace{1mm}
\If{$\mathbf{T}_{i,j,0,1}^{(d-1)} \leq \mathbf{T}_{m,n,0,1}^{(d)}$} \Comment{Get smallest value}\vspace{1mm}
\State $ \mathbf{T}_{m,n,0,1}^{(d)} \gets \mathbf{T}_{i,j,0,1}^{(d-1)}$ \Comment{Update left}
\EndIf\vspace{1mm}
\If{$\mathbf{T}_{i,j,1,0}^{(d-1)} \geq \mathbf{T}_{m,n,1,0}^{(d)}$} \Comment{Get largest value}\vspace{1mm}
\State $\mathbf{T}_{m,n,1,0}^{(d)} \gets \mathbf{T}_{i,j,1,0}^{(d-1)}$ \Comment{Update bottom}
\EndIf\vspace{1mm}
\If{$\mathbf{T}_{i,j,1,1}^{(d-1)} \geq \mathbf{T}_{m,n,1,1}^{(d)}$} \Comment{Get largest value}\vspace{1mm}
\State $\mathbf{T}_{m,n,1,1}^{(d)} \gets \mathbf{T}_{i,j,1,1}^{(d-1)}$ \Comment{Update right}
\EndIf
\EndFor
\EndFor
\end{algorithmic}
\end{algorithm}

\subsubsection{Concatenations}
Within the U-Net architecture, skip connections from layer $d-c$ are integrated into the decoder blocks by concatenating them with the output of the upsampling layer $d-1$ \cite{skip}. To achieve this, the TRF of the tensors being concatenated, denoted as $\mathbf{T}_{i,j}^{(d-c)}$ and $\mathbf{T}_{i,j}^{(d-1)}$, must first be determined. The TRF of each pixel after concatenation, denoted as $\mathbf{T}_{i,j}^{(d)}$, is obtained by selecting the lowest indices for the left and top of both TRFs, and the highest indices for the right and bottom of both TRFs. This approach ensures that the largest possible TRF is obtained.
\begin{align*}
    \mathbf{T}_{i,j,0,0}^{(d)} &= \min\left(\left\{\mathbf{T}_{i,j,0,0}^{(d-1)}, \mathbf{T}_{i,j,0,0}^{(d-c)}\right\}\right)\\[1mm]
    \mathbf{T}_{i,j,0,1}^{(d)} &= \min\left(\left\{\mathbf{T}_{i,j,0,1}^{(d-1)}, \mathbf{T}_{i,j,0,1}^{(d-c)}\right\}\right)\\[1mm]
    \mathbf{T}_{i,j,1,0}^{(d)} &= \max\left(\left\{\mathbf{T}_{i,j,1,0}^{(d-1)}, \mathbf{T}_{i,j,1,0}^{(d-c)}\right\}\right)\\[1mm]
    \mathbf{T}_{i,j,1,1}^{(d)} &= \max\left(\left\{\mathbf{T}_{i,j,1,1}^{(d-1)}, \mathbf{T}_{i,j,1,1}^{(d-c)}\right\}\right)
\end{align*}

\subsubsection{Activation functions}
While nonlinear activation functions like ReLU and sigmoid do affect the ERF by potentially reducing its size when certain parameters are set to zero \cite{Kuo2016}, they have no effect on the TRF, as these functions act element-wise on the previous layer. Therefore, in a layer $d$ with an activation function, it can be concluded that $\mathbf{T}_{i,j}^{(d)} = \mathbf{T}_{i,j}^{(d-1)}$.

\subsubsection{Attention gates}
Attention gates, a key component of the Attention U-Net architecture (illustrated in Figure \ref{var-unet}), receive input features from a layer denoted as $x'$, and a gating signal from a layer $g$ \cite{oktay2018attentionunet}. The inputs are then subjected to $1\times 1$ convolutions, followed by element-wise addition. At this point, the TRF is equivalent to the maximum range of the TRF of either input, as the TRF is not modified by the $1\times 1$ convolution. Next, a ReLU and sigmoid function are applied, which leave the TRF unchanged, as described in the previous section. Finally, element-wise multiplication is performed on the output, which results in the TRF being equivalent to the maximum range of the TRF of either input. As a result, the TRF size of a U-Net and Attention U-Net with the same depth and convolution kernel sizes are equivalent.

Therefore, similar to concatenations, the TRF of an attention gate $a$ is the maximal range of the TRF from its input features of layer $x'$ and the gating signal of layer $g$.
\begin{align*}
    \mathbf{T}_{i,j,0,0}^{(a)} &= \min\left(\left\{\mathbf{T}_{i,j,0,0}^{(x')}, \mathbf{T}_{i,j,0,0}^{(g)}\right\}\right)\\[1mm]
    \mathbf{T}_{i,j,0,1}^{(a)} &= \min\left(\left\{\mathbf{T}_{i,j,0,1}^{(x')}, \mathbf{T}_{i,j,0,1}^{(g)}\right\}\right)\\[1mm]
    \mathbf{T}_{i,j,1,0}^{(a)} &= \max\left(\left\{\mathbf{T}_{i,j,1,0}^{(x')}, \mathbf{T}_{i,j,1,0}^{(g)}\right\}\right)\\[1mm]
    \mathbf{T}_{i,j,1,1}^{(a)} &= \max\left(\left\{\mathbf{T}_{i,j,1,1}^{(x')}, \mathbf{T}_{i,j,1,1}^{(g)}\right\}\right)
\end{align*}

\subsection{Computing the ERF}
For each pixel $x_{i,j}$ in the input image $\mathbf{X}\in[0,1]^{h\times w}$, its impact on the center pixel of the output image $y_{h/2,w/2}$ is measured by computing the partial derivative of the center output pixel with respect to each input pixel $\partial y_{h/2,w/2}/\partial x_{i,j}$. This method quantifies how much $y_{h/2,w/2}$ changes if $x_{i,j}$ is changed by a small amount \cite{luo2017understanding}. For a TRF, the corresponding ERF ($\mathbf{E}\in\mathds{R}^{m\times n}$) can be expressed as a matrix as shown in equation \ref{eq:erf}.

\begin{equation}\label{eq:erf}
    \mathbf{E}=  
\begin{bmatrix}
\dfrac{\partial y_{h/2,w/2}}{\partial x_{t,l}} & \hdots & \dfrac{\partial y_{h/2,w/2}}{\partial x_{t,r}} \\
\vdots & \ddots & \vdots \\
\dfrac{\partial y_{h/2,w/2}}{\partial x_{b,l}} & \hdots & \dfrac{\partial y_{h/2,w/2}}{\partial x_{b,r}} 
\end{bmatrix}
\end{equation}
The actual computation of the ERF can be done easily with most deep learning frameworks by back-propagating the value of one certain output pixel to the entire input, and taking the $m\times n$ slice of the input at the position of the TRF.

\section{Experiment}

%%%%%%%%%%%%%%%%
% TRAINING %
%%%%%%%%%%%%%%%%
\subsection{Training protocol}
All the models were trained on a high-performance computing node featuring two Intel Xeon Platinum 8360Y CPUs and an NVIDIA A100 GPU with 40 GB of HBM2 memory. We used the PyTorch framework \cite{paszke2019pytorch} and employed Binary Cross-Entropy with Logits Loss as our loss function, with the Adam optimizer facilitating training due to its efficiency and minimal memory requirements \cite{zhang2018improved}.

The initial learning rate was set at $10^{-4}$, and a learning rate scheduling strategy was implemented to optimize learning. This strategy reduces the learning rate by 0.1 when the validation loss plateaus for four epochs, enabling more substantial updates in early training phases and smaller updates as the model nears convergence. Training lasted up to 200 epochs, with early stopping \cite{prechelt2002early} implemented to prevent overfitting. If the validation loss remained static over 20 consecutive epochs, training was ceased, and the parameters that achieved the lowest validation loss were saved.

\begin{figure*}[!ht]
\centering
\begin{tabular}{ccccccc}
 A & A large & A contour & A large contour & B & B large & B contour\\
 \includegraphics[width=2cm]{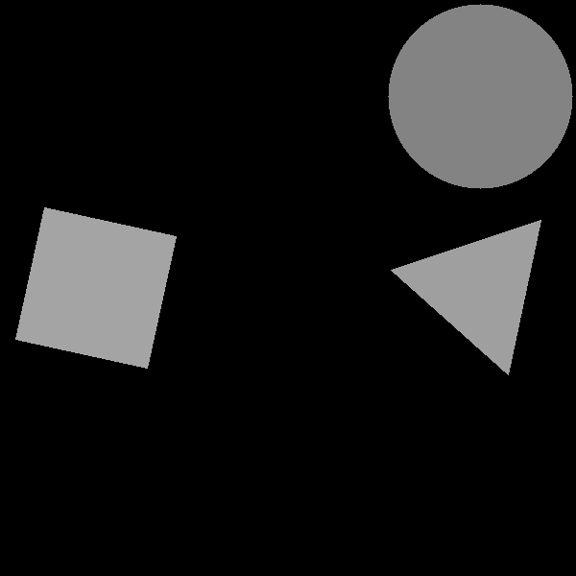} & \includegraphics[width=2cm]{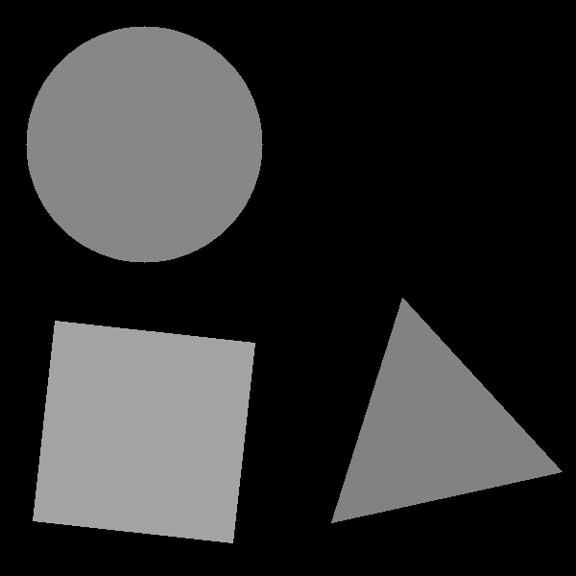} & \includegraphics[width=2cm]{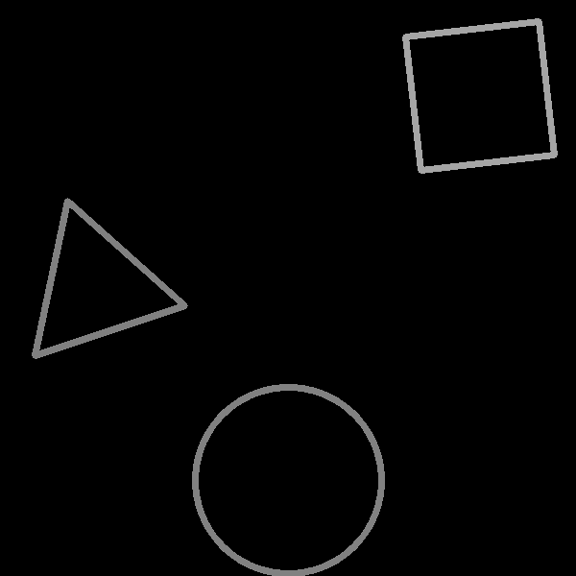} & \includegraphics[width=2cm]{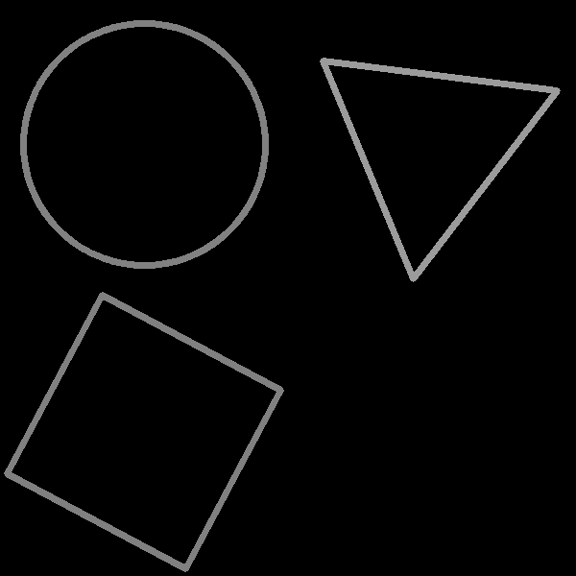} & \includegraphics[width=2cm]{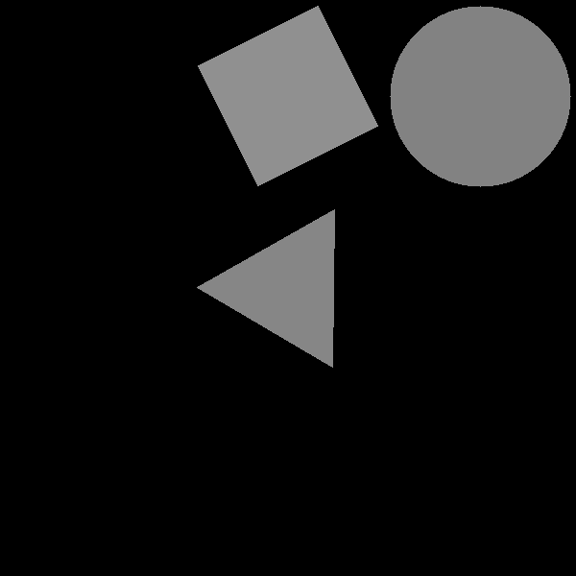} & \includegraphics[width=2cm]{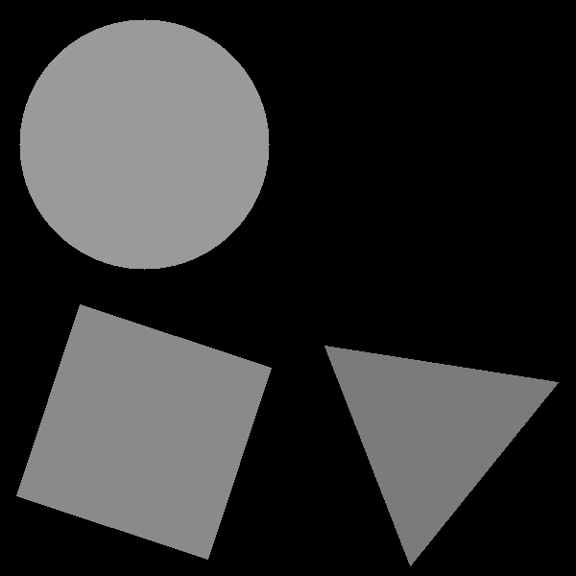} & \includegraphics[width=2cm]{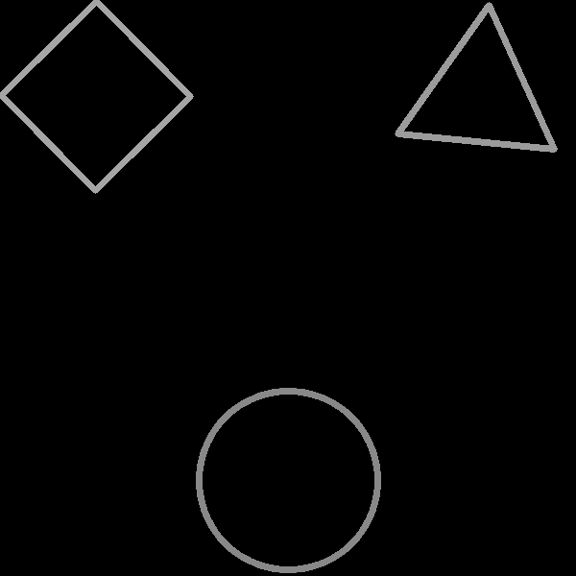} \\
\includegraphics[width=2cm]{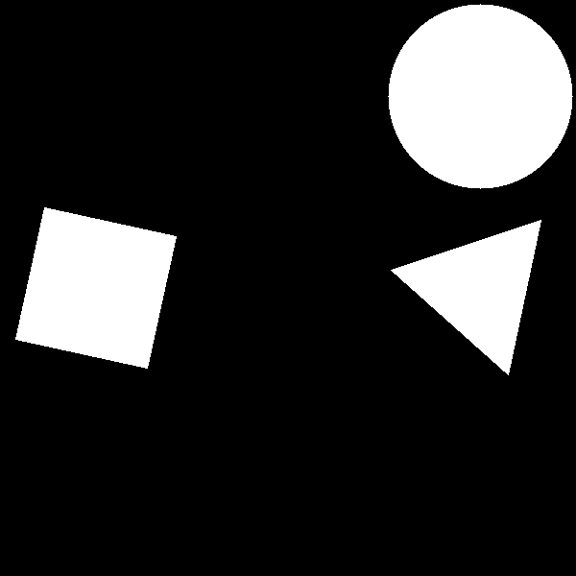} & \includegraphics[width=2cm]{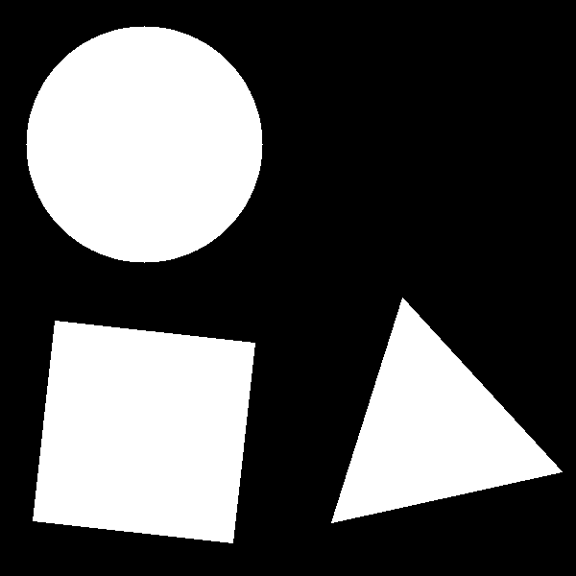} & \includegraphics[width=2cm]{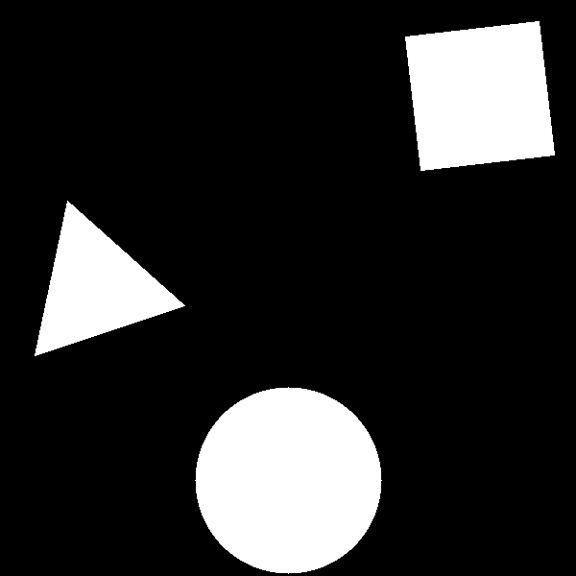} & \includegraphics[width=2cm]{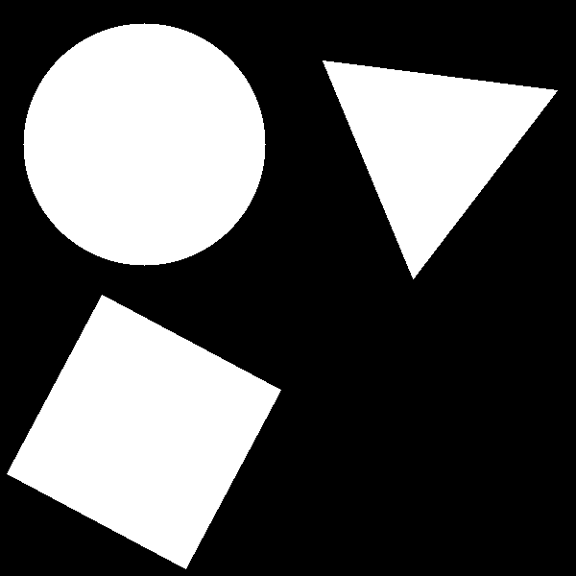} & \includegraphics[width=2cm]{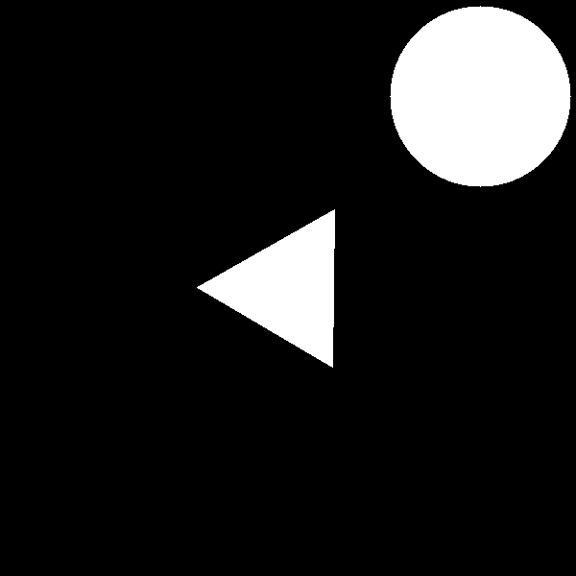} & \includegraphics[width=2cm]{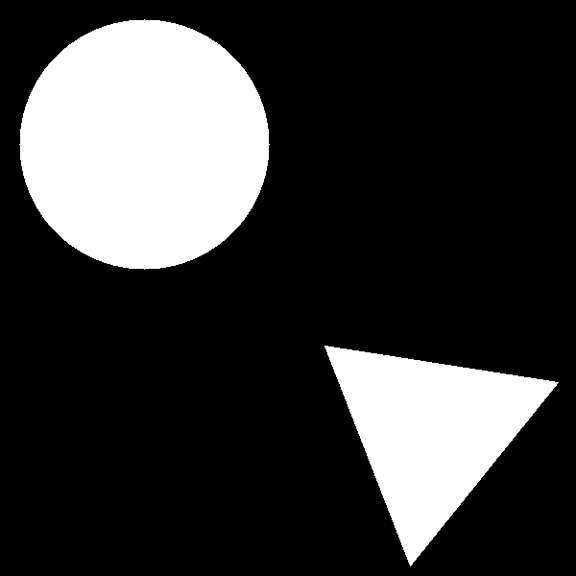} & \includegraphics[width=2cm]{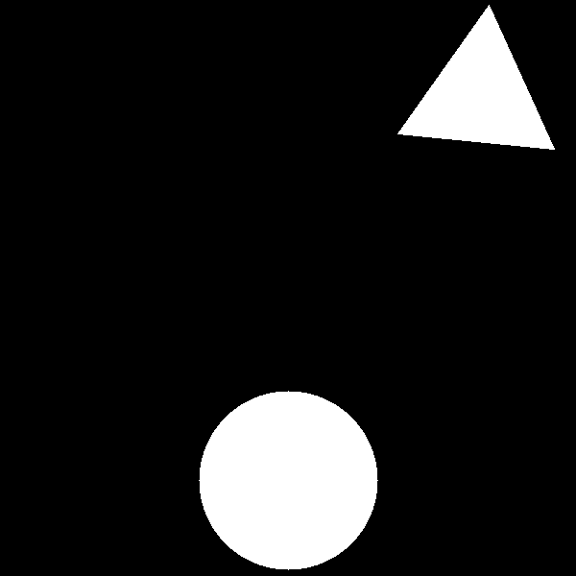}\\\\

B large contour & Fetal head & Fetal head 2 & Kidneys & Lungs & Thyroid & Nerve\\
 \includegraphics[width=2cm]{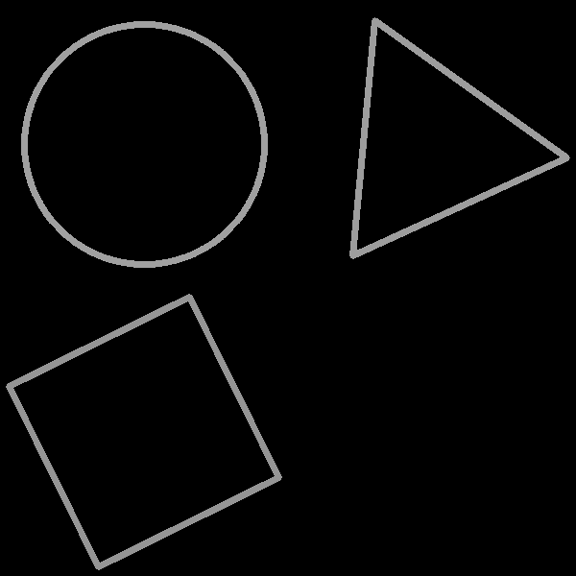} & \includegraphics[width=2cm]{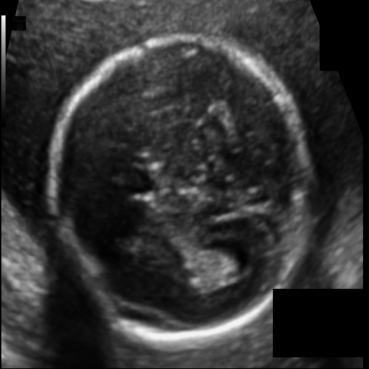} & \includegraphics[width=2cm]{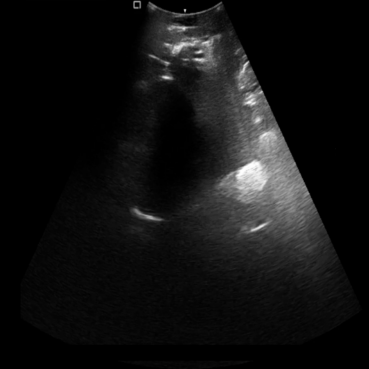} & \includegraphics[width=2cm]{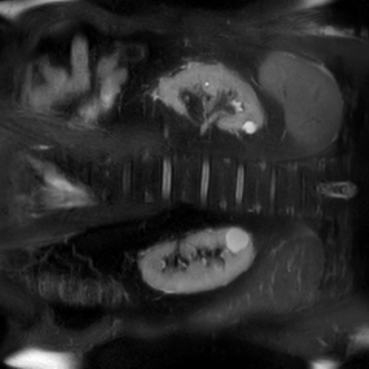} & \includegraphics[width=2cm]{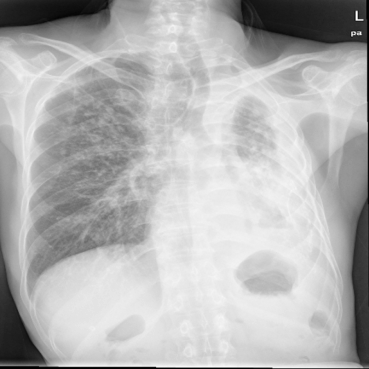} & \includegraphics[width=2cm]{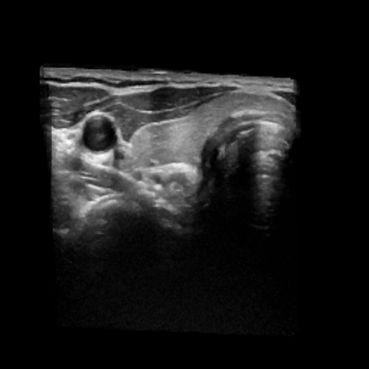} & \includegraphics[width=2cm]{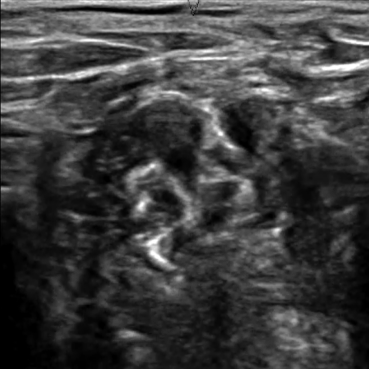}\\
 \includegraphics[width=2cm]{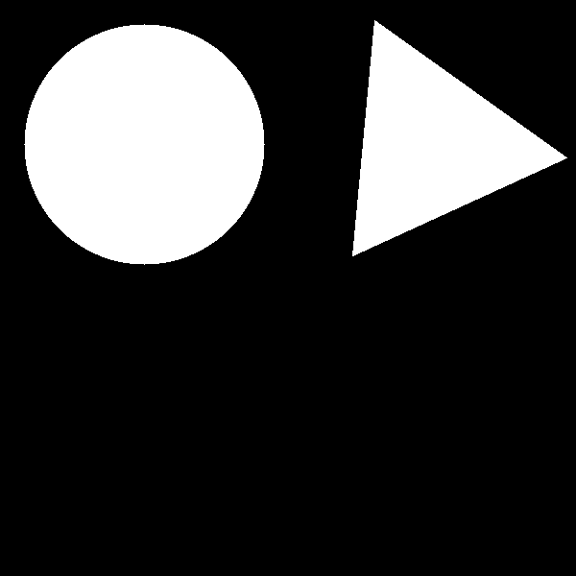} & \includegraphics[width=2cm]{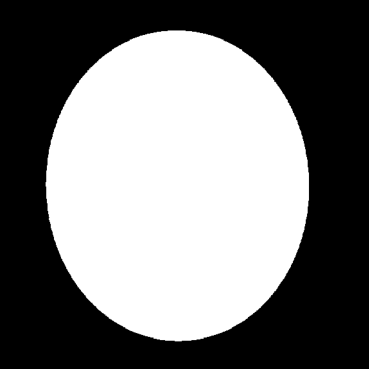} & \includegraphics[width=2cm]{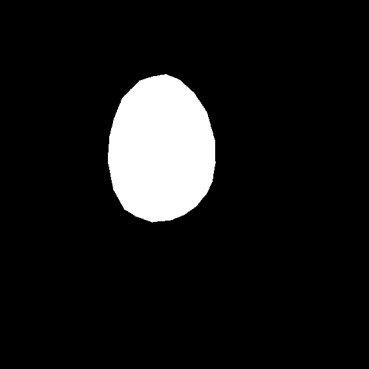} & \includegraphics[width=2cm]{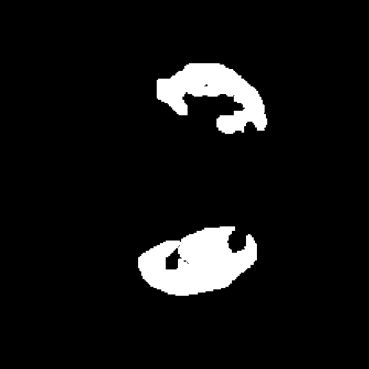} & \includegraphics[width=2cm]{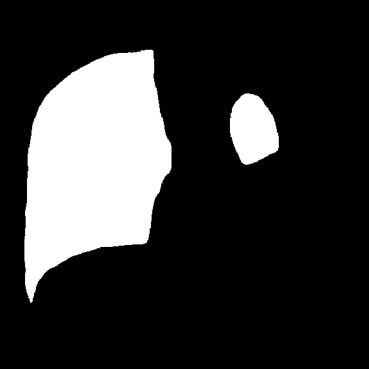} & \includegraphics[width=2cm]{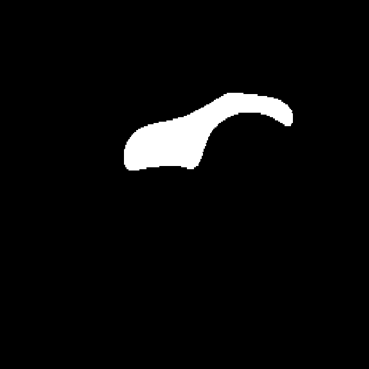} & \includegraphics[width=2cm]{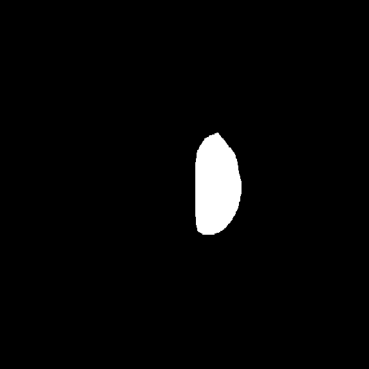}\\
\end{tabular}
\caption{Typical images and segmentation masks for the synthetic datasets (A and B) and medical datasets (Fetal head, Fetal head 2, Kidneys, Lungs, Thyroid, Nerve).}
\label{fig:dataset-examples}
\end{figure*}

\subsection{Datasets}
Our study utilized a wide array of datasets, both synthetic and real-world medical images. The synthetic datasets were specifically designed to evaluate certain hypotheses under controlled conditions. Following this, we applied our hypotheses to medical imaging datasets, which encompassed a variety of imaging techniques and anatomical structures, adding a layer of complexity and realism to our evaluations. Illustrative examples of images and corresponding masks from each dataset can be found in Figure \ref{fig:dataset-examples}.

\subsubsection{Synthetic Datasets}
The synthetic shape datasets are designed to provide a controlled environment for investigating the impact of the TRF on the performance of the models. The datasets consist of synthetic images with predefined shapes and configurations, allowing for a systematic exploration of the models' behaviour under different conditions.

There are a total of 8 datasets with generated images. These are of two types, referred to as Type A and Type B. Both types include three non-overlapping shapes -- a circle, a triangle, and a square -- that are randomly placed and rotated, with a random gray value assigned to each shape. For Type A images, the masks are identical to the shapes in the images. For Type B images, the masks are the same, but the mask of the square is omitted, adding an additional level of complexity to the segmentation task.

For each type, four datasets are created. Two of them contain small shapes placed on an invisible $3\times 3$ grid, and two of them contain large shapes placed on an invisible $2\times 2$ grid. For both the small and large datasets one of them contains images with filled shapes and filled masks, and the other one contains images with only the contours of the shapes with filled masks. Each dataset contains a total of 1000 images of which 700 are used for training, 150 for validation, and 150 for testing.

These synthetic shape datasets offer valuable insights into the role of the TRF on the performance of the models. By comparing the performance of the models on images with small shapes versus large shapes, we can assess how the TRF size affects the model's ability to capture features of different scales. Specifically, it allows us to determine to what degree it matters if the TRF is smaller than the shape, or if the shape fits into the TRF.

The comparison between images with filled shapes and those with contour shapes allows us to determine what happens if the TRF does not capture the entire shape, but only a part of it, such as the part which is completely black in the image but is filled in the mask because it is within the contours. This is particularly relevant for real-world applications, where the images often contain complex structures that the model needs to accurately segment.

Furthermore, the use of Type B images, where the mask of the square is omitted, enables us to examine how the models handle irrelevant features in the images. This is particularly relevant for real-world applications, where the images often contain irrelevant or distracting features that the model needs to ignore to perform the task effectively.

\subsubsection{Medical Datasets}
The experiments were carried out using below listed 6 medical datasets. The datasets are classified into two categories: \textit{high-contrast}, where the RoI can be visually distinguished solely based on its contrast with the background, and \textit{low-contrast}, which requires additional details like the RoI's contour or shape to  distinguish it from the background.

\textit{1. Fetal Head} -- This low-contrast dataset consists of 2D ultrasound images of fetal heads [dataset] \cite{fetal_head}. It includes 350 training images, 74 validation images, and 76 test images. The images were obtained using a standard clinical ultrasound system, and the fetal head circumference was manually annotated by expert sonographers. 

\textit{2. Fetal Head 2} -- This low-contrast dataset is another set of 2D ultrasound images of fetal heads, with a larger number of images [dataset] \cite{fetal_head_2.1, fetal_head_2.2}. It includes 14560 training images, 3240 validation images, and 2875 test images. The images in this dataset were collected from multiple hospitals and were annotated by experienced radiologists.

\textit{3. Kidneys} -- This low-contrast dataset consists of 3D MRI images of kidneys  [dataset] \cite{kidney.1, kidney.2}. It includes 454 training images, 91 validation images, and 104 test images. The images were acquired using a 3T MRI scanner and the kidney regions were manually segmented by radiologists.

\textit{4. Lungs} -- This high-contrast dataset consists of 2D X-Ray images of lungs [dataset] \cite{lungs.1, lungs.2}. It includes 396 training images, 84 validation images, and 86 test images. The images were collected from a variety of patients with different lung conditions, providing a diverse dataset for training and testing.

\textit{5. Thyroid} -- This low-contrast dataset consists of 3D ultrasound images of the thyroid [dataset] \cite{thyroid}. It includes 3160 training images, 439 validation images, and 510 test images. The images were acquired using a high-frequency linear array transducer and the thyroid regions were manually segmented by experienced clinicians.

\textit{6. Nerve} -- This low-contrast dataset consists of 2D ultrasound images of nerves [dataset] \cite{nerve}. It includes 1610 training images, 364 validation images, and 349 test images. The images were collected from a variety of patients and the nerve structures were manually annotated by expert radiologists.

% \begin{table}[htbp]
% \centering
% \caption{Number of samples for the different medical datasets for training, validation, and testing.}
% \label{tbl:medicaldata}
% \resizebox{0.48\textwidth}{!}{
% \begin{tabular}{|l|c|c|c|}
% \hline
% \textbf{Dataset} & \textbf{\# Train} & \textbf{\# Validation} & \textbf{\# Test} \\
% \hline
% Fetal Head \cite{fetal_head} & 350 & 74 & 76 \\
% \hline
% Fetal Head 2 \cite{fetal_head_2.1, fetal_head_2.2} & 14,560 & 3,240 & 2,875 \\
% \hline
% Kidneys \cite{kidney.1, kidney.2} & 454 & 91 & 104 \\
% \hline
% Lungs \cite{lungs.1, lungs.2} & 396 & 84 & 86 \\
% \hline
% Thyroid \cite{thyroid} & 3,160 & 439 & 510 \\
% \hline
% Nerve \cite{nerve} & 1,610 & 364 & 349 \\
% \hline
% \end{tabular}
% }
% \end{table}

\subsection{Data Pre-processing}
All images in the datasets were pre-processed to ensure consistency and optimal performance of the models. The pre-processing steps included resizing all images to a uniform size of 576 $\times$ 576 pixels. For the 3D datasets, all 2D slices were extracted and used as separate images.

The datasets were split into training, validation, and test sets, with approximately 70\% of the images used for training, 15\% for validation, and 15\% for testing. However, to prevent overfitting, slices from one 3D volume or 2D images from the same patient were included in only one of the train, validation, or test sets. This means that the split is not always exactly in these ratios.

Finally, on some of the smaller datasets random data augmentation was applied in order to improve the absolute results. On each sample, four random combinations of a horizontal flip, vertical flip and rotation with 90, 180 or 270 degrees were applied, where each part of the combination is applied with a probability of 0.5.

%%%%%%%%%%%%%%%%%%%%%%%
% EVALUATION MEASURES %
%%%%%%%%%%%%%%%%%%%%%%%
\subsection{Evaluation Measures}
In the realm of image segmentation, five principal metrics are typically utilized to assess performance \cite{metrics1, metrics2}. The Dice Similarity Coefficient (DSC) serves as a statistical metric, measuring the similarity between two sets by calculating the ratio of twice the intersection area to the total size of both sets. Sensitivity, or recall, appraises the model's ability to accurately identify positive instances, hence providing insight into the model's efficacy in segmenting intended areas. Specificity evaluates the model's proficiency in correctly recognizing negative instances, or in other words, its capability to exclude regions not meant to be segmented. Accuracy gauges the model's overall correctness in assigning classifications. Lastly, the Jaccard Index (JI) is an intersection-over-union measure that quantifies the similarity between the predicted and actual segmentations, providing a rigorous assessment of model performance in segmenting images. 

% These metrics are employed in this study to evaluate our models by calculating the average value for each metric across the entirety of the individual test sets for each dataset. 
Moreover, to understand fully the impact of TRF and ERF on model performance, two additional metrics are proposed in this work: ERF rate and object rate. We also factor in the training time (epochs) as a metric, quantifying the epochs needed to attain the lowest validation loss. This allows us to comparatively analyze the training cost across various models.

\subsubsection{ERF rate}
% A novel metric, termed as the ERF rate, is introduced to assess the distribution of the ERF. It calculates the total area of the substantially contributing pixels within the ERF in proportion to the area of the TRF, taking into account the absolute value of the ERF pixels. For an output image with an ERF $\mathbf{E}\in\mathds{R}^{m\times n}$, the ERF rate is represented in equation \ref{erf_rate}. The ERF rate considers all the pixels within the ERF that make a significant contribution and provides additional weight to the pixels with higher values. The final result is normalized by dividing it with the area of the TRF. During model evaluation, the ERF rate is calculated for each image in the test set, and the mean ERF rate is reported as the overall metric value.

We propose a new metric called ERF rate to measure the ERF distribution. It quantifies the fraction of significantly contributing pixels within the ERF against the TRF area, utilizing the absolute value of the ERF pixels. The ERF rate (equation \ref{erf_rate}) accounts for all the meaningful pixels above a certain threshold ($\varepsilon$) in the ERF, giving more weight to pixels with higher values and normalizing the result with the TRF area. The metric is computed for each test image, reporting the mean ERF rate as the overall score.

\begin{equation}\label{erf_rate}
r = \frac{\sum_{y\in \mathbf{E}}[|y|>\varepsilon ] \cdot(1+|y|)}{m\cdot n}  
\end{equation}

% Kernel density estimation (KDE) can be used to determine the threshold ($\varepsilon$) for significantly contributing pixels. KDE is a method for estimating the probability density function (PDF) of a continuous random variable based on a sample of observations \cite{kdestimation}. The basic idea of KDE is to estimate the PDF by placing a kernel function centered at each observation and summing the contributions of all kernels.

% Formally, the density function $f(x)$ of the values $y$ in the ERF $\mathbf{E}$ can be estimated by using the following formula:

We use Kernel Density Estimation (KDE) to find the threshold ($\varepsilon$) for key contributing pixels, estimating the Probability Density Function (PDF) of a continuous variable based on observed samples \cite{kdestimation}.

The density function $f(x)$ of ERF values can be calculated using the formula in equation \ref{eq:kde}, where $\hat{f}(x)$ is the estimated PDF, $K(x)$ is a kernel function with bandwidth $h$, and $m\cdot n$ is the number of observations in $\mathbf{E}$. It is centered at each observation $y$.

\begin{equation}\label{eq:kde}
    \hat{f}(x) = \frac{1}{m\cdot n} \cdot \sum_{y\in\mathbf{E}} \frac{1}{h} K\left(\frac{x - |y|}{h}\right)
\end{equation}

To identify the ideal parameters for KDE, we examined the ERF absolute value histogram for a large dataset sample. It reveals two different types of distributions: (i) ERFs with both contributing and non-contributing pixels have a bimodal distribution with a left peak representing non-contributing pixels and a right peak representing contributing pixels, and (ii) ERFs with mostly non-contributing pixels have a highly positively skewed distribution. The first parameter, the bandwidth ($h$), controls the kernel width and PDF smoothing level. Silverman's rule-of-thumb \cite{Silverman86} was used to automatically determine $h$ ($h=1.06\cdot\hat{\sigma}mn$, where $\hat{\sigma}$ is the standard deviation of sample of size $m\cdot n$), because it performs well on both bimodal and skewed distributions \cite{Harpole2014}.

\begin{figure} [h]
    \centering
    \includegraphics[width=1\linewidth]{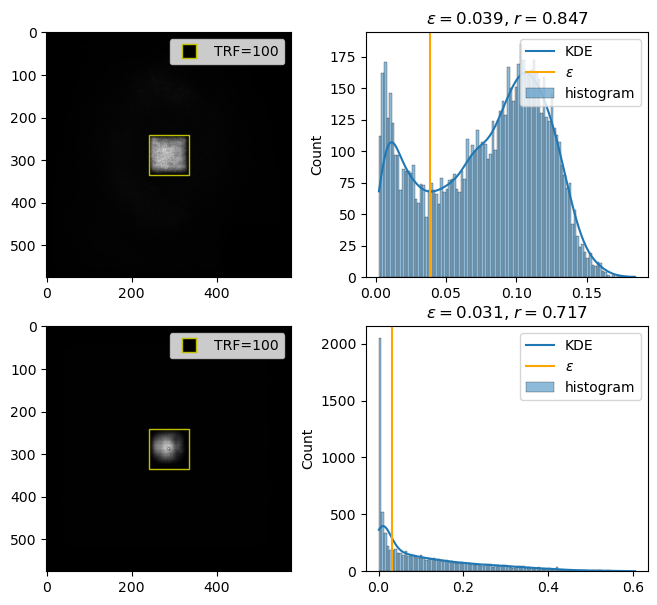}
    \caption{Examples of determining the threshold ($\varepsilon$) for the ERF rate with KDE for bimodally distributed ERF pixel values (top row) and positively skewed distributed ERF pixel values (bottom row).}
    \label{fig:erf-rate}
\end{figure}

Finally, the threshold ($\varepsilon$) was selected based on the trough in bimodal distributions or the stopping point of decrease in skewed distributions (Figure \ref{fig:erf-rate}). To reduce the number of troughs, thus making it easier to find the optimal threshold, a Gaussian kernel function was used to smooth the estimated PDF \cite{kdestimation}.

\subsubsection{Object rate}
In order to assess the relative size of the object to be segmented in comparison to the TRF size, a new metric denoted as object rate is proposed. This metric is computed by dividing the total area of a rectangle encompassing the edges of the object by the total area of the TRF size, or $TRF^2$ as defined in equation \ref{eq:trf_size}. Thus, for an object with its highest point at $t$, lowest at $b$, leftmost at $l$, and rightmost at $r$, the object rate can be calculated as follows:
\begin{equation}
    \text{OR} = \frac{(b-t)\cdot(r-l)}{\text{TRF}^2}
\end{equation}

%%%%%%%%%%%
% RESULTS %
%%%%%%%%%%%
\section{Results \& Discussion}
Detailed results of the performance of the U-Net model for
the different metrics on all medical datasets can be found in Table \ref{tbl:all_results}. Results of the Attention U-Net on the medical datasets and the U-Net on the synthetic datasets of type A and B can be found in \ref{app:medical_results_attention}, \ref{app:shapes_results_a}, and \ref{app:shapes_results_b} respectively. In the following section we present different plots to interpret and discuss these results.

\begin{table*}[ht!]\centering
\caption{All the results for the different evaluation measures on the medical datasets (fetal head, fetal head 2, kidneys, lungs, nerve, and thyroid) for the U-Net.}

\label{tbl:all_results}
\def\arraystretch{1}

\resizebox{0.8\textwidth}{!}{
\begin{tabular}{|lrrrrrrrrrr|}
\hline
\textbf{TRF size} & \textbf{trf54} & \textbf{trf100} & \textbf{trf146} & \textbf{trf204} & \textbf{trf230} & \textbf{trf298} & \textbf{trf360} & \textbf{trf412} & \textbf{trf486} & \textbf{trf570} \\ \hline
& & & & & & & & & & \\
\textit{Fetal head} & & & & & & & & & & \\ \hline
Training time (epochs) & 38 & 31 & 21 & 26 & 34 & 21 & 24 & 26 & 28 & 30 \\
ERF rate before training & 0.0135 & 0.0097 & 0.0047 & 0.0011 & 0.0046 & 0.0014 & 0.0044 & 0.0005 & 0.0012 & 0.0009 \\
ERF rate & 0.8898 & 0.9380 & 0.9153 & 0.8614 & 0.8196 & 0.5970 & 0.5785 & 0.4618 & 0.4175 & 0.2309 \\
Dice score & 0.7752 & 0.8866 & 0.9224 & 0.9527 & 0.9506 & 0.9526 & 0.9623 & 0.9614 & 0.9650 & 0.9665 \\
Object rate & 67.3558 & 16.8389 & 7.4840 & 4.0325 & 2.6942 & 1.7922 & 1.0997 & 0.9870 & 0.6452 & 0.4481 \\
Accuracy & 0.8687 & 0.9300 & 0.9518 & 0.9690 & 0.9693 & 0.9704 & 0.9749 & 0.9735 & 0.9761 & 0.9773 \\
Sensitivity & 0.8458 & 0.8914 & 0.9395 & 0.9508 & 0.9707 & 0.9709 & 0.9675 & 0.9570 & 0.9627 & 0.9680 \\
Specificity & 0.8831 & 0.9530 & 0.9612 & 0.9828 & 0.9739 & 0.9745 & 0.9844 & 0.9878 & 0.9887 & 0.9873 \\
Jaccard index & 0.6577 & 0.8152 & 0.8723 & 0.9212 & 0.9210 & 0.9248 & 0.9390 & 0.9371 & 0.9435 & 0.9466 \\
& & & & & & & & & & \\
\textit{Fetal head 2} & & & & & & & & & & \\ \hline
Training time (epochs) & 1 & 1 & 1 & 6 & 1 & 8 & 7 & 7 & 25 & 14 \\
ERF rate before training & 0.1531 & 0.0330 & 0.0300 & 0.0097 & 0.0245 & 0.0073 & 0.0163 & 0.0019 & 0.0051 & 0.0054 \\
ERF rate & 1.0005 & 0.0944 & 0.2275 & 0.0095 & 0.0168 & 0.0051 & 0.2761 & 0.0010 & 0.3159 & 0.1735 \\
Dice score & 0.6009 & 0.6261 & 0.7582 & 0.8745 & 0.7950 & 0.8588 & 0.9028 & 0.9071 & 0.9116 & 0.9214 \\
Object rate & 35.5599 & 8.8900 & 3.9511 & 2.1289 & 1.4224 & 0.9462 & 0.5806 & 0.5211 & 0.3406 & 0.2365 \\
Accuracy & 0.9889 & 0.8888 & 0.9207 & 0.9557 & 0.9315 & 0.9525 & 0.9654 & 0.9675 & 0.9686 & 0.9725 \\
Sensitivity & 0.7058 & 0.7818 & 0.8512 & 0.9133 & 0.9284 & 0.9502 & 0.9567 & 0.9524 & 0.9479 & 0.9616 \\
Specificity & 0.9913 & 0.9004 & 0.9303 & 0.9627 & 0.9316 & 0.9514 & 0.9645 & 0.9679 & 0.9708 & 0.9723 \\
Jaccard index & 0.5154 & 0.4764 & 0.6265 & 0.7876 & 0.6841 & 0.7805 & 0.8365 & 0.8445 & 0.8513 & 0.8656 \\
& & & & & & & & & & \\
\textit{Kidneys} & & & & & & & & & & \\ \hline
Training time (epochs) & 21 & 34 & 31 & 32 & 38 & 47 & 54 & 47 & 48 & 54 \\
ERF rate before training & 0.1715 & 0.0496 & 0.0303 & 0.0126 & 0.0287 & 0.0058 & 0.0217 & 0.0024 & 0.0059 & 0.0065 \\
ERF rate & 0.0123 & 0.0341 & 0.0227 & 0.0057 & 0.0162 & 0.0035 & 0.0088 & 0.0012 & 0.0038 & 0.0038 \\
Dice score & 0.7560 & 0.8367 & 0.8477 & 0.8524 & 0.8617 & 0.8364 & 0.8865 & 0.8657 & 0.8439 & 0.8802 \\
Object rate & 27.4954 & 6.8738 & 3.0550 & 1.6461 & 1.0998 & 0.7316 & 0.4489 & 0.4029 & 0.2634 & 0.1829 \\
Accuracy & 0.9832 & 0.9904 & 0.9911 & 0.9911 & 0.9917 & 0.9889 & 0.9923 & 0.9902 & 0.9900 & 0.9918 \\
Sensitivity & 0.7808 & 0.8856 & 0.8762 & 0.8783 & 0.8645 & 0.8814 & 0.8993 & 0.8914 & 0.8647 & 0.8884 \\
Specificity & 0.9892 & 0.9935 & 0.9945 & 0.9948 & 0.9963 & 0.9916 & 0.9961 & 0.9935 & 0.9942 & 0.9954 \\
Jaccard index & 0.6320 & 0.7509 & 0.7679 & 0.7717 & 0.7853 & 0.7490 & 0.8119 & 0.7836 & 0.7607 & 0.8055 \\
& & & & & & & & & & \\
\textit{Lungs} & & & & & & & & & & \\ \hline
Training time (epochs) & 15 & 20 & 29 & 26 & 36 & 26 & 51 & 28 & 33 & 40 \\
ERF rate before training & 0.0417 & 0.0419 & 0.0216 & 0.0127 & 0.0122 & 0.0069 & 0.0134 & 0.0009 & 0.0036 & 0.0061 \\
ERF rate & 0.0614 & 0.1298 & 0.0315 & 0.0040 & 0.0137 & 0.0031 & 0.0245 & 0.0012 & 0.0011 & 0.0039 \\
Dice score & 0.9601 & 0.9673 & 0.9687 & 0.9686 & 0.9683 & 0.9666 & 0.9689 & 0.9683 & 0.9662 & 0.9673 \\
Object rate & 84.4219 & 21.1055 & 9.3802 & 5.0542 & 3.3769 & 2.2463 & 1.3784 & 1.2371 & 0.8087 & 0.5616 \\
Accuracy & 0.9784 & 0.9823 & 0.9830 & 0.9829 & 0.9829 & 0.9820 & 0.9830 & 0.9827 & 0.9818 & 0.9824 \\
Sensitivity & 0.9650 & 0.9677 & 0.9697 & 0.9681 & 0.9695 & 0.9654 & 0.9694 & 0.9654 & 0.9746 & 0.9776 \\
Specificity & 0.9825 & 0.9870 & 0.9872 & 0.9878 & 0.9869 & 0.9873 & 0.9875 & 0.9886 & 0.9834 & 0.9834 \\
Jaccard index & 0.9240 & 0.9371 & 0.9398 & 0.9396 & 0.9391 & 0.9361 & 0.9402 & 0.9389 & 0.9353 & 0.9373 \\
& & & & & & & & & & \\
\textit{Nerve} & & & & & & & & & & \\ \hline
Training time (epochs) & 7 & 13 & 15 & 8 & 14 & 17 & 10 & 8 & 10 & 12 \\
ERF rate before training & 0.1425 & 0.0444 & 0.0381 & 0.0107 & 0.0287 & 0.0096 & 0.0213 & 0.0019 & 0.0104 & 0.0078 \\
ERF rate & 0.9312 & 0.7345 & 0.6953 & 0.0057 & 0.3348 & 0.0224 & 0.1363 & 0.0008 & 0.1244 & 0.0399 \\
Dice score & 0.4685 & 0.7329 & 0.7531 & 0.7745 & 0.7792 & 0.7863 & 0.7965 & 0.7951 & 0.7960 & 0.7947 \\
Object rate & 7.3183 & 1.8296 & 0.8131 & 0.4381 & 0.2927 & 0.1947 & 0.1195 & 0.1072 & 0.0701 & 0.0487 \\
Accuracy & 0.9758 & 0.9848 & 0.9859 & 0.9868 & 0.9873 & 0.9872 & 0.9876 & 0.9878 & 0.9881 & 0.9880 \\
Sensitivity & 0.6442 & 0.7519 & 0.7637 & 0.7803 & 0.7978 & 0.7990 & 0.7901 & 0.8068 & 0.8301 & 0.8289 \\
Specificity & 0.9808 & 0.9914 & 0.9923 & 0.9930 & 0.9928 & 0.9930 & 0.9940 & 0.9934 & 0.9927 & 0.9927 \\
Jaccard index & 0.3281 & 0.6030 & 0.6321 & 0.6572 & 0.6614 & 0.6701 & 0.6827 & 0.6800 & 0.6786 & 0.6785 \\
& & & & & & & & & & \\
\textit{Thyroid} & & & & & & & & & & \\ \hline
Training time (epochs) & 1 & 1 & 3 & 2 & 2 & 5 & 3 & 2 & 7 & 4 \\
ERF rate before training & 0.1652 & 0.0434 & 0.0280 & 0.0094 & 0.0227 & 0.0093 & 0.0180 & 0.0021 & 0.0089 & 0.0054 \\
ERF rate & 0.1041 & 0.1439 & 0.1823 & 0.0124 & 0.0482 & 0.0240 & 0.0269 & 0.0038 & 0.0152 & 0.0054 \\
Dice score & 0.5155 & 0.5829 & 0.6456 & 0.7043 & 0.7124 & 0.6907 & 0.6680 & 0.6667 & 0.7457 & 0.7284 \\
Object rate & 14.8609 & 3.7152 & 1.6512 & 0.8897 & 0.5944 & 0.3954 & 0.2426 & 0.2178 & 0.1424 & 0.0989 \\
Accuracy & 0.9718 & 0.9807 & 0.9840 & 0.9860 & 0.9822 & 0.9864 & 0.9854 & 0.9859 & 0.9871 & 0.9837 \\
Sensitivity & 0.6309 & 0.7427 & 0.7374 & 0.7516 & 0.7021 & 0.7563 & 0.7449 & 0.7705 & 0.7602 & 0.7268 \\
Specificity & 0.9788 & 0.9827 & 0.9879 & 0.9912 & 0.9928 & 0.9905 & 0.9884 & 0.9879 & 0.9935 & 0.9933 \\
Jaccard index & 0.3881 & 0.4779 & 0.5481 & 0.6107 & 0.6168 & 0.6036 & 0.5785 & 0.5746 & 0.6526 & 0.6246 \\ \hline
\end{tabular}
}
\end{table*}

\begin{figure}[!ht]
    \centering
    \resizebox{1\linewidth}{!}{\input{images/analysis-shapes.pgf}}
    \caption{Performance of the shapes datasets (A and B) for the regular U-Net}
    \label{fig:analysis-shapes}
\end{figure}
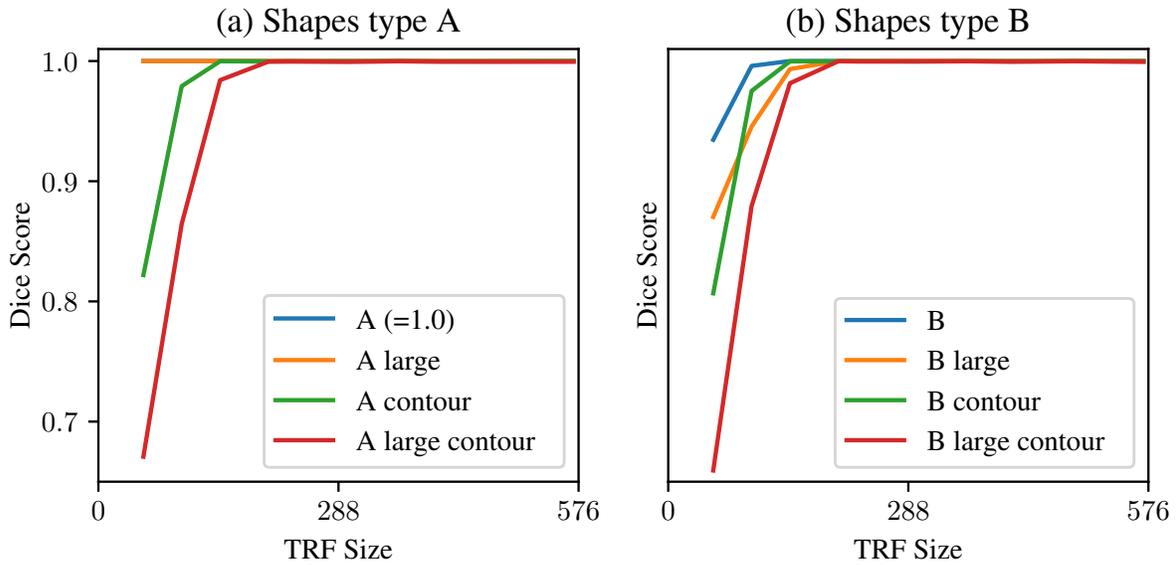

\begin{figure*}[!ht]
    \centering
    \resizebox{\linewidth}{!}{\input{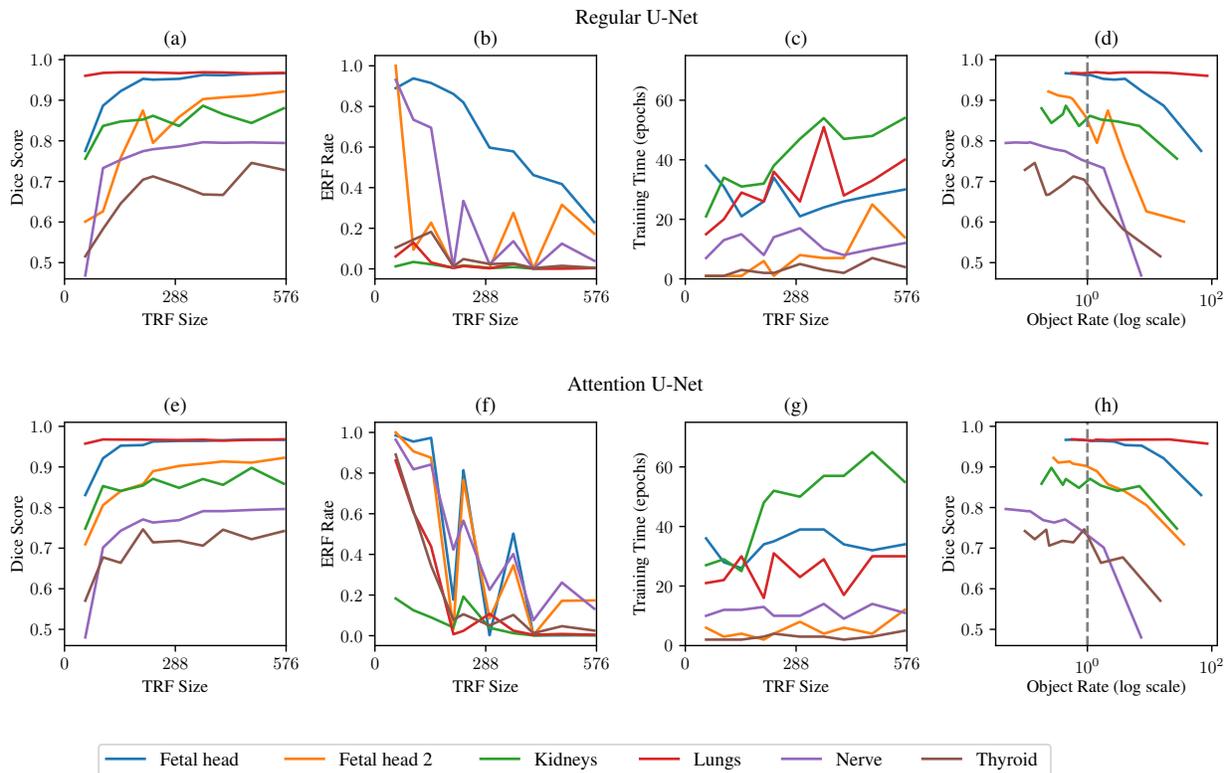}}
    \caption{Various plots for the analyses of the medical datasets for U-Net (a) DSC Vs TRF size, (b) ERF rate Vs TRF size, (c) Training time (epochs) Vs TRF size, (d) Dice Score Vs TRF, and for Attention U-Net (e) DSC Vs TRF size, (f) ERF rate Vs TRF size, (g) Training time (epochs) Vs TRF size, (h) Dice Score Vs TRF.}
    \label{fig:analysis}
\end{figure*}

\begin{figure*}[!ht]
\centering

\begin{subfigure}{0.45\linewidth}
  \centering
  \begin{tabular}{ccccccc}
   TRF = 54 & TRF = 230 & TRF = 486\\
   \includegraphics[width=2cm]{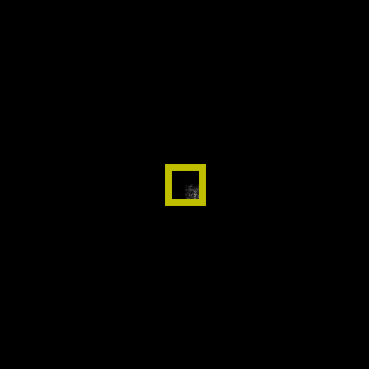} & \includegraphics[width=2cm]{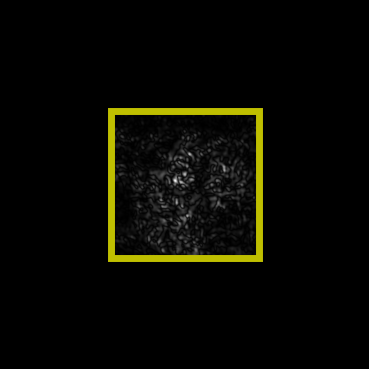} & \includegraphics[width=2cm]{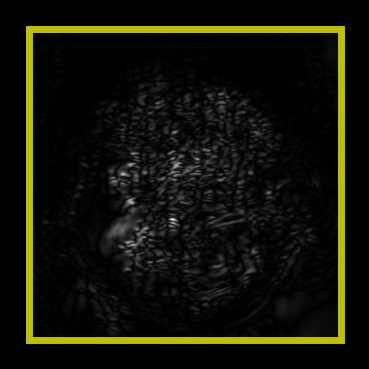} \\
  \includegraphics[width=2cm]{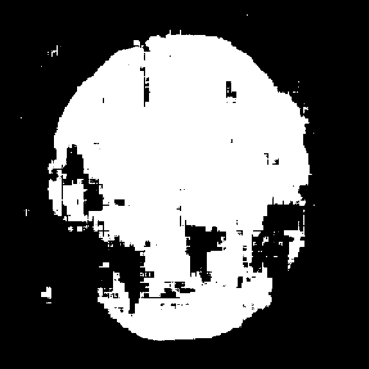} & \includegraphics[width=2cm]{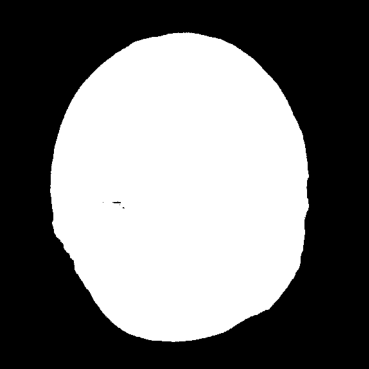} & \includegraphics[width=2cm]{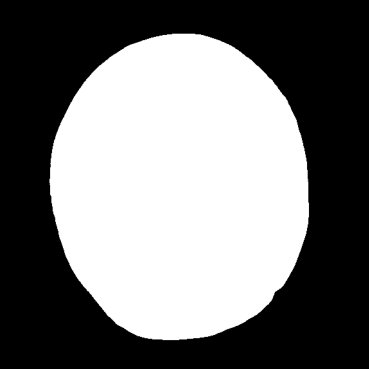}
  \end{tabular}
  \caption{Examples on the fetal head dataset.}
  \label{fig:rf-with-segmentation-fetal-head}
\end{subfigure}
\begin{subfigure}{0.45\linewidth}
    \centering
    \begin{tabular}{ccccccc}
     TRF = 54 & TRF = 230 & TRF = 486\\
     \includegraphics[width=2cm]{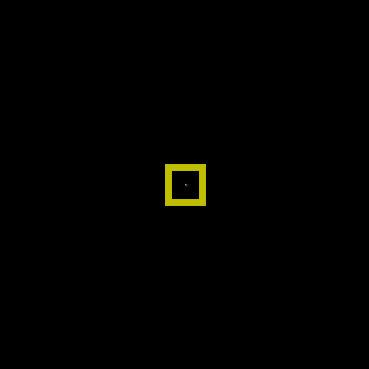} & \includegraphics[width=2cm]{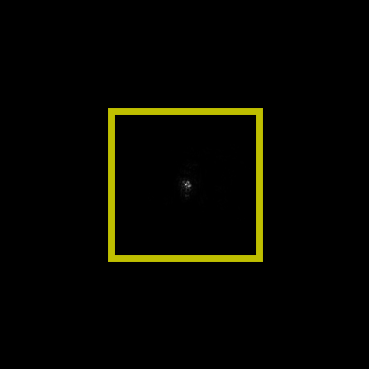} & \includegraphics[width=2cm]{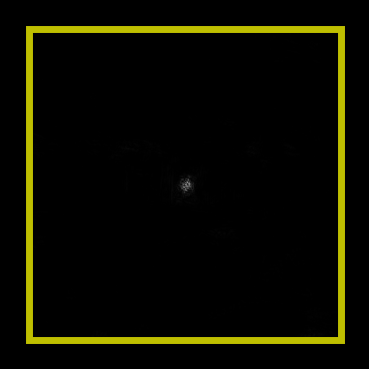} \\
    \includegraphics[width=2cm]{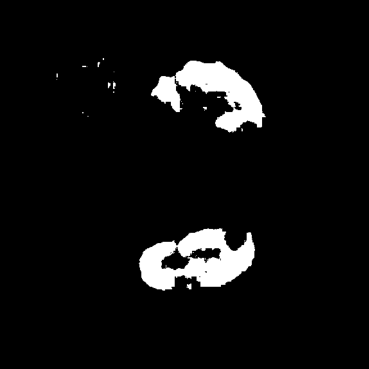} & \includegraphics[width=2cm]{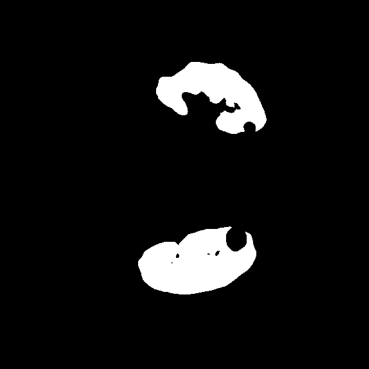} & \includegraphics[width=2cm]{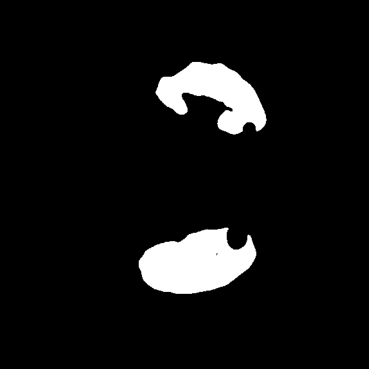}
    \end{tabular}
    \caption{Examples on the kidneys dataset.}
    \label{fig:rf-with-segmentation-kidneys}
\end{subfigure}

\vspace{0.5cm}

\begin{subfigure}{0.45\linewidth}
    \centering
    \begin{tabular}{ccccccc}
     TRF = 54 & TRF = 230 & TRF = 486\\
     \includegraphics[width=2cm]{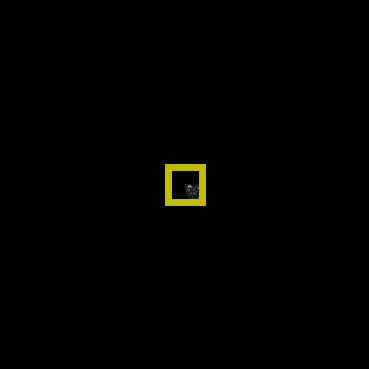} & \includegraphics[width=2cm]{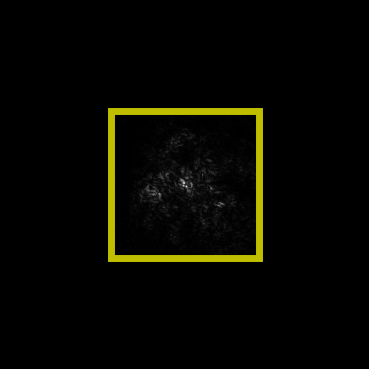} & \includegraphics[width=2cm]{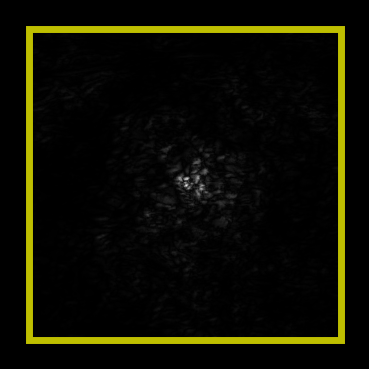} \\
    \includegraphics[width=2cm]{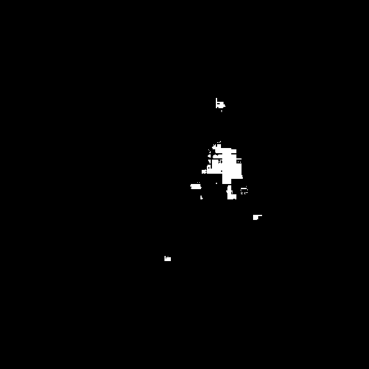} & \includegraphics[width=2cm]{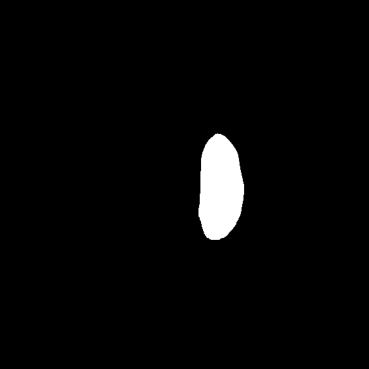} & \includegraphics[width=2cm]{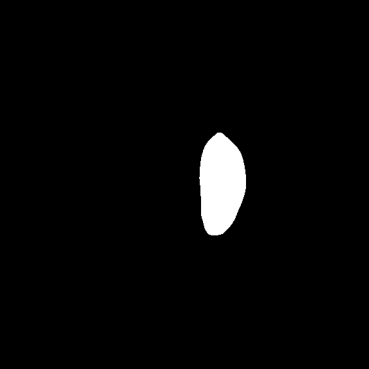}
    \end{tabular}
    \caption{Examples on the nerve dataset.}
    \label{fig:rf-with-segmentation-nerve}
\end{subfigure}
\begin{subfigure}{0.45\linewidth}
    \centering
    \begin{tabular}{ccccccc}
     TRF = 54 & TRF = 230 & TRF = 486\\
     \includegraphics[width=2cm]{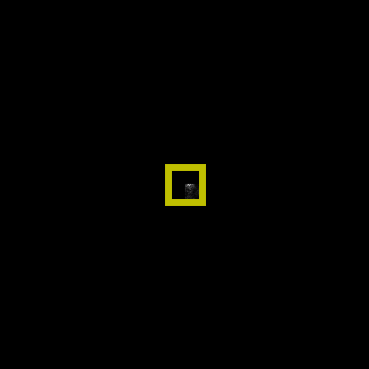} & \includegraphics[width=2cm]{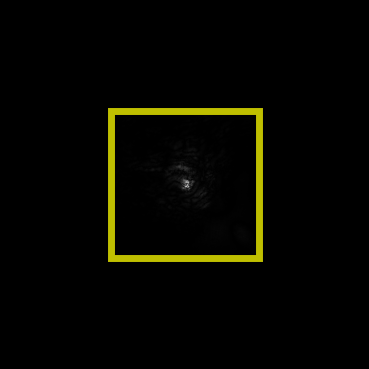} & \includegraphics[width=2cm]{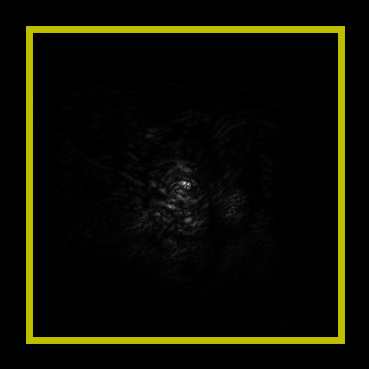} \\
    \includegraphics[width=2cm]{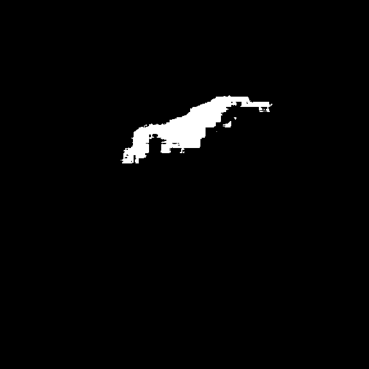} & \includegraphics[width=2cm]{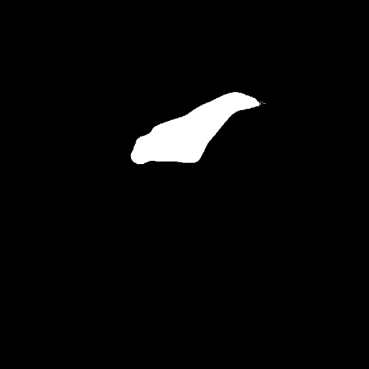} & \includegraphics[width=2cm]{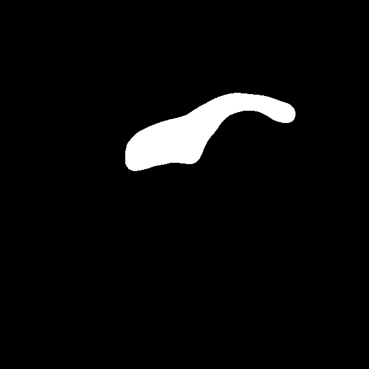}
    \end{tabular}
    \caption{Examples on the thyroid dataset.}
    \label{fig:rf-with-segmentation-thyroid}
\end{subfigure}

\caption{Examples of the TRF (yellow square), ERF (pixels within the TRF) in the top row in each subfigure and the predicted segmentation for various TRF sizes in the bottom row, on the samples from the datasets in Figure \ref{fig:dataset-examples}.}
\label{fig:combined-figure}
\end{figure*}

\begin{figure}[!ht]
\centering
\begin{tabular}{ccccccc}
 TRF = 54 & TRF = 230 & TRF = 486\\
 \includegraphics[width=2cm]{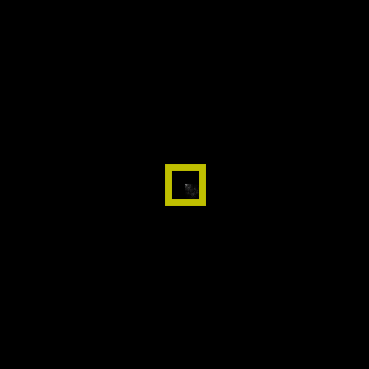} & \includegraphics[width=2cm]{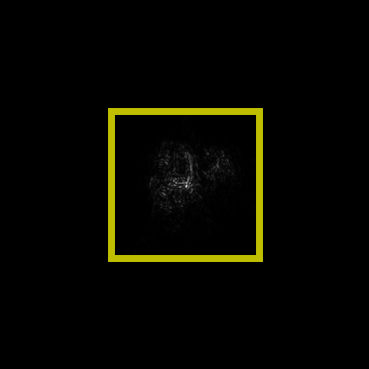} & \includegraphics[width=2cm]{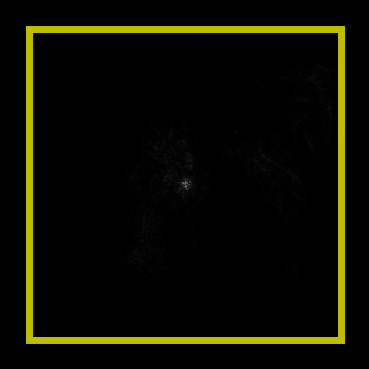} \\
\includegraphics[width=2cm]{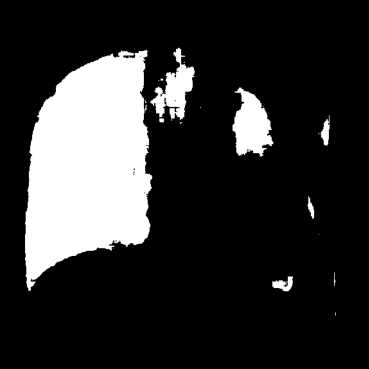} & \includegraphics[width=2cm]{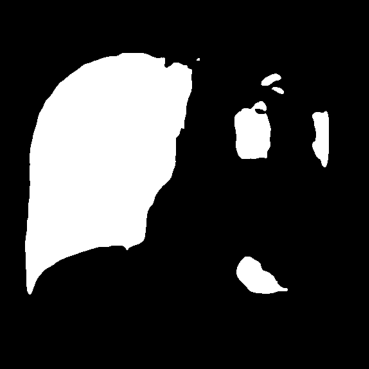} & \includegraphics[width=2cm]{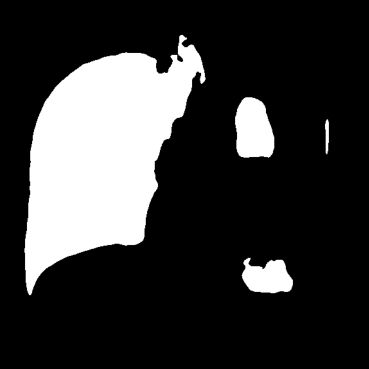}
\end{tabular}
\caption{Examples of the TRF (yellow square), ERF (pixels within the TRF) in the top row and the predicted segmentation for various TRF sizes in the bottom row, on the sample from the lungs dataset from Figure \ref{fig:dataset-examples}.}
\label{fig:rf-with-segmentation-lungs}
\end{figure}
\subsection{The Role of Contrast}
Figure \ref{fig:analysis-shapes} illustrates the relationship between the TRF size and the DSC for all synthetic shape datasets, encompassing both Type A and Type B for the U-Net model. For all datasets that can be segmented solely based on contrast (\textit{A, A large}), the model attains perfect performance even at the smallest TRF size (Figure \ref{fig:analysis-shapes} (a)). For datasets, that present an added layer of complexity by either representing only contours of RoI in input images (\textit{A contour, A large contour}) or by excluding the square from the mask (\textit{Type B}), require a larger TRF to reach peak performance (Figure \ref{fig:analysis-shapes} (a), Figure \ref{fig:analysis-shapes} (b)). These datasets with an added complexity in segmentation show a model performance trend where DSC starts at a lower point for a small TRF and requires a larger TRF to reach peak performance unlike the consistent perfect performance in the contrast-based datasets.

The same pattern is present in the medical datasets: all datasets which have a low-contrast RoI show a trend of increasing DSC as the TRF size grows, whereas the high-contrast lung dataset attains peak performance starting at the lowest TRF (Figs. \ref{fig:analysis}(a) and \ref{fig:analysis}(e)). 
The segmentation output for the datasets of fetal head, kidneys, nerve and thyroid for different TRF's are shown in Figs. \ref{fig:combined-figure} (a), \ref{fig:combined-figure} (b), \ref{fig:combined-figure} (c), and \ref{fig:combined-figure} (d) respectively and the combined results for the U-net model are shown in Table-\ref{tbl:all_results}. Fig. \ref{fig:rf-with-segmentation-lungs} shows the results for the lung dataset for different TRF's. Since, in the lung dataset, the RoI can be identified visually using the contrast, the DSC attains close to peak value even for a very small TRF.  
% A visual representation of these trends can be seen in Figs. \ref{fig:rf-with-segmentation-fetal-head-2} and \ref{fig:combined-figure} for the low-contrast datasets, and Fig. \ref{fig:rf-with-segmentation-lungs} for the high-contrast lungs dataset. 
It is clear that the predicted segmentation improves significantly with increasing TRF for all datasets except the lung dataset.
% as the TRF sizes grows in the former, and not as much in the latter.

This pattern is further highlighted in Table \ref{tbl:all_results}, where all low-contrast datasets consistently show a trend of increasing DSC with TRF, and all high-contrast datasets do not show the same trend.

% \begin{figure}[h]
% \centering
% \caption{Examples of the TRF (yellow square), ERF (pixels within the TRF) in the top row and the predicted segmentation for various TRF sizes in the bottom row, on the sample from the fetal head 2 dataset from Figure \ref{fig:dataset-examples}.}
% \label{fig:rf-with-segmentation-fetal-head-2}
% \begin{tabular}{ccccccc}
%  TRF = 54 & TRF = 230 & TRF = 486\\
%  \includegraphics[width=2cm]{images/rfs-with-segmentations/fetal_head_2-trf54-rf.png} & \includegraphics[width=2cm]{images/rfs-with-segmentations/fetal_head_2-trf230-rf.png} & \includegraphics[width=2cm]{images/rfs-with-segmentations/fetal_head_2-trf486-rf.png} \\
% \includegraphics[width=2cm]{images/rfs-with-segmentations/fetal_head_2-trf54-prediction.png} & \includegraphics[width=2cm]{images/rfs-with-segmentations/fetal_head_2-trf230-prediction.png} & \includegraphics[width=2cm]{images/rfs-with-segmentations/fetal_head_2-trf486-prediction.png}
% \end{tabular}
% \end{figure}

\begin{figure*}[h]
    \centering
    \begin{tabular}{ccccccccccc}
     TRF = 54 & 100 & 146 & 204 & 230 & 298 & 360 & 412 & 486 & 570\\
     \includegraphics[width=1.2cm]{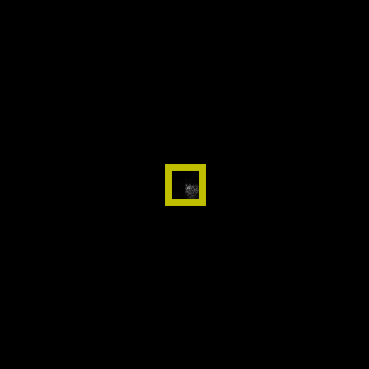} & \includegraphics[width=1.2cm]{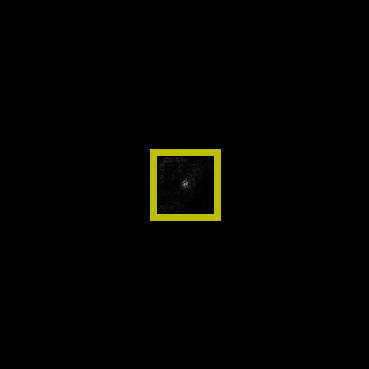} & \includegraphics[width=1.2cm]{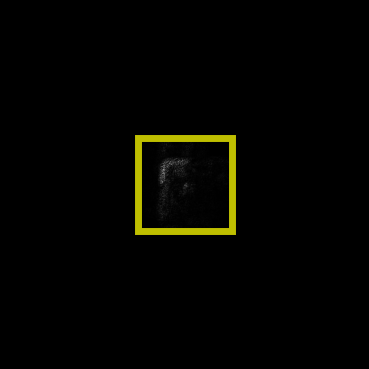} & \includegraphics[width=1.2cm]{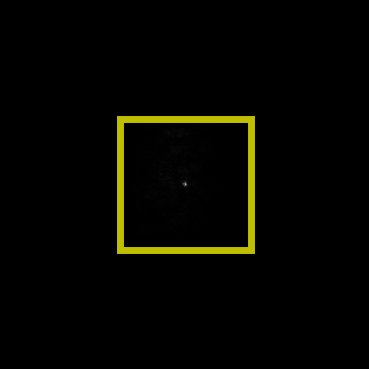} & \includegraphics[width=1.2cm]{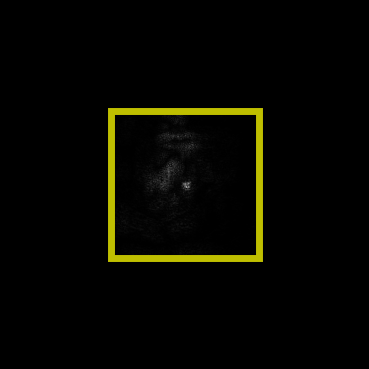} & \includegraphics[width=1.2cm]{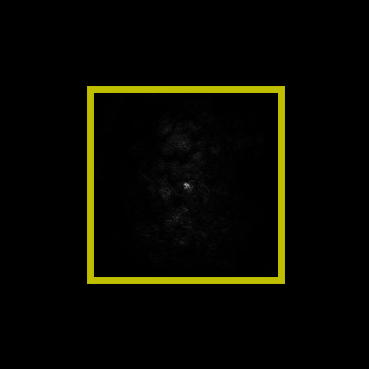} & \includegraphics[width=1.2cm]{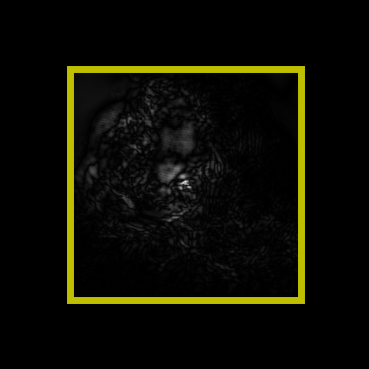} & \includegraphics[width=1.2cm]{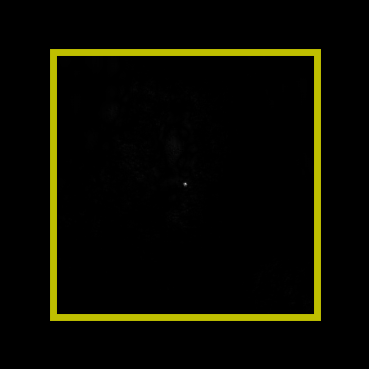} & \includegraphics[width=1.2cm]{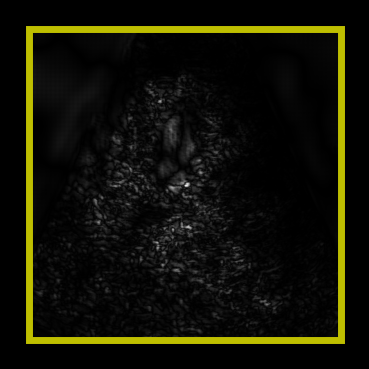} & \includegraphics[width=1.2cm]{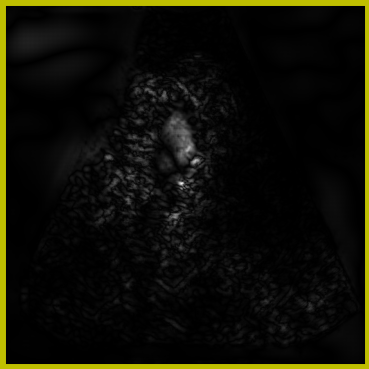} \\
    \includegraphics[width=1.2cm]{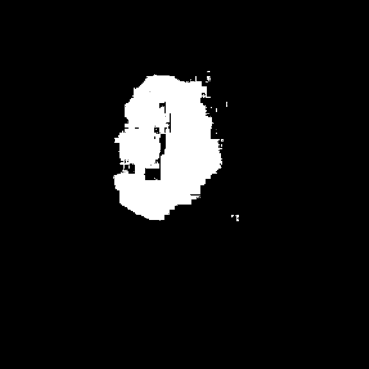} & \includegraphics[width=1.2cm]{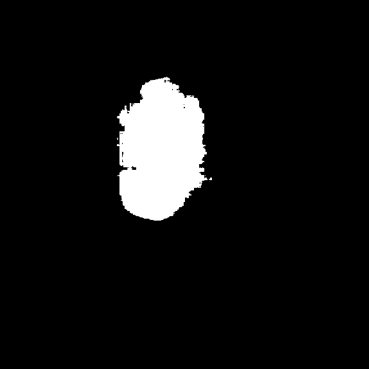} & \includegraphics[width=1.2cm]{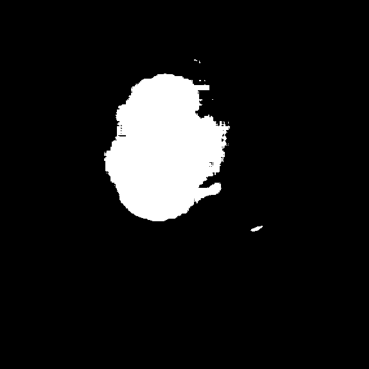} & \includegraphics[width=1.2cm]{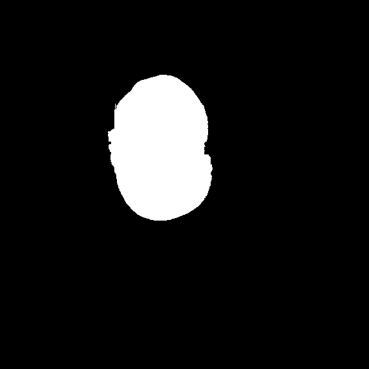} & \includegraphics[width=1.2cm]{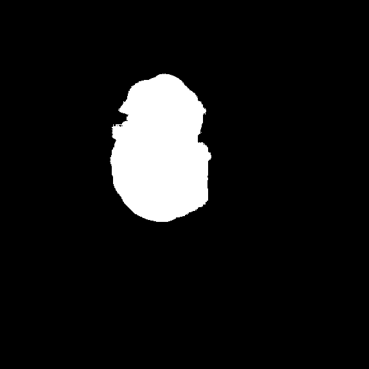} & \includegraphics[width=1.2cm]{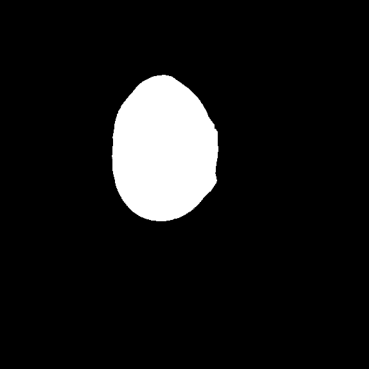} & \includegraphics[width=1.2cm]{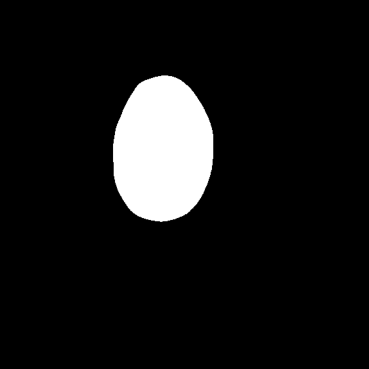} & \includegraphics[width=1.2cm]{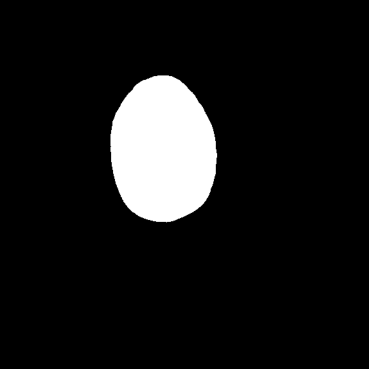} & \includegraphics[width=1.2cm]{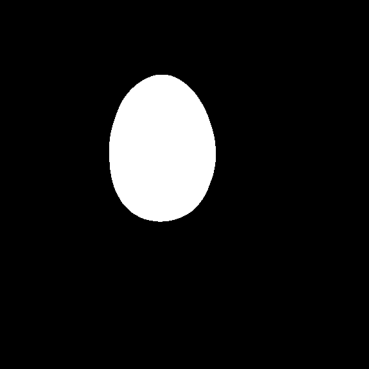} & \includegraphics[width=1.2cm]{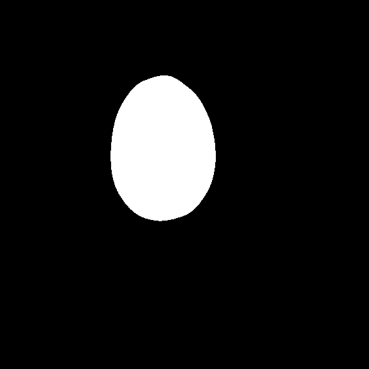}
    \end{tabular}
    \caption{Examples of the TRF (yellow square), ERF (pixels within the TRF) in the top row and the predicted segmentation for various TRF sizes in the bottom row, on the sample from the fetal head 2 dataset from Figure \ref{fig:dataset-examples}.}
    \label{fig:rf-with-segmentation-fetal-head-2}
\end{figure*}

\subsection{Optimal TRF Size}
% discuss erf rate decreasing and object rate and training time
In Figs. \ref{fig:analysis}(b) and \ref{fig:analysis}(f), a trend is visible which shows that the ERF rate shrinks with the enlargement of the TRF size for the U-Net and the Attention U-Net respectively. This suggests that as the TRF size increases, a smaller proportion of pixels actually contribute to the predicted segmentation. Moreover, as the TRF size increases, the training time (epochs) also tends to increase as displayed in Figs. \ref{fig:analysis}(c) and \ref{fig:analysis}(g) for the U-Net and the Attention U-Net respectively. This finding implies that an excessively large TRF size may lead to unnecessary computations, potentially explaining the observed increase in training time (epochs) with the expansion of the TRF size.

In this context, the object rate, plotted against the DSC in Figs. \ref{fig:analysis}(d) (U-Net) and \ref{fig:analysis}(h) (Attention U-Net), also seems to play a role. When the object rate, i.e. the size of the RoI relative to the TRF, increases, the DSC degrades. This is corroborated by the two rightmost columns in Table \ref{tbl:summary}, where for most low-contrast datasets where the TRF size plays a major role, the optimal TRF size, i.e. the TRF size at which the DSC saturates, is usually only slightly smaller than the size of the RoI.

Despite the overall trend of increasing DSC with expanding TRF size, we observe slight drops. This can be interpreted in light of the concept of variability in neural networks, as discussed by \cite{yu2023multilayer}. Variability, as they define it, represents the richness of landscape patterns in the data space with respect to well-scaled random weights. As the TRF size increases, the model starts to incorporate more global context into its predictions. While this can be beneficial for capturing larger-scale structures in the image, it may also introduce more noise into the model's predictions, especially if the larger TRF includes irrelevant or distracting features. This could result in slight decreases in the DSC.

\subsection{Attention Mechanism and TRF Size}
In Figs. \ref{fig:analysis}(a) and \ref{fig:analysis}(e) the TRF size is plotted against the DSC for the U-Net and Attention U-Net respectively for all the medical datasets. In both instances, the trend of an increasing DSC as the TRF grows is present. Fig. \ref{fig:rf-with-segmentation-fetal-head-2} shows the segmentation for the fetal head 2 dataset for all the TRF's and the corresponding TRF and ERF. As the TRF size increases, the segmentation accuracy increases and the overall trend can be seen in Table-\ref{tbl:all_results}. The same trend is also visible for the Attention U-Net model, the results of which are shown in \ref{app:medical_results_attention}. However, all absolute scores are higher in the case of the Attention U-Net. Table-\ref{tbl:summary} column 4 shows the summary of results for all the datasets if they follow the pattern of DSC using Attention U-Net model. All the medical imaging datasets except the lung dataset follow the pattern for both U-Net as well as Attention U-Net model, with a higher absolute scores for attention counterpart. Hence, it can be said that the attention mechanism will consistently improve the performance, regardless of TRF size. Even with attention mechanism, TRF plays an important role and a larger TRF might further improve the performance of Attention U-net models.

\subsection{Designing Efficient Architectures}

In this work, we performed the experiments for different TRF's having same number of total parameters for different datasets as can be seen in Table-\ref{tbl:configs}. Detailed results of the performance of the U-Net model for
the different metrics on all medical datasets can be found in Table \ref{tbl:all_results}. Results of the Attention U-Net on the medical datasets and the U-Net on the synthetic datasets of type A and B can be found in \ref{app:medical_results_attention}, \ref{app:shapes_results_a}, and \ref{app:shapes_results_b} respectively. These results show that even for the same number of parameters there is a very high effect on the performance of the network if the TRF is changed. Inclusion of TRF size as a parameter for models can lead to a more fair comparison among their performance. It will also help in designing efficient architectures, ones with optimal trade-off between performance and number of parameters.

\begin{table*}[ht!]\centering
\caption{Summary of the insights from the results. Values with \textit{no*} mean that the RoI can be identified visually by contrast, but that not all regions which have this contrast are also included in the mask.\label{tbl:summary}}
\begin{tabular}{|p{2.5cm}|p{1.5cm}|p{2cm}|p{2cm}|p{2cm}|p{2cm}|p{2cm}|p{2cm}|}\hline
\textbf{Dataset} & \textbf{Dataset Type} & \textbf{RoI can be identified visually only by contrast} & \textbf{Pattern of increasing DSC with TRF} & \textbf{Pattern retained with Attention U-Net, but higher absolute score} & \textbf{Average dimension of RoI} & \textbf{DSC saturates between TRF sizes} \\\hline
Nerve & Clinical & \tikzmarknode{A1}No & Yes & Yes & 159 & 298-360 \\
B contour & Synthetic &  No & Yes & N/A & 168 & 100-146 \\
A contour & Synthetic &  No & Yes & N/A & 169 & 100-146 \\
Thyroid & Clinical &  No & Yes & Yes & 187 & 146-204 \\
B large contour & Synthetic &  No & Yes & N/A & 237 & 146-204 \\
A large contour & Synthetic &  No & Yes & N/A & 242 & 146-204 \\
Fetal head 2 & Clinical &  No & Yes & Yes & 255 & 146-204 \\
Fetal head & Clinical &  No & Yes & Yes & 260 & 146-204 \\
Kidneys & Clinical &  No* & Yes \tikzmarknode{A2}& Yes & 101 & 298-360 \\
B & Synthetic &  No* & Yes & N/A & 168 & 54-100 \\
B large & Synthetic & No* & Yes & N/A & 238 & 100-146 \\
% \vspace{0.00001em} &&&&&&\\
Lungs & Clinical & Yes &  No & N/A & 329 & 0-54 \\
A & Synthetic &  \tikzmarknode{B1}Yes &  No & N/A & 168 & 0-54 \\
A large & Synthetic & Yes & No \tikzmarknode{B2}& N/A & 244 & 0-54 \\\hline
\end{tabular}
\begin{tikzpicture}[overlay,remember picture]
\node[ draw=green,line width=2pt,fit={(A1)(A2)($(A1.north west)+(-2pt,-0pt)$)($(A2.south east)+(45pt,-22pt)$)}]{};
\end{tikzpicture}
\begin{tikzpicture}[overlay,remember picture]
\node[ draw=yellow,line width=2pt,fit={(B1)(B2)($(B1.north west)+(-2pt,10pt)$)($(B2.south east)+(47pt,-0pt)$)}]{};
\end{tikzpicture}
\end{table*}

%%%%%%%%%%%%%%
% CONCLUSION %
%%%%%%%%%%%%%%
\section{Conclusion}\label{conclusion}
This work highlights the essential role of the TRF size in semantic segmentation tasks with U-Net and Attention U-Net architectures across datasets of various modalities. We discovered that an optimal TRF size, the one which balances capturing of global context and computational efficiency, can significantly enhance model performance. This implies that an excessively large TRF size may lead to unnecessary computational costs without corresponding improvements in performance. Additionally, our results also emphasize the added value of the attention mechanism in boosting segmentation accuracy, irrespective of the TRF size.

Our findings suggest that the datasets where RoI can be visually identified by contrast comparison alone, typically attain peak performance with even small TRF. Conversely, this is not the case when additional complexities are present, such as contrast not being only criteria for identifying RoI or contours demarcating RoI. This implies that the model's performance also depends on factors like the complexity of the task and the size of the RoI relative to the TRF size.

Furthermore, our study indicates that the DSC tends to plateau at a certain TRF size depending on the dataset. This suggests that there exists an optimal TRF size for each dataset, beyond which further expansion of the TRF size does not significantly improve the DSC. These findings can have practical implications for the design of segmentation models, suggesting that increasing TRF size may not always be necessary or beneficial.

These insights provide a valuable reference for designing and optimizing U-Net-based architectures for various tasks and datasets in medical imaging. While our study focused on the U-Net and Attention U-Net architectures, there are many other architectures used for semantic segmentation tasks, such as SegNet \cite{segnet}, PSPNet \cite{pspnet}, and DeepLab \cite{deeplab}. Future research could investigate the impact of the TRF size on the performance of these architectures.

% While we have made significant findings in understanding the impact of the RF size on the performance of U-Net and Attention U-Net architectures, several aspects warrant further exploration.

% One promising area for future research involves a more detailed investigation into how adjusting the model's hyperparameters to tune the RF affects performance. For instance, modifying the convolutional kernel sizes of certain layers could significantly impact the TRF and, consequently, the model's performance. A systematic study exploring the effects of these hyperparameters could yield valuable insights into the optimal configuration of these architectures for different tasks and datasets.

% \section{Resources and Tools}
% The complete source code utilized in this work can be accessed via our GitHub repository at \url{https://github.com/vinloo/u-net-receptive-field-study}. In addition to this, we have developed an open-source tool designed to calculate and suggest an appropriate TRF size based on a specified U-Net configuration and dataset. This tool is intended to aid researchers and practitioners in the field and is included in the repository.

\section*{Acknowledgements}
This research did not receive any specific grant from funding agencies in the public, commercial, or not-for-profit sectors.

%%Harvard
% \bibliographystyle{model2-names.bst}\biboptions{authoryear}
% \bibliography{refs}

\appendix
% \clearpage

\begin{table*}[b!]\centering
\def\arraystretch{1}
\section*{Appendices}
\section{All results for the different evaluation measures on the medical datasets (fetal head, fetal head 2, kidneys, lungs, nerve, and thyroid) for the Attention- U-Net.}
\label{app:medical_results_attention}
% \caption{Overview of all results for the Attention U-Net on the medical datasets\label{tbl:all_attention_results}}
\resizebox{0.8\textwidth}{!}{
\begin{tabular}{|lrrrrrrrrrr|}
\hline
\textbf{TRF size} & \textbf{trf54} & \textbf{trf100} & \textbf{trf146} & \textbf{trf204} & \textbf{trf230} & \textbf{trf298} & \textbf{trf360} & \textbf{trf412} & \textbf{trf486} & \textbf{trf570} \\ \hline
& & & & & & & & & & \\
\textit{Fetal head} & & & & & & & & & & \\ \hline
Training time (epochs) & 36 & 28 & 26 & 34 & 35 & 39 & 39 & 34 & 32 & 34 \\
ERF rate before training & 0.8256 & 0.5080 & 0.4090 & 0.4592 & 0.3215 & 0.3601 & 0.2955 & 0.4388 & 0.2972 & 0.2690 \\
ERF rate & 0.9854 & 0.9548 & 0.9732 & 0.1771 & 0.8143 & 0.0024 & 0.5025 & 0.0013 & 0.0016 & 0.0025 \\
Dice score & 0.8307 & 0.9213 & 0.9524 & 0.9538 & 0.9625 & 0.9640 & 0.9642 & 0.9655 & 0.9675 & 0.9667 \\
Object rate & 67.3558 & 16.8389 & 7.4840 & 4.0325 & 2.6942 & 1.7922 & 1.0997 & 0.9870 & 0.6452 & 0.4481 \\
Accuracy & 0.8984 & 0.9512 & 0.9708 & 0.9715 & 0.9769 & 0.9781 & 0.9780 & 0.9791 & 0.9807 & 0.9803 \\
Sensitivity & 0.9085 & 0.9553 & 0.9528 & 0.9727 & 0.9591 & 0.9665 & 0.9637 & 0.9621 & 0.9638 & 0.9671 \\
Specificity & 0.8920 & 0.9476 & 0.9799 & 0.9711 & 0.9869 & 0.9848 & 0.9865 & 0.9891 & 0.9905 & 0.9879 \\
Jaccard index & 0.7200 & 0.8656 & 0.9206 & 0.9231 & 0.9388 & 0.9417 & 0.9420 & 0.9445 & 0.9482 & 0.9469 \\
& & & & & & & & & & \\
\textit{Fetal head 2} & & & & & & & & & & \\ \hline
Training time (epochs) & 6 & 3 & 4 & 2 & 4 & 8 & 4 & 6 & 4 & 12 \\
ERF rate before training & 0.8459 & 0.5388 & 0.4327 & 0.5030 & 0.3511 & 0.4051 & 0.3217 & 0.4826 & 0.3293 & 0.3049 \\
ERF rate & 1.0000 & 0.9070 & 0.8755 & 0.0325 & 0.7660 & 0.0845 & 0.3467 & 0.0030 & 0.1717 & 0.1735 \\
Dice score & 0.7094 & 0.8058 & 0.8403 & 0.8574 & 0.8895 & 0.9025 & 0.9082 & 0.9135 & 0.9106 & 0.9224 \\
Object rate & 35.5599 & 8.8900 & 3.9511 & 2.1289 & 1.4224 & 0.9462 & 0.5806 & 0.5211 & 0.3406 & 0.2850 \\
Accuracy & 0.9155 & 0.9380 & 0.9476 & 0.9508 & 0.9621 & 0.9658 & 0.9673 & 0.9699 & 0.9698 & 0.9518 \\
Sensitivity & 0.8614 & 0.8797 & 0.9172 & 0.9281 & 0.9419 & 0.9521 & 0.9465 & 0.9424 & 0.9513 & 0.9395 \\
Specificity & 0.9232 & 0.9479 & 0.9516 & 0.9529 & 0.9648 & 0.9668 & 0.9688 & 0.9739 & 0.9705 & 0.9612 \\
Jaccard index & 0.5716 & 0.6892 & 0.7341 & 0.7616 & 0.8130 & 0.8350 & 0.8441 & 0.8542 & 0.8517 & 0.8723 \\
& & & & & & & & & & \\
\textit{Kidneys} & & & & & & & & & & \\ \hline
Training time (epochs) & 27 & 29 & 25 & 48 & 52 & 50 & 57 & 57 & 65 & 55 \\
ERF rate before training & 0.8694 & 0.5124 & 0.4018 & 0.4715 & 0.3262 & 0.3916 & 0.2979 & 0.4270 & 0.3122 & 0.2975 \\
ERF rate & 0.1832 & 0.1251 & 0.0902 & 0.0406 & 0.1925 & 0.0388 & 0.0122 & 0.0015 & 0.0049 & 0.0019 \\
Dice score & 0.7481 & 0.8529 & 0.8410 & 0.8542 & 0.8709 & 0.8484 & 0.8703 & 0.8558 & 0.8979 & 0.8586 \\
Object rate & 27.4954 & 6.8738 & 3.0550 & 1.6461 & 1.0998 & 0.7316 & 0.4489 & 0.4029 & 0.2634 & 0.1829 \\
Accuracy & 0.9829 & 0.9911 & 0.9896 & 0.9911 & 0.9917 & 0.9901 & 0.9918 & 0.9913 & 0.9930 & 0.9903 \\
Sensitivity & 0.8210 & 0.8484 & 0.8500 & 0.8813 & 0.8878 & 0.8360 & 0.8813 & 0.8849 & 0.8920 & 0.8742 \\
Specificity & 0.9867 & 0.9959 & 0.9951 & 0.9942 & 0.9951 & 0.9952 & 0.9954 & 0.9952 & 0.9965 & 0.9936 \\
Jaccard index & 0.6197 & 0.7720 & 0.7505 & 0.7747 & 0.7967 & 0.7687 & 0.7990 & 0.7798 & 0.8300 & 0.7759 \\
& & & & & & & & & & \\
\textit{Lungs} & & & & & & & & & & \\ \hline
Training time (epochs) & 21 & 22 & 30 & 16 & 31 & 23 & 29 & 17 & 30 & 30 \\
ERF rate before training & 0.8377 & 0.5446 & 0.4295 & 0.5266 & 0.3453 & 0.3745 & 0.3063 & 0.4558 & 0.3053 & 0.3143 \\
ERF rate & 0.8615 & 0.6073 & 0.4391 & 0.0071 & 0.0240 & 0.1075 & 0.0245 & 0.0044 & 0.0084 & 0.0045 \\
Dice score & 0.9574 & 0.9677 & 0.9673 & 0.9672 & 0.9668 & 0.9662 & 0.9671 & 0.9649 & 0.9666 & 0.9681 \\
Object rate & 84.4219 & 21.1055 & 9.3802 & 5.0542 & 3.3769 & 2.2463 & 1.3784 & 1.2371 & 0.8087 & 0.5616 \\
Accuracy & 0.9769 & 0.9824 & 0.9822 & 0.9823 & 0.9819 & 0.9817 & 0.9822 & 0.9810 & 0.9819 & 0.9827 \\
Sensitivity & 0.9488 & 0.9738 & 0.9629 & 0.9665 & 0.9728 & 0.9636 & 0.9711 & 0.9618 & 0.9685 & 0.9684 \\
Specificity & 0.9867 & 0.9850 & 0.9886 & 0.9872 & 0.9846 & 0.9875 & 0.9856 & 0.9871 & 0.9862 & 0.9874 \\
Jaccard index & 0.9191 & 0.9378 & 0.9371 & 0.9370 & 0.9363 & 0.9352 & 0.9370 & 0.9331 & 0.9362 & 0.9388 \\
& & & & & & & & & & \\
\textit{Nerve} & & & & & & & & & & \\ \hline
Training time (epochs) & 10 & 12 & 12 & 13 & 10 & 10 & 14 & 9 & 14 & 11 \\
ERF rate before training & 0.8478 & 0.5326 & 0.4261 & 0.4985 & 0.3459 & 0.3965 & 0.3144 & 0.4639 & 0.3263 & 0.3110 \\
ERF rate & 0.9638 & 0.8183 & 0.8422 & 0.4236 & 0.5661 & 0.2254 & 0.4018 & 0.0762 & 0.2615 & 0.1329 \\
Dice score & 0.4801 & 0.7014 & 0.7428 & 0.7708 & 0.7631 & 0.7689 & 0.7911 & 0.7911 & 0.7941 & 0.7964 \\
Object rate & 7.3183 & 1.8296 & 0.8131 & 0.4381 & 0.2927 & 0.1947 & 0.1195 & 0.1072 & 0.0701 & 0.0487 \\
Accuracy & 0.9746 & 0.9849 & 0.9860 & 0.9872 & 0.9869 & 0.9867 & 0.9881 & 0.9875 & 0.9880 & 0.9881 \\
Sensitivity & 0.5738 & 0.8122 & 0.7848 & 0.7959 & 0.8150 & 0.7957 & 0.8210 & 0.8139 & 0.8343 & 0.8228 \\
Specificity & 0.9824 & 0.9885 & 0.9911 & 0.9924 & 0.9914 & 0.9922 & 0.9927 & 0.9929 & 0.9924 & 0.9930 \\
Jaccard index & 0.3427 & 0.5687 & 0.6203 & 0.6541 & 0.6437 & 0.6490 & 0.6761 & 0.6765 & 0.6770 & 0.6805 \\
& & & & & & & & & & \\
\textit{Thyroid} & & & & & & & & & & \\ \hline
Training time (epochs) & 2 & 2 & 2 & 3 & 4 & 3 & 3 & 2 & 3 & 5 \\
ERF rate before training & 0.8760 & 0.5268 & 0.4179 & 0.4842 & 0.3426 & 0.4035 & 0.3087 & 0.4485 & 0.2973 & 0.3025 \\
ERF rate & 0.8913 & 0.6137 & 0.3449 & 0.0786 & 0.1057 & 0.0504 & 0.1024 & 0.0125 & 0.0468 & 0.0253 \\
Dice score & 0.5706 & 0.6773 & 0.6638 & 0.7464 & 0.7142 & 0.7181 & 0.7060 & 0.7455 & 0.7219 & 0.7420 \\
Object rate & 14.8609 & 3.7152 & 1.6512 & 0.8897 & 0.5944 & 0.3954 & 0.2426 & 0.2178 & 0.1424 & 0.0989 \\
Accuracy & 0.9704 & 0.9811 & 0.9819 & 0.9844 & 0.9818 & 0.9835 & 0.9839 & 0.9832 & 0.9802 & 0.9836 \\
Sensitivity & 0.6093 & 0.7146 & 0.7767 & 0.7933 & 0.7155 & 0.7677 & 0.7581 & 0.7960 & 0.7262 & 0.7863 \\
Specificity & 0.9832 & 0.9897 & 0.9871 & 0.9904 & 0.9914 & 0.9903 & 0.9899 & 0.9902 & 0.9922 & 0.9901 \\
Jaccard index & 0.4402 & 0.5736 & 0.5609 & 0.6372 & 0.6085 & 0.6137 & 0.6084 & 0.6343 & 0.6114 & 0.6333 \\ \hline
\end{tabular}
}
\end{table*}

% \clearpage
\begin{table*}[!ht]\centering
\def\arraystretch{1}
\section{Results of the regular U-Net model on the Type A shapes datasets}
\label{app:shapes_results_a}
% \caption{Overview of all results for the shapes dataset on the U-Net\label{tbl:all_attention_results}}
\resizebox{0.83\textwidth}{!}{
\begin{tabular}{|lrrrrrrrrrr|}
\hline
\textbf{TRF size} & \textbf{54} & \textbf{100} & \textbf{146} & \textbf{204} & \textbf{230} & \textbf{298} & \textbf{360} & \textbf{412} & \textbf{486} & \textbf{570} \\ \hline
& & & & & & & & & & \\
\textit{A} & & & & & & & & & & \\ \hline
Training time (epochs) & 63 & 66 & 111 & 71 & 69 & 71 & 200 & 74 & 192 & 72 \\
ERF rate before training & 0.0514 & 0.0131 & 0.0104 & 0.0033 & 0.0078 & 0.0043 & 0.0080 & 0.0005 & 0.0015 & 0.0044 \\
ERF rate & 0.0028 & 0.0527 & 0.0009 & 0.0008 & 0.0008 & 0.0006 & 0.0002 & 0.0002 & 0.0002 & 0.0002 \\
Dice score & 1.0000 & 1.0000 & 1.0000 & 1.0000 & 1.0000 & 1.0000 & 1.0000 & 1.0000 & 1.0000 & 1.0000 \\
Object rate & 100.5928 & 25.1482 & 11.1770 & 6.0224 & 4.0237 & 2.6766 & 1.6424 & 1.4741 & 0.9636 & 0.6692 \\
Accuracy & 1.0000 & 1.0000 & 1.0000 & 1.0000 & 1.0000 & 1.0000 & 1.0000 & 1.0000 & 1.0000 & 1.0000 \\
Sensitivity & 1.0000 & 1.0000 & 1.0000 & 1.0000 & 1.0000 & 1.0000 & 1.0000 & 1.0000 & 1.0000 & 1.0000 \\
Specificity & 1.0000 & 1.0000 & 1.0000 & 1.0000 & 1.0000 & 1.0000 & 1.0000 & 1.0000 & 1.0000 & 1.0000 \\
Jaccard index & 1.0000 & 1.0000 & 1.0000 & 1.0000 & 1.0000 & 1.0000 & 1.0000 & 1.0000 & 1.0000 & 1.0000 \\
& & & & & & & & & & \\
\textit{A contour} & & & & & & & & & & \\ \hline
Training time (epochs) & 14 & 17 & 87 & 18 & 200 & 47 & 200 & 39 & 49 & 117 \\
ERF rate before training & 0.0353 & 0.0138 & 0.0088 & 0.0038 & 0.0059 & 0.0020 & 0.0040 & 0.0004 & 0.0010 & 0.0011 \\
ERF rate & 0.4710 & 0.0502 & 0.0392 & 0.0048 & 0.0047 & 0.0029 & 0.0036 & 0.0005 & 0.0013 & 0.0017 \\
Dice score & 0.8219 & 0.9791 & 0.9998 & 0.9996 & 0.9996 & 0.9997 & 0.9998 & 0.9996 & 0.9997 & 0.9999 \\
Object rate & 98.8970 & 24.7242 & 10.9886 & 5.9208 & 3.9559 & 2.6315 & 1.6147 & 1.4492 & 0.9473 & 0.6579 \\
Accuracy & 0.9469 & 0.9930 & 0.9999 & 0.9999 & 0.9999 & 0.9999 & 0.9999 & 0.9999 & 0.9999 & 1.0000 \\
Sensitivity & 0.9505 & 0.9933 & 0.9999 & 0.9997 & 0.9996 & 0.9998 & 0.9998 & 0.9997 & 0.9997 & 0.9999 \\
Specificity & 0.9465 & 0.9929 & 1.0000 & 0.9999 & 0.9999 & 0.9999 & 1.0000 & 0.9999 & 0.9999 & 1.0000 \\
Jaccard index & 0.6983 & 0.9591 & 0.9996 & 0.9992 & 0.9992 & 0.9995 & 0.9997 & 0.9992 & 0.9994 & 0.9997 \\
& & & & & & & & & & \\
\textit{A large} & & & & & & & & & & \\ \hline
Training time (epochs) & 50 & 64 & 57 & 70 & 97 & 68 & 105 & 67 & 99 & 69 \\
ERF rate before training & 0.0326 & 0.0097 & 0.0115 & 0.0019 & 0.0058 & 0.0008 & 0.0047 & 0.0008 & 0.0020 & 0.0015 \\
ERF rate & 0.0028 & 0.0007 & 0.0004 & 0.0003 & 0.0002 & 0.0002 & 0.0001 & 0.0001 & 0.0001 & 0.0000 \\
Dice score & 1.0000 & 1.0000 & 1.0000 & 1.0000 & 1.0000 & 1.0000 & 1.0000 & 1.0000 & 1.0000 & 1.0000 \\
Object rate & 137.8993 & 34.4748 & 15.3221 & 8.2559 & 5.5160 & 3.6693 & 2.2515 & 2.0207 & 1.3209 & 0.9173 \\
Accuracy & 1.0000 & 1.0000 & 1.0000 & 1.0000 & 1.0000 & 1.0000 & 1.0000 & 1.0000 & 1.0000 & 1.0000 \\
Sensitivity & 1.0000 & 1.0000 & 1.0000 & 1.0000 & 1.0000 & 1.0000 & 1.0000 & 1.0000 & 1.0000 & 1.0000 \\
Specificity & 1.0000 & 1.0000 & 1.0000 & 1.0000 & 1.0000 & 1.0000 & 1.0000 & 1.0000 & 1.0000 & 1.0000 \\
Jaccard index & 1.0000 & 1.0000 & 1.0000 & 1.0000 & 1.0000 & 1.0000 & 1.0000 & 1.0000 & 1.0000 & 1.0000 \\
& & & & & & & & & & \\
\textit{A large contour} & & & & & & & & & & \\ \hline
Training time (epochs) & 25 & 31 & 56 & 13 & 24 & 199 & 19 & 12 & 25 & 33 \\
ERF rate before training & 0.0290 & 0.0148 & 0.0081 & 0.0037 & 0.0055 & 0.0031 & 0.0052 & 0.0006 & 0.0013 & 0.0014 \\
ERF rate & 0.4730 & 0.2601 & 0.8689 & 0.0028 & 0.0034 & 0.0015 & 0.0025 & 0.0003 & 0.0006 & 0.0007 \\
Dice score & 0.6708 & 0.8646 & 0.9841 & 0.9995 & 0.9997 & 0.9993 & 0.9998 & 0.9995 & 0.9994 & 0.9994 \\
Object rate & 138.7451 & 34.6863 & 15.4161 & 8.3065 & 5.5498 & 3.6918 & 2.2653 & 2.0331 & 1.3290 & 0.9229 \\
Accuracy & 0.8197 & 0.9119 & 0.9892 & 0.9997 & 0.9998 & 0.9995 & 0.9998 & 0.9997 & 0.9996 & 0.9996 \\
Sensitivity & 0.8949 & 0.9112 & 0.9912 & 0.9995 & 0.9997 & 0.9993 & 0.9998 & 0.9995 & 0.9995 & 0.9993 \\
Specificity & 0.8005 & 0.9123 & 0.9881 & 0.9998 & 0.9998 & 0.9996 & 0.9999 & 0.9998 & 0.9996 & 0.9998 \\
Jaccard index & 0.5056 & 0.7626 & 0.9688 & 0.9991 & 0.9994 & 0.9986 & 0.9995 & 0.9990 & 0.9987 & 0.9989 \\ \hline
\end{tabular}
}
\end{table*}

% \newpage
\begin{table*}[ht!]\centering
\def\arraystretch{1}
\section{Results of the regular U-Net model on the Type B shapes datasets}
\label{app:shapes_results_b}
\resizebox{0.83\textwidth}{!}{
% \caption{Overview of all results for the shapes dataset on the U-Net\label{tbl:all_attention_results}}
\begin{tabular}{|lrrrrrrrrrr|}
\hline
\textbf{TRF size} & \textbf{54} & \textbf{100} & \textbf{146} & \textbf{204} & \textbf{230} & \textbf{298} & \textbf{360} & \textbf{412} & \textbf{486} & \textbf{570} \\ \hline
& & & & & & & & & & \\
\textit{B large} & & & & & & & & & & \\ \hline
Training time (epochs) & 46 & 52 & 173 & 190 & 123 & 47 & 45 & 44 & 45 & 198 \\
ERF rate before training & 0.0387 & 0.0113 & 0.0128 & 0.0022 & 0.0105 & 0.0022 & 0.0054 & 0.0007 & 0.0019 & 0.0019 \\
ERF rate & 0.0918 & 0.0556 & 0.0016 & 0.0010 & 0.0005 & 0.0006 & 0.0002 & 0.0002 & 0.0001 & 0.0001 \\
Dice score & 0.9345 & 0.9959 & 1.0000 & 1.0000 & 1.0000 & 1.0000 & 0.9998 & 1.0000 & 0.9999 & 1.0000 \\
Object rate & 57.3475 & 14.3369 & 6.3719 & 3.4333 & 2.2939 & 1.5259 & 0.9363 & 0.8404 & 0.5493 & 0.3815 \\
Accuracy & 0.9843 & 0.9991 & 1.0000 & 1.0000 & 1.0000 & 1.0000 & 1.0000 & 1.0000 & 1.0000 & 1.0000 \\
Sensitivity & 0.8810 & 0.9921 & 1.0000 & 1.0000 & 1.0000 & 0.9999 & 0.9998 & 1.0000 & 0.9999 & 1.0000 \\
Specificity & 0.9994 & 1.0000 & 1.0000 & 1.0000 & 1.0000 & 1.0000 & 1.0000 & 1.0000 & 1.0000 & 1.0000 \\
Jaccard index & 0.8772 & 0.9920 & 1.0000 & 1.0000 & 1.0000 & 0.9999 & 0.9996 & 1.0000 & 0.9999 & 1.0000 \\
& & & & & & & & & & \\
\textit{B contour} & & & & & & & & & & \\ \hline
Training time (epochs) & 20 & 12 & 81 & 125 & 137 & 42 & 194 & 139 & 144 & 182 \\
ERF rate before training & 0.0277 & 0.0063 & 0.0059 & 0.0018 & 0.0061 & 0.0019 & 0.0046 & 0.0005 & 0.0013 & 0.0012 \\
ERF rate & 0.5588 & 0.3507 & 0.0383 & 0.0018 & 0.0042 & 0.0012 & 0.0033 & 0.0006 & 0.0022 & 0.0009 \\
Dice score & 0.8067 & 0.9751 & 0.9997 & 0.9998 & 0.9998 & 0.9995 & 0.9998 & 0.9997 & 0.9998 & 0.9998 \\
Object rate & 62.1263 & 15.5316 & 6.9029 & 3.7194 & 2.4851 & 1.6531 & 1.0143 & 0.9104 & 0.5951 & 0.4133 \\
Accuracy & 0.9620 & 0.9945 & 0.9999 & 1.0000 & 1.0000 & 0.9999 & 1.0000 & 0.9999 & 1.0000 & 1.0000 \\
Sensitivity & 0.9490 & 0.9961 & 0.9998 & 0.9998 & 0.9998 & 0.9994 & 0.9998 & 0.9997 & 0.9998 & 0.9998 \\
Specificity & 0.9633 & 0.9943 & 1.0000 & 1.0000 & 1.0000 & 0.9999 & 1.0000 & 1.0000 & 1.0000 & 1.0000 \\
Jaccard index & 0.6771 & 0.9515 & 0.9995 & 0.9996 & 0.9996 & 0.9989 & 0.9997 & 0.9995 & 0.9996 & 0.9997 \\
& & & & & & & & & & \\
\textit{B large} & & & & & & & & & & \\ \hline
Training time (epochs) & 34 & 67 & 63 & 184 & 28 & 59 & 143 & 35 & 110 & 199 \\
ERF rate before training & 0.0231 & 0.0077 & 0.0087 & 0.0023 & 0.0048 & 0.0030 & 0.0049 & 0.0005 & 0.0019 & 0.0011 \\
ERF rate & 0.0021 & 0.0016 & 0.0006 & 0.0002 & 0.0002 & 0.0002 & 0.0001 & 0.0001 & 0.0001 & 0.0001 \\
Dice score & 0.8703 & 0.9454 & 0.9933 & 1.0000 & 1.0000 & 1.0000 & 1.0000 & 1.0000 & 1.0000 & 1.0000 \\
Object rate & 85.1332 & 21.2833 & 9.4592 & 5.0968 & 3.4053 & 2.2653 & 1.3900 & 1.2475 & 0.8155 & 0.5663 \\
Accuracy & 0.9368 & 0.9756 & 0.9971 & 1.0000 & 1.0000 & 1.0000 & 1.0000 & 1.0000 & 1.0000 & 1.0000 \\
Sensitivity & 0.7825 & 0.9161 & 0.9933 & 1.0000 & 0.9999 & 1.0000 & 1.0000 & 1.0000 & 1.0000 & 1.0000 \\
Specificity & 0.9943 & 0.9937 & 0.9982 & 1.0000 & 1.0000 & 1.0000 & 1.0000 & 1.0000 & 1.0000 & 1.0000 \\
Jaccard index & 0.7706 & 0.8974 & 0.9868 & 1.0000 & 0.9999 & 0.9999 & 1.0000 & 0.9999 & 1.0000 & 1.0000 \\
& & & & & & & & & & \\
\textit{B large contour} & & & & & & & & & & \\ \hline
Training time (epochs) & 27 & 55 & 59 & 135 & 32 & 180 & 198 & 30 & 200 & 35 \\
ERF rate before training & 0.0119 & 0.0088 & 0.0064 & 0.0040 & 0.0049 & 0.0019 & 0.0040 & 0.0006 & 0.0016 & 0.0019 \\
ERF rate & 0.6198 & 0.2014 & 0.6863 & 0.0011 & 0.0062 & 0.0013 & 0.0027 & 0.0004 & 0.0008 & 0.0008 \\
Dice score & 0.6593 & 0.8793 & 0.9815 & 0.9998 & 0.9996 & 0.9998 & 0.9998 & 0.9994 & 0.9998 & 0.9993 \\
Object rate & 89.1812 & 22.2953 & 9.9090 & 5.3392 & 3.5672 & 2.3730 & 1.4561 & 1.3068 & 0.8543 & 0.5932 \\
Accuracy & 0.8838 & 0.9513 & 0.9919 & 0.9999 & 0.9998 & 0.9999 & 0.9999 & 0.9998 & 0.9999 & 0.9997 \\
Sensitivity & 0.9156 & 0.9584 & 0.9922 & 0.9998 & 0.9996 & 0.9998 & 0.9998 & 0.9995 & 0.9998 & 0.9989 \\
Specificity & 0.8795 & 0.9497 & 0.9918 & 0.9999 & 0.9999 & 0.9999 & 1.0000 & 0.9998 & 0.9999 & 0.9999 \\
Jaccard index & 0.4939 & 0.7860 & 0.9637 & 0.9996 & 0.9991 & 0.9996 & 0.9996 & 0.9989 & 0.9996 & 0.9986 \\ \hline
\end{tabular}
}
\end{table*}

\subsection{Disclosures}
No conflicts of interest. 

\subsection{Code, Data, and Materials Availability}
The complete source code utilized in this work can be accessed via our GitHub repository at \url{https://github.com/vinloo/u-net-receptive-field-study}. In addition to this, we have developed an open-source tool designed to calculate and suggest an appropriate TRF size based on a specified U-Net configuration and dataset. This tool is intended to aid researchers and practitioners in the field and is included in the repository.  

% \subsection{Acknowledgments}
% Acknowledgments and funding information should be added after the conclusion, and before references. Include grant numbers and the full name of the funding body. The acknowledgments section does not have a section number.

% \disclosures 
% \subsection*{Disclosures}
% No Disclosures.

% \subsection* {Code, Data, and Materials Availability} 
% In support of open scientific exchange, SPIE journals require Code, Data, and Materials Availability Statements in all accepted papers. This requirement went into effect on 1 May 2023. These statements should describe how to access any data that would be required to replicate or interpret the findings reported in the paper. Authors are encouraged to make the data and code related to the manuscript publicly available whenever possible, and utilize repositories that are well-known to the field (FigShare, Github, CodeOcean, etc.). If the data or code cannot be made publicly available, the authors should state the reason and explain how it can be obtained. Likewise, if data sharing is not applicable, the statement must say so. Example statements may be found in the Author Guidelines for the journal.

% \subsection* {Acknowledgments}
% This unnumbered section is used to identify those who have aided the authors in understanding or accomplishing the work presented and to acknowledge sources of funding. 

%%%%% References %%%%%

\bibliography{report}   % bibliography data in report.bib
\bibliographystyle{spiejour}   % makes bibtex use spiejour.bst

%%%%% Biographies of authors %%%%%

% \vspace{2ex}\noindent\textbf{First Author} is an assistant professor at the University of Optical Engineering. He received his BS and MS degrees in physics from the University of Optics in 1985 and 1987, respectively, and his PhD degree in optics from the Institute of Technology in 1991.  He is the author of more than 50 journal papers and has written three book chapters. His current research interests include optical interconnects, holography, and optoelectronic systems. He is a member of SPIE.

\vspace{1ex}
% \noindent Biographies and photographs of the other authors are not available.

\listoffigures
\listoftables

\end{spacing}
\end{document}

%% file: images/analysis-shapes.pgf
%% Creator: Matplotlib, PGF backend
%%
%% To include the figure in your LaTeX document, write
%%   \input{<filename>.pgf}
%%
%% Make sure the required packages are loaded in your preamble
%%   \usepackage{pgf}
%%
%% Also ensure that all the required font packages are loaded; for instance,
%% the lmodern package is sometimes necessary when using math font.
%%   \usepackage{lmodern}
%%
%% Figures using additional raster images can only be included by \input if
%% they are in the same directory as the main LaTeX file. For loading figures
%% from other directories you can use the `import` package
%%   \usepackage{import}
%%
%% and then include the figures with
%%   \import{<path to file>}{<filename>.pgf}
%%
%% Matplotlib used the following preamble
%%   
%%   \makeatletter\@ifpackageloaded{underscore}{}{\usepackage[strings]{underscore}}\makeatother
%%
\begingroup%
\makeatletter%
\begin{pgfpicture}%
\pgfpathrectangle{\pgfpointorigin}{\pgfqpoint{6.000000in}{3.000000in}}%
\pgfusepath{use as bounding box, clip}%
\begin{pgfscope}%
\pgfsetbuttcap%
\pgfsetmiterjoin%
\definecolor{currentfill}{rgb}{1.000000,1.000000,1.000000}%
\pgfsetfillcolor{currentfill}%
\pgfsetlinewidth{0.000000pt}%
\definecolor{currentstroke}{rgb}{1.000000,1.000000,1.000000}%
\pgfsetstrokecolor{currentstroke}%
\pgfsetdash{}{0pt}%
\pgfpathmoveto{\pgfqpoint{0.000000in}{0.000000in}}%
\pgfpathlineto{\pgfqpoint{6.000000in}{0.000000in}}%
\pgfpathlineto{\pgfqpoint{6.000000in}{3.000000in}}%
\pgfpathlineto{\pgfqpoint{0.000000in}{3.000000in}}%
\pgfpathlineto{\pgfqpoint{0.000000in}{0.000000in}}%
\pgfpathclose%
\pgfusepath{fill}%
\end{pgfscope}%
\begin{pgfscope}%
\pgfsetbuttcap%
\pgfsetmiterjoin%
\definecolor{currentfill}{rgb}{1.000000,1.000000,1.000000}%
\pgfsetfillcolor{currentfill}%
\pgfsetlinewidth{0.000000pt}%
\definecolor{currentstroke}{rgb}{0.000000,0.000000,0.000000}%
\pgfsetstrokecolor{currentstroke}%
\pgfsetstrokeopacity{0.000000}%
\pgfsetdash{}{0pt}%
\pgfpathmoveto{\pgfqpoint{0.673149in}{0.549691in}}%
\pgfpathlineto{\pgfqpoint{2.992901in}{0.549691in}}%
\pgfpathlineto{\pgfqpoint{2.992901in}{2.641667in}}%
\pgfpathlineto{\pgfqpoint{0.673149in}{2.641667in}}%
\pgfpathlineto{\pgfqpoint{0.673149in}{0.549691in}}%
\pgfpathclose%
\pgfusepath{fill}%
\end{pgfscope}%
\begin{pgfscope}%
\pgfsetbuttcap%
\pgfsetroundjoin%
\definecolor{currentfill}{rgb}{0.000000,0.000000,0.000000}%
\pgfsetfillcolor{currentfill}%
\pgfsetlinewidth{0.803000pt}%
\definecolor{currentstroke}{rgb}{0.000000,0.000000,0.000000}%
\pgfsetstrokecolor{currentstroke}%
\pgfsetdash{}{0pt}%
\pgfsys@defobject{currentmarker}{\pgfqpoint{0.000000in}{-0.048611in}}{\pgfqpoint{0.000000in}{0.000000in}}{%
\pgfpathmoveto{\pgfqpoint{0.000000in}{0.000000in}}%
\pgfpathlineto{\pgfqpoint{0.000000in}{-0.048611in}}%
\pgfusepath{stroke,fill}%
}%
\begin{pgfscope}%
\pgfsys@transformshift{0.673149in}{0.549691in}%
\pgfsys@useobject{currentmarker}{}%
\end{pgfscope}%
\end{pgfscope}%
\begin{pgfscope}%
\definecolor{textcolor}{rgb}{0.000000,0.000000,0.000000}%
\pgfsetstrokecolor{textcolor}%
\pgfsetfillcolor{textcolor}%
\pgftext[x=0.673149in,y=0.452469in,,top]{\color{textcolor}\rmfamily\fontsize{10.000000}{12.000000}\selectfont \(\displaystyle {0}\)}%
\end{pgfscope}%
\begin{pgfscope}%
\pgfsetbuttcap%
\pgfsetroundjoin%
\definecolor{currentfill}{rgb}{0.000000,0.000000,0.000000}%
\pgfsetfillcolor{currentfill}%
\pgfsetlinewidth{0.803000pt}%
\definecolor{currentstroke}{rgb}{0.000000,0.000000,0.000000}%
\pgfsetstrokecolor{currentstroke}%
\pgfsetdash{}{0pt}%
\pgfsys@defobject{currentmarker}{\pgfqpoint{0.000000in}{-0.048611in}}{\pgfqpoint{0.000000in}{0.000000in}}{%
\pgfpathmoveto{\pgfqpoint{0.000000in}{0.000000in}}%
\pgfpathlineto{\pgfqpoint{0.000000in}{-0.048611in}}%
\pgfusepath{stroke,fill}%
}%
\begin{pgfscope}%
\pgfsys@transformshift{1.833025in}{0.549691in}%
\pgfsys@useobject{currentmarker}{}%
\end{pgfscope}%
\end{pgfscope}%
\begin{pgfscope}%
\definecolor{textcolor}{rgb}{0.000000,0.000000,0.000000}%
\pgfsetstrokecolor{textcolor}%
\pgfsetfillcolor{textcolor}%
\pgftext[x=1.833025in,y=0.452469in,,top]{\color{textcolor}\rmfamily\fontsize{10.000000}{12.000000}\selectfont \(\displaystyle {288}\)}%
\end{pgfscope}%
\begin{pgfscope}%
\pgfsetbuttcap%
\pgfsetroundjoin%
\definecolor{currentfill}{rgb}{0.000000,0.000000,0.000000}%
\pgfsetfillcolor{currentfill}%
\pgfsetlinewidth{0.803000pt}%
\definecolor{currentstroke}{rgb}{0.000000,0.000000,0.000000}%
\pgfsetstrokecolor{currentstroke}%
\pgfsetdash{}{0pt}%
\pgfsys@defobject{currentmarker}{\pgfqpoint{0.000000in}{-0.048611in}}{\pgfqpoint{0.000000in}{0.000000in}}{%
\pgfpathmoveto{\pgfqpoint{0.000000in}{0.000000in}}%
\pgfpathlineto{\pgfqpoint{0.000000in}{-0.048611in}}%
\pgfusepath{stroke,fill}%
}%
\begin{pgfscope}%
\pgfsys@transformshift{2.992901in}{0.549691in}%
\pgfsys@useobject{currentmarker}{}%
\end{pgfscope}%
\end{pgfscope}%
\begin{pgfscope}%
\definecolor{textcolor}{rgb}{0.000000,0.000000,0.000000}%
\pgfsetstrokecolor{textcolor}%
\pgfsetfillcolor{textcolor}%
\pgftext[x=2.992901in,y=0.452469in,,top]{\color{textcolor}\rmfamily\fontsize{10.000000}{12.000000}\selectfont \(\displaystyle {576}\)}%
\end{pgfscope}%
\begin{pgfscope}%
\definecolor{textcolor}{rgb}{0.000000,0.000000,0.000000}%
\pgfsetstrokecolor{textcolor}%
\pgfsetfillcolor{textcolor}%
\pgftext[x=1.833025in,y=0.273457in,,top]{\color{textcolor}\rmfamily\fontsize{10.000000}{12.000000}\selectfont TRF Size}%
\end{pgfscope}%
\begin{pgfscope}%
\pgfsetbuttcap%
\pgfsetroundjoin%
\definecolor{currentfill}{rgb}{0.000000,0.000000,0.000000}%
\pgfsetfillcolor{currentfill}%
\pgfsetlinewidth{0.803000pt}%
\definecolor{currentstroke}{rgb}{0.000000,0.000000,0.000000}%
\pgfsetstrokecolor{currentstroke}%
\pgfsetdash{}{0pt}%
\pgfsys@defobject{currentmarker}{\pgfqpoint{-0.048611in}{0.000000in}}{\pgfqpoint{-0.000000in}{0.000000in}}{%
\pgfpathmoveto{\pgfqpoint{-0.000000in}{0.000000in}}%
\pgfpathlineto{\pgfqpoint{-0.048611in}{0.000000in}}%
\pgfusepath{stroke,fill}%
}%
\begin{pgfscope}%
\pgfsys@transformshift{0.673149in}{0.840243in}%
\pgfsys@useobject{currentmarker}{}%
\end{pgfscope}%
\end{pgfscope}%
\begin{pgfscope}%
\definecolor{textcolor}{rgb}{0.000000,0.000000,0.000000}%
\pgfsetstrokecolor{textcolor}%
\pgfsetfillcolor{textcolor}%
\pgftext[x=0.398457in, y=0.792018in, left, base]{\color{textcolor}\rmfamily\fontsize{10.000000}{12.000000}\selectfont \(\displaystyle {0.7}\)}%
\end{pgfscope}%
\begin{pgfscope}%
\pgfsetbuttcap%
\pgfsetroundjoin%
\definecolor{currentfill}{rgb}{0.000000,0.000000,0.000000}%
\pgfsetfillcolor{currentfill}%
\pgfsetlinewidth{0.803000pt}%
\definecolor{currentstroke}{rgb}{0.000000,0.000000,0.000000}%
\pgfsetstrokecolor{currentstroke}%
\pgfsetdash{}{0pt}%
\pgfsys@defobject{currentmarker}{\pgfqpoint{-0.048611in}{0.000000in}}{\pgfqpoint{-0.000000in}{0.000000in}}{%
\pgfpathmoveto{\pgfqpoint{-0.000000in}{0.000000in}}%
\pgfpathlineto{\pgfqpoint{-0.048611in}{0.000000in}}%
\pgfusepath{stroke,fill}%
}%
\begin{pgfscope}%
\pgfsys@transformshift{0.673149in}{1.421348in}%
\pgfsys@useobject{currentmarker}{}%
\end{pgfscope}%
\end{pgfscope}%
\begin{pgfscope}%
\definecolor{textcolor}{rgb}{0.000000,0.000000,0.000000}%
\pgfsetstrokecolor{textcolor}%
\pgfsetfillcolor{textcolor}%
\pgftext[x=0.398457in, y=1.373122in, left, base]{\color{textcolor}\rmfamily\fontsize{10.000000}{12.000000}\selectfont \(\displaystyle {0.8}\)}%
\end{pgfscope}%
\begin{pgfscope}%
\pgfsetbuttcap%
\pgfsetroundjoin%
\definecolor{currentfill}{rgb}{0.000000,0.000000,0.000000}%
\pgfsetfillcolor{currentfill}%
\pgfsetlinewidth{0.803000pt}%
\definecolor{currentstroke}{rgb}{0.000000,0.000000,0.000000}%
\pgfsetstrokecolor{currentstroke}%
\pgfsetdash{}{0pt}%
\pgfsys@defobject{currentmarker}{\pgfqpoint{-0.048611in}{0.000000in}}{\pgfqpoint{-0.000000in}{0.000000in}}{%
\pgfpathmoveto{\pgfqpoint{-0.000000in}{0.000000in}}%
\pgfpathlineto{\pgfqpoint{-0.048611in}{0.000000in}}%
\pgfusepath{stroke,fill}%
}%
\begin{pgfscope}%
\pgfsys@transformshift{0.673149in}{2.002452in}%
\pgfsys@useobject{currentmarker}{}%
\end{pgfscope}%
\end{pgfscope}%
\begin{pgfscope}%
\definecolor{textcolor}{rgb}{0.000000,0.000000,0.000000}%
\pgfsetstrokecolor{textcolor}%
\pgfsetfillcolor{textcolor}%
\pgftext[x=0.398457in, y=1.954227in, left, base]{\color{textcolor}\rmfamily\fontsize{10.000000}{12.000000}\selectfont \(\displaystyle {0.9}\)}%
\end{pgfscope}%
\begin{pgfscope}%
\pgfsetbuttcap%
\pgfsetroundjoin%
\definecolor{currentfill}{rgb}{0.000000,0.000000,0.000000}%
\pgfsetfillcolor{currentfill}%
\pgfsetlinewidth{0.803000pt}%
\definecolor{currentstroke}{rgb}{0.000000,0.000000,0.000000}%
\pgfsetstrokecolor{currentstroke}%
\pgfsetdash{}{0pt}%
\pgfsys@defobject{currentmarker}{\pgfqpoint{-0.048611in}{0.000000in}}{\pgfqpoint{-0.000000in}{0.000000in}}{%
\pgfpathmoveto{\pgfqpoint{-0.000000in}{0.000000in}}%
\pgfpathlineto{\pgfqpoint{-0.048611in}{0.000000in}}%
\pgfusepath{stroke,fill}%
}%
\begin{pgfscope}%
\pgfsys@transformshift{0.673149in}{2.583556in}%
\pgfsys@useobject{currentmarker}{}%
\end{pgfscope}%
\end{pgfscope}%
\begin{pgfscope}%
\definecolor{textcolor}{rgb}{0.000000,0.000000,0.000000}%
\pgfsetstrokecolor{textcolor}%
\pgfsetfillcolor{textcolor}%
\pgftext[x=0.398457in, y=2.535331in, left, base]{\color{textcolor}\rmfamily\fontsize{10.000000}{12.000000}\selectfont \(\displaystyle {1.0}\)}%
\end{pgfscope}%
\begin{pgfscope}%
\definecolor{textcolor}{rgb}{0.000000,0.000000,0.000000}%
\pgfsetstrokecolor{textcolor}%
\pgfsetfillcolor{textcolor}%
\pgftext[x=0.342901in,y=1.595679in,,bottom,rotate=90.000000]{\color{textcolor}\rmfamily\fontsize{10.000000}{12.000000}\selectfont Dice Score}%
\end{pgfscope}%
\begin{pgfscope}%
\pgfpathrectangle{\pgfqpoint{0.673149in}{0.549691in}}{\pgfqpoint{2.319753in}{2.091976in}}%
\pgfusepath{clip}%
\pgfsetrectcap%
\pgfsetroundjoin%
\pgfsetlinewidth{1.505625pt}%
\definecolor{currentstroke}{rgb}{0.121569,0.466667,0.705882}%
\pgfsetstrokecolor{currentstroke}%
\pgfsetdash{}{0pt}%
\pgfpathmoveto{\pgfqpoint{0.890626in}{2.583494in}}%
\pgfpathlineto{\pgfqpoint{1.075884in}{2.583555in}}%
\pgfpathlineto{\pgfqpoint{1.261142in}{2.583556in}}%
\pgfpathlineto{\pgfqpoint{1.494728in}{2.583554in}}%
\pgfpathlineto{\pgfqpoint{1.599439in}{2.583554in}}%
\pgfpathlineto{\pgfqpoint{1.873298in}{2.583549in}}%
\pgfpathlineto{\pgfqpoint{2.122994in}{2.583553in}}%
\pgfpathlineto{\pgfqpoint{2.332416in}{2.583552in}}%
\pgfpathlineto{\pgfqpoint{2.630440in}{2.583552in}}%
\pgfpathlineto{\pgfqpoint{2.968737in}{2.583545in}}%
\pgfusepath{stroke}%
\end{pgfscope}%
\begin{pgfscope}%
\pgfpathrectangle{\pgfqpoint{0.673149in}{0.549691in}}{\pgfqpoint{2.319753in}{2.091976in}}%
\pgfusepath{clip}%
\pgfsetrectcap%
\pgfsetroundjoin%
\pgfsetlinewidth{1.505625pt}%
\definecolor{currentstroke}{rgb}{1.000000,0.498039,0.054902}%
\pgfsetstrokecolor{currentstroke}%
\pgfsetdash{}{0pt}%
\pgfpathmoveto{\pgfqpoint{0.890626in}{2.583556in}}%
\pgfpathlineto{\pgfqpoint{1.075884in}{2.583556in}}%
\pgfpathlineto{\pgfqpoint{1.261142in}{2.583556in}}%
\pgfpathlineto{\pgfqpoint{1.494728in}{2.583556in}}%
\pgfpathlineto{\pgfqpoint{1.599439in}{2.583556in}}%
\pgfpathlineto{\pgfqpoint{1.873298in}{2.583556in}}%
\pgfpathlineto{\pgfqpoint{2.122994in}{2.583556in}}%
\pgfpathlineto{\pgfqpoint{2.332416in}{2.583556in}}%
\pgfpathlineto{\pgfqpoint{2.630440in}{2.583556in}}%
\pgfpathlineto{\pgfqpoint{2.968737in}{2.583556in}}%
\pgfusepath{stroke}%
\end{pgfscope}%
\begin{pgfscope}%
\pgfpathrectangle{\pgfqpoint{0.673149in}{0.549691in}}{\pgfqpoint{2.319753in}{2.091976in}}%
\pgfusepath{clip}%
\pgfsetrectcap%
\pgfsetroundjoin%
\pgfsetlinewidth{1.505625pt}%
\definecolor{currentstroke}{rgb}{0.172549,0.627451,0.172549}%
\pgfsetstrokecolor{currentstroke}%
\pgfsetdash{}{0pt}%
\pgfpathmoveto{\pgfqpoint{0.890626in}{1.548609in}}%
\pgfpathlineto{\pgfqpoint{1.075884in}{2.462105in}}%
\pgfpathlineto{\pgfqpoint{1.261142in}{2.582394in}}%
\pgfpathlineto{\pgfqpoint{1.494728in}{2.581232in}}%
\pgfpathlineto{\pgfqpoint{1.599439in}{2.581232in}}%
\pgfpathlineto{\pgfqpoint{1.873298in}{2.581813in}}%
\pgfpathlineto{\pgfqpoint{2.122994in}{2.582394in}}%
\pgfpathlineto{\pgfqpoint{2.332416in}{2.581232in}}%
\pgfpathlineto{\pgfqpoint{2.630440in}{2.581813in}}%
\pgfpathlineto{\pgfqpoint{2.968737in}{2.582975in}}%
\pgfusepath{stroke}%
\end{pgfscope}%
\begin{pgfscope}%
\pgfpathrectangle{\pgfqpoint{0.673149in}{0.549691in}}{\pgfqpoint{2.319753in}{2.091976in}}%
\pgfusepath{clip}%
\pgfsetrectcap%
\pgfsetroundjoin%
\pgfsetlinewidth{1.505625pt}%
\definecolor{currentstroke}{rgb}{0.839216,0.152941,0.156863}%
\pgfsetstrokecolor{currentstroke}%
\pgfsetdash{}{0pt}%
\pgfpathmoveto{\pgfqpoint{0.890626in}{0.670561in}}%
\pgfpathlineto{\pgfqpoint{1.075884in}{1.796741in}}%
\pgfpathlineto{\pgfqpoint{1.261142in}{2.491161in}}%
\pgfpathlineto{\pgfqpoint{1.494728in}{2.580651in}}%
\pgfpathlineto{\pgfqpoint{1.599439in}{2.581813in}}%
\pgfpathlineto{\pgfqpoint{1.873298in}{2.579489in}}%
\pgfpathlineto{\pgfqpoint{2.122994in}{2.582394in}}%
\pgfpathlineto{\pgfqpoint{2.332416in}{2.580651in}}%
\pgfpathlineto{\pgfqpoint{2.630440in}{2.580070in}}%
\pgfpathlineto{\pgfqpoint{2.968737in}{2.580070in}}%
\pgfusepath{stroke}%
\end{pgfscope}%
\begin{pgfscope}%
\pgfsetrectcap%
\pgfsetmiterjoin%
\pgfsetlinewidth{0.803000pt}%
\definecolor{currentstroke}{rgb}{0.000000,0.000000,0.000000}%
\pgfsetstrokecolor{currentstroke}%
\pgfsetdash{}{0pt}%
\pgfpathmoveto{\pgfqpoint{0.673149in}{0.549691in}}%
\pgfpathlineto{\pgfqpoint{0.673149in}{2.641667in}}%
\pgfusepath{stroke}%
\end{pgfscope}%
\begin{pgfscope}%
\pgfsetrectcap%
\pgfsetmiterjoin%
\pgfsetlinewidth{0.803000pt}%
\definecolor{currentstroke}{rgb}{0.000000,0.000000,0.000000}%
\pgfsetstrokecolor{currentstroke}%
\pgfsetdash{}{0pt}%
\pgfpathmoveto{\pgfqpoint{2.992901in}{0.549691in}}%
\pgfpathlineto{\pgfqpoint{2.992901in}{2.641667in}}%
\pgfusepath{stroke}%
\end{pgfscope}%
\begin{pgfscope}%
\pgfsetrectcap%
\pgfsetmiterjoin%
\pgfsetlinewidth{0.803000pt}%
\definecolor{currentstroke}{rgb}{0.000000,0.000000,0.000000}%
\pgfsetstrokecolor{currentstroke}%
\pgfsetdash{}{0pt}%
\pgfpathmoveto{\pgfqpoint{0.673149in}{0.549691in}}%
\pgfpathlineto{\pgfqpoint{2.992901in}{0.549691in}}%
\pgfusepath{stroke}%
\end{pgfscope}%
\begin{pgfscope}%
\pgfsetrectcap%
\pgfsetmiterjoin%
\pgfsetlinewidth{0.803000pt}%
\definecolor{currentstroke}{rgb}{0.000000,0.000000,0.000000}%
\pgfsetstrokecolor{currentstroke}%
\pgfsetdash{}{0pt}%
\pgfpathmoveto{\pgfqpoint{0.673149in}{2.641667in}}%
\pgfpathlineto{\pgfqpoint{2.992901in}{2.641667in}}%
\pgfusepath{stroke}%
\end{pgfscope}%
\begin{pgfscope}%
\definecolor{textcolor}{rgb}{0.000000,0.000000,0.000000}%
\pgfsetstrokecolor{textcolor}%
\pgfsetfillcolor{textcolor}%
\pgftext[x=1.833025in,y=2.725000in,,base]{\color{textcolor}\rmfamily\fontsize{12.000000}{14.400000}\selectfont (a) Shapes type A}%
\end{pgfscope}%
\begin{pgfscope}%
\pgfsetbuttcap%
\pgfsetmiterjoin%
\definecolor{currentfill}{rgb}{1.000000,1.000000,1.000000}%
\pgfsetfillcolor{currentfill}%
\pgfsetfillopacity{0.800000}%
\pgfsetlinewidth{1.003750pt}%
\definecolor{currentstroke}{rgb}{0.800000,0.800000,0.800000}%
\pgfsetstrokecolor{currentstroke}%
\pgfsetstrokeopacity{0.800000}%
\pgfsetdash{}{0pt}%
\pgfpathmoveto{\pgfqpoint{1.501387in}{0.619136in}}%
\pgfpathlineto{\pgfqpoint{2.895679in}{0.619136in}}%
\pgfpathquadraticcurveto{\pgfqpoint{2.923457in}{0.619136in}}{\pgfqpoint{2.923457in}{0.646913in}}%
\pgfpathlineto{\pgfqpoint{2.923457in}{1.422376in}}%
\pgfpathquadraticcurveto{\pgfqpoint{2.923457in}{1.450154in}}{\pgfqpoint{2.895679in}{1.450154in}}%
\pgfpathlineto{\pgfqpoint{1.501387in}{1.450154in}}%
\pgfpathquadraticcurveto{\pgfqpoint{1.473610in}{1.450154in}}{\pgfqpoint{1.473610in}{1.422376in}}%
\pgfpathlineto{\pgfqpoint{1.473610in}{0.646913in}}%
\pgfpathquadraticcurveto{\pgfqpoint{1.473610in}{0.619136in}}{\pgfqpoint{1.501387in}{0.619136in}}%
\pgfpathlineto{\pgfqpoint{1.501387in}{0.619136in}}%
\pgfpathclose%
\pgfusepath{stroke,fill}%
\end{pgfscope}%
\begin{pgfscope}%
\pgfsetrectcap%
\pgfsetroundjoin%
\pgfsetlinewidth{1.505625pt}%
\definecolor{currentstroke}{rgb}{0.121569,0.466667,0.705882}%
\pgfsetstrokecolor{currentstroke}%
\pgfsetdash{}{0pt}%
\pgfpathmoveto{\pgfqpoint{1.529165in}{1.339043in}}%
\pgfpathlineto{\pgfqpoint{1.668054in}{1.339043in}}%
\pgfpathlineto{\pgfqpoint{1.806943in}{1.339043in}}%
\pgfusepath{stroke}%
\end{pgfscope}%
\begin{pgfscope}%
\definecolor{textcolor}{rgb}{0.000000,0.000000,0.000000}%
\pgfsetstrokecolor{textcolor}%
\pgfsetfillcolor{textcolor}%
\pgftext[x=1.918054in,y=1.290432in,left,base]{\color{textcolor}\rmfamily\fontsize{10.000000}{12.000000}\selectfont A (=1.0)}%
\end{pgfscope}%
\begin{pgfscope}%
\pgfsetrectcap%
\pgfsetroundjoin%
\pgfsetlinewidth{1.505625pt}%
\definecolor{currentstroke}{rgb}{1.000000,0.498039,0.054902}%
\pgfsetstrokecolor{currentstroke}%
\pgfsetdash{}{0pt}%
\pgfpathmoveto{\pgfqpoint{1.529165in}{1.137654in}}%
\pgfpathlineto{\pgfqpoint{1.668054in}{1.137654in}}%
\pgfpathlineto{\pgfqpoint{1.806943in}{1.137654in}}%
\pgfusepath{stroke}%
\end{pgfscope}%
\begin{pgfscope}%
\definecolor{textcolor}{rgb}{0.000000,0.000000,0.000000}%
\pgfsetstrokecolor{textcolor}%
\pgfsetfillcolor{textcolor}%
\pgftext[x=1.918054in,y=1.089043in,left,base]{\color{textcolor}\rmfamily\fontsize{10.000000}{12.000000}\selectfont A large}%
\end{pgfscope}%
\begin{pgfscope}%
\pgfsetrectcap%
\pgfsetroundjoin%
\pgfsetlinewidth{1.505625pt}%
\definecolor{currentstroke}{rgb}{0.172549,0.627451,0.172549}%
\pgfsetstrokecolor{currentstroke}%
\pgfsetdash{}{0pt}%
\pgfpathmoveto{\pgfqpoint{1.529165in}{0.943981in}}%
\pgfpathlineto{\pgfqpoint{1.668054in}{0.943981in}}%
\pgfpathlineto{\pgfqpoint{1.806943in}{0.943981in}}%
\pgfusepath{stroke}%
\end{pgfscope}%
\begin{pgfscope}%
\definecolor{textcolor}{rgb}{0.000000,0.000000,0.000000}%
\pgfsetstrokecolor{textcolor}%
\pgfsetfillcolor{textcolor}%
\pgftext[x=1.918054in,y=0.895370in,left,base]{\color{textcolor}\rmfamily\fontsize{10.000000}{12.000000}\selectfont A contour}%
\end{pgfscope}%
\begin{pgfscope}%
\pgfsetrectcap%
\pgfsetroundjoin%
\pgfsetlinewidth{1.505625pt}%
\definecolor{currentstroke}{rgb}{0.839216,0.152941,0.156863}%
\pgfsetstrokecolor{currentstroke}%
\pgfsetdash{}{0pt}%
\pgfpathmoveto{\pgfqpoint{1.529165in}{0.750308in}}%
\pgfpathlineto{\pgfqpoint{1.668054in}{0.750308in}}%
\pgfpathlineto{\pgfqpoint{1.806943in}{0.750308in}}%
\pgfusepath{stroke}%
\end{pgfscope}%
\begin{pgfscope}%
\definecolor{textcolor}{rgb}{0.000000,0.000000,0.000000}%
\pgfsetstrokecolor{textcolor}%
\pgfsetfillcolor{textcolor}%
\pgftext[x=1.918054in,y=0.701697in,left,base]{\color{textcolor}\rmfamily\fontsize{10.000000}{12.000000}\selectfont A large contour}%
\end{pgfscope}%
\begin{pgfscope}%
\pgfsetbuttcap%
\pgfsetmiterjoin%
\definecolor{currentfill}{rgb}{1.000000,1.000000,1.000000}%
\pgfsetfillcolor{currentfill}%
\pgfsetlinewidth{0.000000pt}%
\definecolor{currentstroke}{rgb}{0.000000,0.000000,0.000000}%
\pgfsetstrokecolor{currentstroke}%
\pgfsetstrokeopacity{0.000000}%
\pgfsetdash{}{0pt}%
\pgfpathmoveto{\pgfqpoint{3.426080in}{0.549691in}}%
\pgfpathlineto{\pgfqpoint{5.745833in}{0.549691in}}%
\pgfpathlineto{\pgfqpoint{5.745833in}{2.641667in}}%
\pgfpathlineto{\pgfqpoint{3.426080in}{2.641667in}}%
\pgfpathlineto{\pgfqpoint{3.426080in}{0.549691in}}%
\pgfpathclose%
\pgfusepath{fill}%
\end{pgfscope}%
\begin{pgfscope}%
\pgfsetbuttcap%
\pgfsetroundjoin%
\definecolor{currentfill}{rgb}{0.000000,0.000000,0.000000}%
\pgfsetfillcolor{currentfill}%
\pgfsetlinewidth{0.803000pt}%
\definecolor{currentstroke}{rgb}{0.000000,0.000000,0.000000}%
\pgfsetstrokecolor{currentstroke}%
\pgfsetdash{}{0pt}%
\pgfsys@defobject{currentmarker}{\pgfqpoint{0.000000in}{-0.048611in}}{\pgfqpoint{0.000000in}{0.000000in}}{%
\pgfpathmoveto{\pgfqpoint{0.000000in}{0.000000in}}%
\pgfpathlineto{\pgfqpoint{0.000000in}{-0.048611in}}%
\pgfusepath{stroke,fill}%
}%
\begin{pgfscope}%
\pgfsys@transformshift{3.426080in}{0.549691in}%
\pgfsys@useobject{currentmarker}{}%
\end{pgfscope}%
\end{pgfscope}%
\begin{pgfscope}%
\definecolor{textcolor}{rgb}{0.000000,0.000000,0.000000}%
\pgfsetstrokecolor{textcolor}%
\pgfsetfillcolor{textcolor}%
\pgftext[x=3.426080in,y=0.452469in,,top]{\color{textcolor}\rmfamily\fontsize{10.000000}{12.000000}\selectfont \(\displaystyle {0}\)}%
\end{pgfscope}%
\begin{pgfscope}%
\pgfsetbuttcap%
\pgfsetroundjoin%
\definecolor{currentfill}{rgb}{0.000000,0.000000,0.000000}%
\pgfsetfillcolor{currentfill}%
\pgfsetlinewidth{0.803000pt}%
\definecolor{currentstroke}{rgb}{0.000000,0.000000,0.000000}%
\pgfsetstrokecolor{currentstroke}%
\pgfsetdash{}{0pt}%
\pgfsys@defobject{currentmarker}{\pgfqpoint{0.000000in}{-0.048611in}}{\pgfqpoint{0.000000in}{0.000000in}}{%
\pgfpathmoveto{\pgfqpoint{0.000000in}{0.000000in}}%
\pgfpathlineto{\pgfqpoint{0.000000in}{-0.048611in}}%
\pgfusepath{stroke,fill}%
}%
\begin{pgfscope}%
\pgfsys@transformshift{4.585957in}{0.549691in}%
\pgfsys@useobject{currentmarker}{}%
\end{pgfscope}%
\end{pgfscope}%
\begin{pgfscope}%
\definecolor{textcolor}{rgb}{0.000000,0.000000,0.000000}%
\pgfsetstrokecolor{textcolor}%
\pgfsetfillcolor{textcolor}%
\pgftext[x=4.585957in,y=0.452469in,,top]{\color{textcolor}\rmfamily\fontsize{10.000000}{12.000000}\selectfont \(\displaystyle {288}\)}%
\end{pgfscope}%
\begin{pgfscope}%
\pgfsetbuttcap%
\pgfsetroundjoin%
\definecolor{currentfill}{rgb}{0.000000,0.000000,0.000000}%
\pgfsetfillcolor{currentfill}%
\pgfsetlinewidth{0.803000pt}%
\definecolor{currentstroke}{rgb}{0.000000,0.000000,0.000000}%
\pgfsetstrokecolor{currentstroke}%
\pgfsetdash{}{0pt}%
\pgfsys@defobject{currentmarker}{\pgfqpoint{0.000000in}{-0.048611in}}{\pgfqpoint{0.000000in}{0.000000in}}{%
\pgfpathmoveto{\pgfqpoint{0.000000in}{0.000000in}}%
\pgfpathlineto{\pgfqpoint{0.000000in}{-0.048611in}}%
\pgfusepath{stroke,fill}%
}%
\begin{pgfscope}%
\pgfsys@transformshift{5.745833in}{0.549691in}%
\pgfsys@useobject{currentmarker}{}%
\end{pgfscope}%
\end{pgfscope}%
\begin{pgfscope}%
\definecolor{textcolor}{rgb}{0.000000,0.000000,0.000000}%
\pgfsetstrokecolor{textcolor}%
\pgfsetfillcolor{textcolor}%
\pgftext[x=5.745833in,y=0.452469in,,top]{\color{textcolor}\rmfamily\fontsize{10.000000}{12.000000}\selectfont \(\displaystyle {576}\)}%
\end{pgfscope}%
\begin{pgfscope}%
\definecolor{textcolor}{rgb}{0.000000,0.000000,0.000000}%
\pgfsetstrokecolor{textcolor}%
\pgfsetfillcolor{textcolor}%
\pgftext[x=4.585957in,y=0.273457in,,top]{\color{textcolor}\rmfamily\fontsize{10.000000}{12.000000}\selectfont TRF Size}%
\end{pgfscope}%
\begin{pgfscope}%
\definecolor{textcolor}{rgb}{0.000000,0.000000,0.000000}%
\pgfsetstrokecolor{textcolor}%
\pgfsetfillcolor{textcolor}%
\pgftext[x=3.370525in,y=1.595679in,,bottom,rotate=90.000000]{\color{textcolor}\rmfamily\fontsize{10.000000}{12.000000}\selectfont Dice Score}%
\end{pgfscope}%
\begin{pgfscope}%
\pgfpathrectangle{\pgfqpoint{3.426080in}{0.549691in}}{\pgfqpoint{2.319752in}{2.091976in}}%
\pgfusepath{clip}%
\pgfsetrectcap%
\pgfsetroundjoin%
\pgfsetlinewidth{1.505625pt}%
\definecolor{currentstroke}{rgb}{0.121569,0.466667,0.705882}%
\pgfsetstrokecolor{currentstroke}%
\pgfsetdash{}{0pt}%
\pgfpathmoveto{\pgfqpoint{3.643557in}{2.202824in}}%
\pgfpathlineto{\pgfqpoint{3.828815in}{2.559947in}}%
\pgfpathlineto{\pgfqpoint{4.014073in}{2.583533in}}%
\pgfpathlineto{\pgfqpoint{4.247659in}{2.583548in}}%
\pgfpathlineto{\pgfqpoint{4.352371in}{2.583513in}}%
\pgfpathlineto{\pgfqpoint{4.626230in}{2.583328in}}%
\pgfpathlineto{\pgfqpoint{4.875926in}{2.582496in}}%
\pgfpathlineto{\pgfqpoint{5.085348in}{2.583501in}}%
\pgfpathlineto{\pgfqpoint{5.383372in}{2.583168in}}%
\pgfpathlineto{\pgfqpoint{5.721669in}{2.583496in}}%
\pgfusepath{stroke}%
\end{pgfscope}%
\begin{pgfscope}%
\pgfpathrectangle{\pgfqpoint{3.426080in}{0.549691in}}{\pgfqpoint{2.319752in}{2.091976in}}%
\pgfusepath{clip}%
\pgfsetrectcap%
\pgfsetroundjoin%
\pgfsetlinewidth{1.505625pt}%
\definecolor{currentstroke}{rgb}{1.000000,0.498039,0.054902}%
\pgfsetstrokecolor{currentstroke}%
\pgfsetdash{}{0pt}%
\pgfpathmoveto{\pgfqpoint{3.643557in}{1.829864in}}%
\pgfpathlineto{\pgfqpoint{3.828815in}{2.266273in}}%
\pgfpathlineto{\pgfqpoint{4.014073in}{2.544622in}}%
\pgfpathlineto{\pgfqpoint{4.247659in}{2.583556in}}%
\pgfpathlineto{\pgfqpoint{4.352371in}{2.583556in}}%
\pgfpathlineto{\pgfqpoint{4.626230in}{2.583556in}}%
\pgfpathlineto{\pgfqpoint{4.875926in}{2.583556in}}%
\pgfpathlineto{\pgfqpoint{5.085348in}{2.583556in}}%
\pgfpathlineto{\pgfqpoint{5.383372in}{2.583556in}}%
\pgfpathlineto{\pgfqpoint{5.721669in}{2.583556in}}%
\pgfusepath{stroke}%
\end{pgfscope}%
\begin{pgfscope}%
\pgfpathrectangle{\pgfqpoint{3.426080in}{0.549691in}}{\pgfqpoint{2.319752in}{2.091976in}}%
\pgfusepath{clip}%
\pgfsetrectcap%
\pgfsetroundjoin%
\pgfsetlinewidth{1.505625pt}%
\definecolor{currentstroke}{rgb}{0.172549,0.627451,0.172549}%
\pgfsetstrokecolor{currentstroke}%
\pgfsetdash{}{0pt}%
\pgfpathmoveto{\pgfqpoint{3.643557in}{1.460282in}}%
\pgfpathlineto{\pgfqpoint{3.828815in}{2.438861in}}%
\pgfpathlineto{\pgfqpoint{4.014073in}{2.581813in}}%
\pgfpathlineto{\pgfqpoint{4.247659in}{2.582394in}}%
\pgfpathlineto{\pgfqpoint{4.352371in}{2.582394in}}%
\pgfpathlineto{\pgfqpoint{4.626230in}{2.580651in}}%
\pgfpathlineto{\pgfqpoint{4.875926in}{2.582394in}}%
\pgfpathlineto{\pgfqpoint{5.085348in}{2.581813in}}%
\pgfpathlineto{\pgfqpoint{5.383372in}{2.582394in}}%
\pgfpathlineto{\pgfqpoint{5.721669in}{2.582394in}}%
\pgfusepath{stroke}%
\end{pgfscope}%
\begin{pgfscope}%
\pgfpathrectangle{\pgfqpoint{3.426080in}{0.549691in}}{\pgfqpoint{2.319752in}{2.091976in}}%
\pgfusepath{clip}%
\pgfsetrectcap%
\pgfsetroundjoin%
\pgfsetlinewidth{1.505625pt}%
\definecolor{currentstroke}{rgb}{0.839216,0.152941,0.156863}%
\pgfsetstrokecolor{currentstroke}%
\pgfsetdash{}{0pt}%
\pgfpathmoveto{\pgfqpoint{3.643557in}{0.603734in}}%
\pgfpathlineto{\pgfqpoint{3.828815in}{1.882163in}}%
\pgfpathlineto{\pgfqpoint{4.014073in}{2.476052in}}%
\pgfpathlineto{\pgfqpoint{4.247659in}{2.582394in}}%
\pgfpathlineto{\pgfqpoint{4.352371in}{2.581232in}}%
\pgfpathlineto{\pgfqpoint{4.626230in}{2.582394in}}%
\pgfpathlineto{\pgfqpoint{4.875926in}{2.582394in}}%
\pgfpathlineto{\pgfqpoint{5.085348in}{2.580070in}}%
\pgfpathlineto{\pgfqpoint{5.383372in}{2.582394in}}%
\pgfpathlineto{\pgfqpoint{5.721669in}{2.579489in}}%
\pgfusepath{stroke}%
\end{pgfscope}%
\begin{pgfscope}%
\pgfsetrectcap%
\pgfsetmiterjoin%
\pgfsetlinewidth{0.803000pt}%
\definecolor{currentstroke}{rgb}{0.000000,0.000000,0.000000}%
\pgfsetstrokecolor{currentstroke}%
\pgfsetdash{}{0pt}%
\pgfpathmoveto{\pgfqpoint{3.426080in}{0.549691in}}%
\pgfpathlineto{\pgfqpoint{3.426080in}{2.641667in}}%
\pgfusepath{stroke}%
\end{pgfscope}%
\begin{pgfscope}%
\pgfsetrectcap%
\pgfsetmiterjoin%
\pgfsetlinewidth{0.803000pt}%
\definecolor{currentstroke}{rgb}{0.000000,0.000000,0.000000}%
\pgfsetstrokecolor{currentstroke}%
\pgfsetdash{}{0pt}%
\pgfpathmoveto{\pgfqpoint{5.745833in}{0.549691in}}%
\pgfpathlineto{\pgfqpoint{5.745833in}{2.641667in}}%
\pgfusepath{stroke}%
\end{pgfscope}%
\begin{pgfscope}%
\pgfsetrectcap%
\pgfsetmiterjoin%
\pgfsetlinewidth{0.803000pt}%
\definecolor{currentstroke}{rgb}{0.000000,0.000000,0.000000}%
\pgfsetstrokecolor{currentstroke}%
\pgfsetdash{}{0pt}%
\pgfpathmoveto{\pgfqpoint{3.426080in}{0.549691in}}%
\pgfpathlineto{\pgfqpoint{5.745833in}{0.549691in}}%
\pgfusepath{stroke}%
\end{pgfscope}%
\begin{pgfscope}%
\pgfsetrectcap%
\pgfsetmiterjoin%
\pgfsetlinewidth{0.803000pt}%
\definecolor{currentstroke}{rgb}{0.000000,0.000000,0.000000}%
\pgfsetstrokecolor{currentstroke}%
\pgfsetdash{}{0pt}%
\pgfpathmoveto{\pgfqpoint{3.426080in}{2.641667in}}%
\pgfpathlineto{\pgfqpoint{5.745833in}{2.641667in}}%
\pgfusepath{stroke}%
\end{pgfscope}%
\begin{pgfscope}%
\definecolor{textcolor}{rgb}{0.000000,0.000000,0.000000}%
\pgfsetstrokecolor{textcolor}%
\pgfsetfillcolor{textcolor}%
\pgftext[x=4.585957in,y=2.725000in,,base]{\color{textcolor}\rmfamily\fontsize{12.000000}{14.400000}\selectfont (b) Shapes type B}%
\end{pgfscope}%
\begin{pgfscope}%
\pgfsetbuttcap%
\pgfsetmiterjoin%
\definecolor{currentfill}{rgb}{1.000000,1.000000,1.000000}%
\pgfsetfillcolor{currentfill}%
\pgfsetfillopacity{0.800000}%
\pgfsetlinewidth{1.003750pt}%
\definecolor{currentstroke}{rgb}{0.800000,0.800000,0.800000}%
\pgfsetstrokecolor{currentstroke}%
\pgfsetstrokeopacity{0.800000}%
\pgfsetdash{}{0pt}%
\pgfpathmoveto{\pgfqpoint{4.260106in}{0.619136in}}%
\pgfpathlineto{\pgfqpoint{5.648611in}{0.619136in}}%
\pgfpathquadraticcurveto{\pgfqpoint{5.676389in}{0.619136in}}{\pgfqpoint{5.676389in}{0.646913in}}%
\pgfpathlineto{\pgfqpoint{5.676389in}{1.407716in}}%
\pgfpathquadraticcurveto{\pgfqpoint{5.676389in}{1.435493in}}{\pgfqpoint{5.648611in}{1.435493in}}%
\pgfpathlineto{\pgfqpoint{4.260106in}{1.435493in}}%
\pgfpathquadraticcurveto{\pgfqpoint{4.232328in}{1.435493in}}{\pgfqpoint{4.232328in}{1.407716in}}%
\pgfpathlineto{\pgfqpoint{4.232328in}{0.646913in}}%
\pgfpathquadraticcurveto{\pgfqpoint{4.232328in}{0.619136in}}{\pgfqpoint{4.260106in}{0.619136in}}%
\pgfpathlineto{\pgfqpoint{4.260106in}{0.619136in}}%
\pgfpathclose%
\pgfusepath{stroke,fill}%
\end{pgfscope}%
\begin{pgfscope}%
\pgfsetrectcap%
\pgfsetroundjoin%
\pgfsetlinewidth{1.505625pt}%
\definecolor{currentstroke}{rgb}{0.121569,0.466667,0.705882}%
\pgfsetstrokecolor{currentstroke}%
\pgfsetdash{}{0pt}%
\pgfpathmoveto{\pgfqpoint{4.287884in}{1.331327in}}%
\pgfpathlineto{\pgfqpoint{4.426773in}{1.331327in}}%
\pgfpathlineto{\pgfqpoint{4.565661in}{1.331327in}}%
\pgfusepath{stroke}%
\end{pgfscope}%
\begin{pgfscope}%
\definecolor{textcolor}{rgb}{0.000000,0.000000,0.000000}%
\pgfsetstrokecolor{textcolor}%
\pgfsetfillcolor{textcolor}%
\pgftext[x=4.676773in,y=1.282716in,left,base]{\color{textcolor}\rmfamily\fontsize{10.000000}{12.000000}\selectfont B}%
\end{pgfscope}%
\begin{pgfscope}%
\pgfsetrectcap%
\pgfsetroundjoin%
\pgfsetlinewidth{1.505625pt}%
\definecolor{currentstroke}{rgb}{1.000000,0.498039,0.054902}%
\pgfsetstrokecolor{currentstroke}%
\pgfsetdash{}{0pt}%
\pgfpathmoveto{\pgfqpoint{4.287884in}{1.137654in}}%
\pgfpathlineto{\pgfqpoint{4.426773in}{1.137654in}}%
\pgfpathlineto{\pgfqpoint{4.565661in}{1.137654in}}%
\pgfusepath{stroke}%
\end{pgfscope}%
\begin{pgfscope}%
\definecolor{textcolor}{rgb}{0.000000,0.000000,0.000000}%
\pgfsetstrokecolor{textcolor}%
\pgfsetfillcolor{textcolor}%
\pgftext[x=4.676773in,y=1.089043in,left,base]{\color{textcolor}\rmfamily\fontsize{10.000000}{12.000000}\selectfont B large}%
\end{pgfscope}%
\begin{pgfscope}%
\pgfsetrectcap%
\pgfsetroundjoin%
\pgfsetlinewidth{1.505625pt}%
\definecolor{currentstroke}{rgb}{0.172549,0.627451,0.172549}%
\pgfsetstrokecolor{currentstroke}%
\pgfsetdash{}{0pt}%
\pgfpathmoveto{\pgfqpoint{4.287884in}{0.943981in}}%
\pgfpathlineto{\pgfqpoint{4.426773in}{0.943981in}}%
\pgfpathlineto{\pgfqpoint{4.565661in}{0.943981in}}%
\pgfusepath{stroke}%
\end{pgfscope}%
\begin{pgfscope}%
\definecolor{textcolor}{rgb}{0.000000,0.000000,0.000000}%
\pgfsetstrokecolor{textcolor}%
\pgfsetfillcolor{textcolor}%
\pgftext[x=4.676773in,y=0.895370in,left,base]{\color{textcolor}\rmfamily\fontsize{10.000000}{12.000000}\selectfont B contour}%
\end{pgfscope}%
\begin{pgfscope}%
\pgfsetrectcap%
\pgfsetroundjoin%
\pgfsetlinewidth{1.505625pt}%
\definecolor{currentstroke}{rgb}{0.839216,0.152941,0.156863}%
\pgfsetstrokecolor{currentstroke}%
\pgfsetdash{}{0pt}%
\pgfpathmoveto{\pgfqpoint{4.287884in}{0.750308in}}%
\pgfpathlineto{\pgfqpoint{4.426773in}{0.750308in}}%
\pgfpathlineto{\pgfqpoint{4.565661in}{0.750308in}}%
\pgfusepath{stroke}%
\end{pgfscope}%
\begin{pgfscope}%
\definecolor{textcolor}{rgb}{0.000000,0.000000,0.000000}%
\pgfsetstrokecolor{textcolor}%
\pgfsetfillcolor{textcolor}%
\pgftext[x=4.676773in,y=0.701697in,left,base]{\color{textcolor}\rmfamily\fontsize{10.000000}{12.000000}\selectfont B large contour}%
\end{pgfscope}%
\end{pgfpicture}%
\makeatother%
\endgroup%

%% file: article.bbl
\begin{thebibliography}{10}

\bibitem{litjens2017}
G.~Litjens, T.~Kooi, B.~E. Bejnordi, {\em et~al.}, ``A survey on deep learning in medical image analysis,'' {\em Medical Image Analysis} {\bf 42}, 60--88  (2017).

\bibitem{hesamian2019deep}
M.~H. Hesamian, W.~Jia, X.~He, {\em et~al.}, ``Deep learning techniques for medical image segmentation: achievements and challenges,'' {\em Journal of digital imaging} {\bf 32}, 582--596  (2019).

\bibitem{ronneberger2015unet}
O.~Ronneberger, P.~Fischer, and T.~Brox, ``U-net: Convolutional networks for biomedical image segmentation,''  (2015).

\bibitem{williams2023unified}
C.~Williams, F.~Falck, G.~Deligiannidis, {\em et~al.}, ``A unified framework for u-net design and analysis,''  (2023).

\bibitem{oktay2018attentionunet}
O.~Oktay, J.~Schlemper, L.~L. Folgoc, {\em et~al.}, ``Attention u-net: Learning where to look for the pancreas,''  (2018).

\bibitem{luo2017understanding}
W.~Luo, Y.~Li, R.~Urtasun, {\em et~al.}, ``Understanding the effective receptive field in deep convolutional neural networks,''  (2017).

\bibitem{araujo2019computing}
A.~Araujo, W.~Norris, and J.~Sim, ``Computing receptive fields of convolutional neural networks,'' {\em Distill}   (2019).
\newblock https://distill.pub/2019/computing-receptive-fields.

\bibitem{behboodi2020}
B.~Behboodi, M.~Fortin, C.~J. Belasso, {\em et~al.}, ``Receptive field size as a key design parameter for ultrasound image segmentation with u-net,'' in {\em 2020 42nd Annual International Conference of the IEEE Engineering in Medicine \& Biology Society (EMBC)},  2117--2120  (2020).

\bibitem{sytwu2022}
K.~Sytwu, C.~Groschner, and M.~C. Scott, ``{Understanding the Influence of Receptive Field and Network Complexity in Neural Network-Guided TEM Image Analysis},'' {\em Microscopy and Microanalysis} {\bf 28}, 1896--1904  (2022).

\bibitem{yu2016multiscale}
F.~Yu and V.~Koltun, ``Multi-scale context aggregation by dilated convolutions,''  (2016).

\bibitem{SARIGUL2019279}
M.~Sarıgül, B.~Ozyildirim, and M.~Avci, ``Differential convolutional neural network,'' {\em Neural Networks} {\bf 116}, 279--287  (2019).

\bibitem{convolutions}
V.~Dumoulin and F.~Visin, ``A guide to convolution arithmetic for deep learning,''  (2018).

\bibitem{skip}
J.~Wu, Y.~Zhang, K.~Wang, {\em et~al.}, ``Skip connection u-net for white matter hyperintensities segmentation from mri,'' {\em IEEE Access} {\bf 7}, 155194--155202  (2019).

\bibitem{Kuo2016}
C.-C.~J. Kuo, ``Understanding convolutional neural networks with a mathematical model,'' {\em Journal of Visual Communication and Image Representation} {\bf 41}, 406--413  (2016).

\bibitem{paszke2019pytorch}
A.~Paszke, S.~Gross, F.~Massa, {\em et~al.}, ``Pytorch: An imperative style, high-performance deep learning library,'' {\em Advances in neural information processing systems} {\bf 32}  (2019).

\bibitem{zhang2018improved}
Z.~Zhang, ``Improved adam optimizer for deep neural networks,'' in {\em 2018 IEEE/ACM 26th international symposium on quality of service (IWQoS)},  1--2, Ieee  (2018).

\bibitem{prechelt2002early}
L.~Prechelt, ``Early stopping-but when?,'' in {\em Neural Networks: Tricks of the trade},  55--69, Springer  (2002).

\bibitem{fetal_head}
T.~L.~A. van~den Heuvel, D.~de~Bruijn, C.~L. de~Korte, {\em et~al.}, ``{Automated measurement of fetal head circumference using 2D ultrasound images},''  (2018).

\bibitem{fetal_head_2.1}
Y.~Lu, J.~Bai, M.~Zhou, {\em et~al.}, ``Jnu-ifm,''  (2022).

\bibitem{fetal_head_2.2}
Y.~Lu, M.~Zhou, D.~Zhi, {\em et~al.}, ``The {JNU}-{IFM} dataset for segmenting pubic symphysis-fetal head,'' {\em Data in Brief} {\bf 41}, 107904  (2022).

\bibitem{kidney.1}
A.~J. Daniel, C.~E. Buchanan, T.~Allcock, {\em et~al.}, ``T2-weighted kidney mri segmentation,''  (2021).

\bibitem{kidney.2}
A.~J. Daniel, C.~E. Buchanan, T.~Allcock, {\em et~al.}, ``Automated renal segmentation in healthy and chronic kidney disease subjects using a convolutional neural network,'' {\em Magnetic Resonance in Medicine} {\bf 86}, 1125--1136  (2021).

\bibitem{lungs.1}
{Viacheslav Danilov}, ``Chest x-ray dataset for lung segmentation,''  (2022).

\bibitem{lungs.2}
R.~H. Kassamali and S.~Jafarieh, ``Passion and hard work produces high quality research in uk: response to focus on china: should clinicians engage in research? and lessons from other countries,'' {\em Quantitative Imaging in Medicine and Surgery} {\bf 4}(6)  (2014).

\bibitem{thyroid}
T.~Wunderling, B.~Golla, P.~Poudel, {\em et~al.}, ``Comparison of thyroid segmentation techniques for 3d ultrasound,'' in {\em Proceedings of SPIE Medical Imaging},  (Orlando, USA)  (2017).

\bibitem{nerve}
W.~C. Anna~Montoya, ``Ultrasound nerve segmentation,''  (2016).

\bibitem{metrics1}
A.~W. Setiawan, ``Image segmentation metrics in skin lesion: Accuracy, sensitivity, specificity, dice coefficient, jaccard index, and matthews correlation coefficient,'' in {\em 2020 International Conference on Computer Engineering, Network, and Intelligent Multimedia (CENIM)},  97--102  (2020).

\bibitem{metrics2}
D.~M\"{u}ller, I.~Soto-Rey, and F.~Kramer, ``Towards a guideline for evaluation metrics in medical image segmentation,'' {\em {BMC} Research Notes} {\bf 15}  (2022).

\bibitem{kdestimation}
{Weglarczyk, Stanislaw}, ``Kernel density estimation and its application,'' {\em ITM Web Conf.} {\bf 23}, 00037  (2018).

\bibitem{Silverman86}
B.~W. Silverman, {\em Density Estimation for Statistics and Data Analysis}, 47--48.
\newblock Chapman \& Hall, London  (1986).

\bibitem{Harpole2014}
J.~K. Harpole, C.~M. Woods, T.~L. Rodebaugh, {\em et~al.}, ``How bandwidth selection algorithms impact exploratory data analysis using kernel density estimation.,'' {\em Psychological Methods} {\bf 19}, 428--443  (2014).

\bibitem{yu2023multilayer}
Y.~Yu and Y.~Zhang, ``Multi-layer perceptron trainability explained via variability,''  (2023).

\bibitem{segnet}
V.~Badrinarayanan, A.~Kendall, and R.~Cipolla, ``Segnet: A deep convolutional encoder-decoder architecture for image segmentation,'' {\em IEEE transactions on pattern analysis and machine intelligence} {\bf 39}(12), 2481--2495  (2017).

\bibitem{pspnet}
H.~Zhao, J.~Shi, X.~Qi, {\em et~al.}, ``Pyramid scene parsing network,'' in {\em Proceedings of the IEEE conference on computer vision and pattern recognition},  2881--2890  (2017).

\bibitem{deeplab}
L.-C. Chen, G.~Papandreou, I.~Kokkinos, {\em et~al.}, ``Deeplab: Semantic image segmentation with deep convolutional nets, atrous convolution, and fully connected crfs,'' {\em IEEE transactions on pattern analysis and machine intelligence} {\bf 40}(4), 834--848  (2017).

\end{thebibliography}
